\documentclass[preprint]{JHEP3} 

\usepackage{epsfig}
\usepackage{amsfonts}
\usepackage{amssymb,amsmath}

\newcommand{\Tr}[1]{\:{\rm Tr}\,#1}

\newcommand{\mbf}[1]{{\boldsymbol {#1} }}
\newcommand{\complex}{{\mathbb C}} 
\newcommand{\zed}{{\mathbb Z}} 
\newcommand{\real}{{\mathbb R}} 

\def\e{{\,\rm e}\,}

\newcommand{\proj}{{\mathbb P}}

\newcommand{\K}{{\rm K}}
\def\ii{{\,{\rm i}\,}}
\def\dd{{\rm d}}

\newcommand{\sn}{\mathrm{sn}}
\newcommand{\cn}{\mathrm{cn}}
\newcommand{\dn}{\mathrm{dn}}

\def\beq{\begin{equation}}
\def\bee{\begin{equation}}
\def\eeq{\end{equation}}
\def\bea{\begin{eqnarray}}
\def\eea{\end{eqnarray}}
\def\bd{\begin{displaymath}}
\def\ed{\end{displaymath}}

\newcommand{\Cint}{\int\kern-10.5pt-\kern7pt}

\newcommand{\be}{\begin{equation}}
\newcommand{\ee}{\end{equation}}

\newcommand\fverb{\setbox\pippobox=\hbox\bgroup\verb}
\newcommand\fverbdo{\egroup\medskip\noindent%
                        \fbox{\unhbox\pippobox}\ }
\newcommand\fverbit{\egroup\item[\fbox{\unhbox\pippobox}]}
\newbox\pippobox

\newtheorem{theorem}{Theorem}

\title{Topological strings and large $\mbf N$ phase transitions I:
Nonchiral expansion of $\mbf q$-deformed Yang-Mills theory
}
\author{Nicola Caporaso$^{(a)}$, Michele Cirafici$^{(b)}$, Luca
Griguolo$^{(c)}$,
 Sara Pasquetti$^{(c)}$,  Domenico~Seminara$^{(a)}$ and
 Richard~J.~Szabo$^{(b)}$ \\
$^{(a)}$ Dipartimento di Fisica, Polo Scientifico Universit\`a di
Firenze,\\ INFN Sezione di Firenze
Via  G. Sansone 1, 50019 Sesto Fiorentino, Italy\\
$^{(b)}$ Department of Mathematics, Heriot-Watt University\\
Colin Maclaurin Building, Riccarton, Edinburgh EH14 4AS, UK\\
$^{(c)}$ Dipartimento di  Fisica, Universit\`a  di Parma,
INFN Gruppo Collegato di Parma\\
Parco Area delle Scienze 7/A, 43100 Parma, Italy\\
\email{caporaso@fi.infn.it , M.Cirafici@ma.hw.ac.uk , griguolo@fis.unipr.it ,
pasquetti@fis.unipr.it , seminara@fi.infn.it , R.J.Szabo@ma.hw.ac.uk } }

\received{\today}               

\accepted{\today}               
\preprint{ {\tt HWM-05-18} \ \ {\tt EMPG-05-15}
\\ \hepth{0509041}} 

\date{data}
\abstract{We examine the problem of counting bound states of BPS black
  holes on local Calabi-Yau threefolds which are fibrations over a Riemann
  surface by computing the partition function of
$q$-deformed Yang-Mills theory on the Riemann surface. We study in
detail the genus zero case and obtain, at finite $N$, the instanton
expansion of the gauge theory. It can be written exactly as the
partition function for $U(N)$ Chern-Simons gauge theory on a Lens
space, summed over all non-trivial vacua, plus a tower of
non-perturbative instanton contributions. The correspondence between
two and three dimensional gauge theories is elucidated by an
explicit mapping between two-dimensional Yang-Mills instantons and
flat connections on the Lens space. In the large $N$ limit we find a
peculiar phase structure in the model. At weak string coupling the
theory reduces exactly to the trivial flat connection sector with
instanton contributions exponentially suppressed, and the
topological string partition function on the resolved conifold is
reproduced in this regime. At a certain critical point all
non-trivial vacua contribute, instantons are enhanced and the theory
appears to undergo a phase transition into a strong coupling regime.
We rederive these results by performing a saddle-point approximation
to the exact partition function. We obtain a $q$-deformed version of
the Douglas-Kazakov equation for two-dimensional Yang-Mills theory
on the sphere, whose one-cut solution below the transition point
reproduces the resolved conifold geometry. Above the critical point
we propose a two-cut solution that should reproduce the
chiral-antichiral dynamics found for black holes on the Calabi-Yau
threefold and the Gross-Taylor string in the undeformed limit. The
transition from the strong coupling phase to the weak coupling phase
appears to be of third order.}

\begin{document}

\section{Introduction}

Topological string amplitudes have been recently connected to the
counting of microstates of four-dimensional BPS black holes in a
novel and highly non-trivial way \cite{Ooguri:2004zv}. This proposal extends
and generalizes a series of beautiful results \cite{LopesCardoso:1998wt,LopesCardoso:1999cv,LopesCardoso:1999ur,Mohaupt:2000mj}
concerning the entropy of BPS black holes arising in
compactifications of Type~II superstrings on Calabi-Yau threefolds.
A precise relation between the mixed ensemble black hole partition function
$Z_{\rm BH}$ and the topological string vacuum amplitude
$Z_{\rm top}$ has been conjectured as $Z_{\rm BH}$=$|Z_{\rm top}|^2$. In
particular, it has been suggested that for a large black hole
charge $N$ the relation should be valid at any order in the
$\frac1N$ expansion, taking into account the {\it perturbative} definition
of $Z_{\rm top}$. The proposal of \cite{Ooguri:2004zv} actually goes further,
pushing forward the possibility that {\it nonperturbative} topological
string amplitudes, including $O(\e^{-1/g_s})$ corrections,
could be defined from the conjectured relation.

\noindent

In order to check this proposal, one should find some Calabi-Yau
backgrounds in which both sides of the relation can be computed
independently. For this task a general class of suitable
non-compact Calabi-Yau threefolds has been recently studied
\cite{Aganagic:2004js}, generalizing the original example
presented in \cite{Vafa:2004qa}. The relevant threefold $X$ is the
total space of a rank 2 holomorphic vector bundle over a genus $g$
Riemann surface $\Sigma_g$, \beq X={\mathcal
O}(p+2g-2)\oplus{\mathcal O}(-p)~\longrightarrow~\Sigma_g \ ,
\label{totalhol}\eeq where ${\mathcal O}(m)$ is a holomorphic line
bundle of degree $m$ over $\Sigma_g$ (in \cite{Vafa:2004qa} the
case $g=1$ was considered). The counting of BPS states on these
geometries has been claimed to reduce to computing the partition
function of a peculiar deformation of Yang-Mills theory on
$\Sigma_g$ called $q$-deformed Yang-Mills theory. Starting from
this result one should ask if the relation with the perturbative
topological string amplitudes holds in this case. Happily the
partition function $Z_{\rm top}$ for these geometries has been
computed very recently \cite{Bryan:2004iq} and the consistency
check amounts to reproducing these amplitudes as the large $N$
limit of $q$-deformed Yang-Mills theory on $\Sigma_g$. In
\cite{Aganagic:2004js} a large $N$ expansion has been performed
and the conjecture was confirmed, but with a couple of important
subtleties. Firstly, one should include in the definition of
$Z_{\rm top}$ a sum over a $U(1)$ degree of freedom identified
with a Ramond-Ramond flux through the Riemann surface. Secondly,
and more importantly, the relevant topological string partition
function implies the presence of $|2g-2|$ stacks of D-branes
inserted in the fibers of $X$. An explanation of this unexpected
feature in terms of extra closed string moduli, related to the
non-compactness of $X$, has been offered in
\cite{Aganagic:2005dh}. Recent works on the conjecture include
\cite{Dabholkar:2004yr}--\cite{Dabholkar:2005dt}.

A central role in all these advances has been played by the $q$-deformed
$U(N)$ Yang-Mills theory. At finite $N$ it should reproduce the
counting of BPS states of a black hole that arises from $N$
D4-branes wrapping the submanifold $C_4={\mathcal O}(-p)\to\Sigma_g$ of $X$
with any number of D0-branes and D2-branes wrapping $\Sigma_g$.
This statement is equivalent to the surprising result that the
partition function of $q$-deformed Yang-Mills theory is the generating
functional of the Euler characteristics of the moduli spaces of
$U(N)$ instantons in the topologically twisted ${\cal N}=4$ Yang-Mills
theory on $C_4$ \cite{Aganagic:2004js}. It therefore provides instanton counting on
very non-trivial non-compact four-manifolds. On the other hand, its
large $N$ limit is described by the topological A-model on $X$, a
closed topological string theory with six-dimensional target
space. That two-dimensional Yang-Mills theory should be related to
a string theory in the large $N$ limit is not entirely
unexpected due to the well-known Gross-Taylor expansion
\cite{Gross:1992tu}--\cite{Gross:1993hu}. At large $N$, the partition function of two-dimensional
Yang-Mills theory on a Riemann surface $\Sigma_g$ almost factorizes into two
copies (called chiral and antichiral) of the same theory of unfolded
branched covering maps, the target space being $\Sigma_g$ itself.
The chiral-antichiral factorization is violated by some
geometrical structures called orientation-reversing tubes.

The emergence of chiral and antichiral sectors was also observed in
\cite{Aganagic:2004js} in studying the $q$-deformed version of two dimensional
Yang-Mills theory and it implies the appearance of the modulus squared
$|Z_{\rm top}|^2$, a crucial ingredient in checking the relation with
black hole physics. Trying to understand the relation between
Gross-Taylor string theory and the topological string theory
underlying $q$-deformed Yang-Mills theory at large $N$ is quite
tempting. Moreover, due to the intimate relation with
four-dimensional gauge theories, it is important to
understand better how topological strings emerge from two-dimensional gauge
degrees of freedom and whether or not the string
description, with its chiral-antichiral behaviour, is restricted to
a limited region of parameter space. It is well-known
that the familiar Yang-Mills theory on the sphere $S^2$ undergoes a
large $N$ phase transition at a particular value of the
coupling constant \cite{Douglas:1993ii}. A strong coupling phase, wherein the theory
admits the Gross-Taylor string description, is separated by a
weak-coupling phase with gaussian field theoretical behaviour.
Instanton configurations induce the transition to strong coupling
\cite{Gross:1994mr}, while the entropy of branch points appears to be responsible for
the divergence of the string expansion above the critical
point~\cite{Taylor:1994zm,Crescimanno:1994eg}.

In this paper we will study in detail $q$-deformed Yang-Mills theory
on $S^2$ and its relation with topological string theory on the
threefold $X={\mathcal O}(p-2)\oplus{\mathcal O}(-p)\to\proj^1$. We
will show that the theory still exhibits a phase transition at large
$N$. A companion paper~\cite{CGPS} will be devoted to a precise
comparison with topological strings on $X$, showing how this
description arises in the strong coupling phase. The present paper
focuses on the gauge theoretical aspects and on the determination of
the phase transition. We study the weak-coupling phase that is still
represented by a topological string but without chiral-antichiral
factorization. We also concentrate on a detailed description of the
strong-coupling regime, discussing both conceptual and technical
aspects.


The structure of the paper is as follows. In Sect.~2 we review the
relation between black hole entropy, topological string theory on
$X$ and $q$-deformed Yang-Mills theory. This section is intended
to be a brief introduction to the background of the subject which
may be skipped by the experts. We describe how the string coupling
$g_s$ produces the deformation and how the integer $p$ is related
to the effective area seen by the string theory. The expectations
of the large $N$ limit are also discussed.  In Sect.~3 we discuss
the gauge theoretical structure of the theory. In Sect.~3.1 we
describe the general properties of the $q$-deformed theory and
point out how in an appropriate double scaling limit,
$p\to\infty$, $g_s\to 0$ with $g_s p$ fixed, the usual Yang-Mills
theory is recovered. In Sect.~3.2 we obtain, at finite $N$, the
instanton expansion of the gauge theory. In this form the
comparison with the ${\cal N}=4$ topologically twisted theory
should be easier and we make some comments about it. In Sect.~3.3
we show that the partition function can be written exactly as the
partition function of $U(N)$ Chern-Simons theory on the Lens space
$L_p=S^3/\zed_p$, summed over all non-trivial vacua, plus a tower
of non-perturbative instanton contributions. The relation between
$q$-deformed Yang-Mills theory and Chern-Simons theory on Seifert
manifolds \cite{Beasley:2005vf} is thereby unveiled in this case.
We make this relation explicit by explicitly constructing the
mapping between instantons on $S^2$ and flat connections on $L_p$.
We also discuss the possible description in terms of open
topological strings, suggested by the presence of the underlying
Chern-Simons theory. In Sect.~4 we observe a peculiar behaviour
dependent on the integer $p$ specifying the Calabi-Yau threefold
and the string coupling constant $g_s$. In the large $N$ limit,
for all $p\geq1$ and $g_sN<p\,\log ({\sec (\frac{\pi }{p})}^2)$,
the theory completely reduces to the trivial flat connection
sector and all instanton contributions are exponentially
suppressed. This behaviour is reminescent of the weak-coupling
phase appearing in the undeformed case at small area. Not
surprisingly, due to the relation with Chern-Simons theory on
$L_p$, the topological string partition function on the resolved
conifold is reproduced in this regime. Instead, for $p>2$ and
$g_sN>p\,\log ({\sec (\frac{\pi }{p})}^2)$ all non-trivial vacua
contribute, instantons are enhanced and the theory undergoes a
phase transition to a strong coupling regime. These results are
obtained by expliciting evaluating the ratio between the
one-instanton and the zero-instanton contributions to the
partition function, in the spirit of~\cite{Gross:1994mr}. In
Sect.~5 we arrive at the same conclusions by performing a saddle
point approximation to the exact partition function at large~$N$.
In Sect.~5.1 a deformed version of the Douglas-Kazakov equation is
derived and is related to the undeformed one as $p\to \infty$. In
Sect.~5.2 we show that its one-cut solution, valid below the
transition point, reproduces the resolved conifold geometry found
in the weak-coupling phase. Sect. 6 contains the most important
results: we study the theory above the critical value. In
Sect.~6.1 we propose an exact two-cut solution of the saddle-point
equation, that should reproduce the chiral-antichiral dynamics
found for black-holes on $X$ and the Gross-Taylor string for
$p\to\infty$. The equations for the end-points of the cuts are
rather complicated, involving elliptic functions of the third
type. Nevertheless they can be written in an elegant way, reducing
to the Douglas-Kazakov equations in the $p\to \infty$ limit
(Sect.~6.2). We show that for $p\leq 2$ they do not admit
solutions, while above the critical value, for $p>2$, they always
do uniquely. The transition curve is then recovered in Sect.~6.3.
A third-order phase transition, generalizing non-trivially the
Douglas-Kazakov result to the $q$-deformed case, is finally found
in Sect.~6.4, by using an expansion in modular functions around
the critical point. We also give evidences of the arising of the
topological string expansion at large 't Hooft parameters in
Sect.~6.5, by using the modular properties of the exact solution.
In Sect.~7 we draw our
conclusions and speculate on possible applications of these results
to black hole physics. Three appendices at the end of the paper
contain technical details of the computations presented in the main
text.

{\it Note added:} as   this manuscript was being completed, Refs
\cite{Arsiwalla:2005jb} and  \cite{Jafferis:2005jd} appeared,
presenting an overlap with the results of this paper.

\section{Black holes, topological strings and $\mbf q$-deformed Yang-Mills
  theory}

We start by reviewing the conjecture presented in
\cite{Ooguri:2004zv}. Consider Type~II superstring theory on
$X\times\real^{3,1}$, where $X$ is a Calabi-Yau threefold. A BPS
black hole can be obtained, in this context, by wrapping D6, D4, D2
and D0 branes around holomorphic cycles in $X$. The charges carried
by the black holes are determined by appropriately choosing the
holomorphic cycles. Usually D6 and D4-brane charges are referred to
as ``magnetic'' while D2 and D0-brane charges are ``electric'', with
the intersection pairings in $X$ giving rise to electric-magnetic
duality in four dimensional space. One can define a partition
function for a mixed ensemble of BPS black hole states by fixing the
magnetic charges $Q_6$ and $Q_4$ and summing over the D2 and D0
charges with fixed chemical potentials $\phi_2$ and $\phi_0$ to get
\begin{equation}
Z_{\rm
  BH}(Q_6,Q_4,\phi_2,\phi_0)=\sum_{Q_2,Q_0}\Omega(Q_6,Q_4,Q_2,Q_0)
~\exp\bigl[-Q_2\phi_2-Q_0\phi_0\bigr] \ , \label{ZBH}
\end{equation}
where $\Omega(Q_6,Q_4,Q_2,Q_0)$ is the contribution from BPS states
with fixed D-brane charges. The conjecture relates the
partition function (\ref{ZBH}) to the topological string vacuum
amplitude as
\begin{equation}Z_{\rm BH}(Q_6,Q_4,\phi_2,\phi_0)=\bigl|Z_{\rm
    top}(g_s,t_s)\bigr|^2 \ ,
\end{equation}
where $Z_{\rm top}(g_s,t_s)$ is the A-model topological string
partition function with the identifications
\begin{equation}
g_s=\frac{4\pi \ii}{\frac{\ii}{\pi}\,\phi_0+Q_6} \ ,
\,\,\,\,\,\,\,\,\,\,t_s=
\mbox{$\frac{1}{2}$}\,g_s\bigl(\mbox{$-\frac{\ii}{\pi}$}\, \phi_2+N
Q_4\bigr)
\end{equation}
for the topological string coupling $g_s$ and the K\"ahler modulus
$t_s$. For recent reviews on topological strings, see
\cite{Neitzke:2004ni,Vonk:2005yv}

This proposal can be considered as an all orders generalization of
some well-known properties of BPS black holes in ${\cal N}=2$
supergravity~\cite{LopesCardoso:1998wt,LopesCardoso:1999cv,LopesCardoso:1999ur,Mohaupt:2000mj}.
Black hole solutions are found in the background of $2 n_v+2$ gauge
fields ($n_v+1$ magnetic duals of the others) and they carry charges
$(P^I,Q_I)$ with respect to them. The gauge fields are organized
into $n_v$ vector multiplets plus one graviphoton (arising from the
supergravity multiplet). The scalar fields of the theory $X^I$, that
may be regarded as position dependent moduli of the Calabi-Yau
threefold on which superstrings have been compactified, have fixed
values at the black hole horizon determined only by the charges.
This phenomenon is called the attractor
mechanism~\cite{Ferrara:1995ih,Strominger:1996kf}. In turn this
relation can be expressed in purely geometric terms by using the
periods of the holomorphic three-form $\Omega$ along the cycles
$A_I,B^I$ of the Calabi-Yau threefold at the horizon as
\begin{equation}P^I={\rm Re}(X^I)=\oint_{A_I}{\rm Re}(\Omega) \ ,
\,\,\,\,Q_I={\rm Re}(F_I)=\oint_{B^I}{\rm Re}(\Omega) \  ,
\label{Peri}
\end{equation}
where $F_I=\frac{\partial F_0}{\partial X^I}$ and $F_0$ is
the prepotential of $X$. The
black hole entropy $S_{\rm BH}$ is a function only of the charges in
the extremal case. By means of the explicit expressions in
eq.~(\ref{Peri}) one obtains
\begin{equation}S_{\rm BH}\simeq Q_I X^I-P^IF_I \ .
\end{equation} The entropy thus appears as a Legendre transform
of the prepotential $F_0$, which is also
known to be the genus zero free energy of topological strings in
the Calabi-Yau background. At the supergravity level it is possible to
include higher-derivative corrections proportional to
$R^2T^{2g-2}$, where $T$ is the graviphoton field-strength, and to
compute the black hole solutions and their entropies
\cite{LopesCardoso:1998wt,LopesCardoso:1999cv,LopesCardoso:1999ur,Mohaupt:2000mj}. At the first non-trivial order ($g=1$) the
relation with the topological string free energy still holds when one
includes quantum corrections to the prepotential coming from one-loop
amplitudes \cite{Antoniadis:1993ze,Bershadsky:1993cx}. The conjecture of \cite{Ooguri:2004zv} is
consistent with these results and cleverly generalizes them at all
orders in the perturbative topological string expansion. However, the
proposal is even somewhat more startling. Because the partition
function $Z_{\rm BH}$ makes sense also at the nonperturbative level,
eq.~(\ref{ZBH}) should provide a nonperturbative definition of the
topological string partition function. In particular the presence of a
square-modulus signals a breaking of holomorphicity at the
nonperturbative level. We will come back to this point later on.

It is of course natural to attempt to check this conjecture in some
explicit examples. One requires a Calabi-Yau threefold $X$ on which
to engineer a BPS black hole whose partition function could be
computed by a counting of microstates, while at the same time being
simple enough to enable the computation of the topological string
partition function to all orders. While for compact manifolds the
task seems out of reach presently, in the non-compact case there is
the general class of threefolds (\ref{totalhol}) on which the
problem has been attacked \cite{Aganagic:2004js,Vafa:2004qa}. The
study of topological strings on these backgrounds and the related
counting of microstates have also produced a number interesting
independent results. As we will see, different gauge theories in
diverse dimensions appear to be related by their common
gravitational ancestor.

\subsection{Counting microstates in ${\cal N}=4$ and $q$-deformed
Yang-Mills theories}

Let us begin by describing the counting of microstates. It consists of
counting bound states of D4, D2 and D0-branes, where the D4 branes
wrap the four cycle $C_4$ which is the total space of the holomorphic
line bundle
\begin{equation}
C_4={\mathcal O}(-p)~\longrightarrow~\Sigma_g
\end{equation}
and the D2-branes wrap the Riemann surface $\Sigma_g$. The number of
D4-branes is fixed to be $N$ and one should count in the ensemble of
bound states on it. The natural way of doing the computation is by studying the
relevant gauge theory on the brane, where the presence of chemical
potentials for D2 and D0-branes corresponds to the
introduction of some interactions. According to the general
framework \cite{Bershadsky:1995qy} the worldvolume gauge theory on the $N$
D4-branes is the ${\cal N}=4$ topologically twisted $U(N)$
Yang-Mills theory on $C_4$. The presence of chemical potentials is
simulated by turning on the observables in the theory given by
\begin{equation} \label{4daction}
S_{c}=\frac{1}{2 g_s}\,\int_{C_4}{\rm Tr}\bigl(F\wedge
F\bigr)+\frac{\theta}{g_s}\,\int_{C_4}{\rm Tr}\bigl(F\wedge K\bigr)
\ ,
\end{equation}
where $F$ is the Yang-Mills field strength and $K$ is the unit volume
form of $\Sigma_g$. The relation between the gauge parameters
$g_s$, $\theta$ and the chemical potentials $\phi_0$, $\phi_2$ is
\begin{equation}
\phi_0=\frac{4\pi^2}{g_s} \ ,\,\,\,\,\,\,
\phi_2=\frac{2\pi\, \theta}{g_s} \ ,\label{chem}
\end{equation}
in accordance with the identifications for the D0 and D2-brane charges
$q_0$, $q_2$ as
\begin{equation}
q_0=\frac{1}{8\pi^2}\,\int_{C_4}{\rm Tr}\bigl(F\wedge F\bigr) \ ,
\,\,\,\,\,\,q_2=\frac{1}{2\pi}\,\int_{C_4}{\rm Tr}\bigl(F\wedge
K\bigr) \ .
\end{equation}

Obtaining $Z_{\rm BH}$ is therefore equivalent to computing the
expectation value in topologically twisted ${\cal N}=4$
Yang-Mills theory given by
\begin{equation}
Z_{\rm BH}=\Bigl\langle\exp\Bigl[-\frac{1}{2 g_s}\,\int_{C_4}{\rm
Tr}\bigl(F\wedge F\bigr)-\frac{\theta}{g_s}\,\int_{C_4}{\rm
Tr}\bigl(F\wedge K\bigr)\Bigr]\Bigr\rangle=Z_{{\cal N}=4} \ .
\end{equation}
The general structure of this functional integral has
been explored in~\cite{Vafa:1994tf}. There it was shown
that with an appropriate gauge fixing the partition function $Z_{{\cal
    N}=4}$ has an expansion of the form
\begin{equation}
Z_{{\cal
N}=4}=\sum_{q_0,q_2}\Omega(q_0,q_2;N)~
\exp\Bigl(-\frac{4\pi^2}{g_s}\,q_0-\frac{2\pi\,\theta}{g_s}\,q_2\Bigr) \
, \label{N4}
\end{equation}
where $\Omega(q_0,q_2;N)$ is (under suitable assumptions) the
Euler characteristic of the moduli space of $U(N)$ instantons on
$C_4$ in the topological sector labelled by the zeroth and second
Chern numbers $q_0$ and $q_2$. We see that the counting of
microstates is equivalent to an instanton counting in the ${\cal
N}=4$ topological gauge theory. This is of course still a
formidable problem, because no general strategy exists in the case
of non-compact manifolds and very few results
\cite{Bruzzo:2002xf,Fucito:2004ry,nakajima} have been obtained in
this context. Moreover, we expect that $C_4$ has a very
complicated instanton moduli space, especially in the higher genus
case.

A key observation~\cite{Vafa:2004qa} allows one to reduce the
computation to a two-dimensional problem. The ${\cal N}=4$
topological gauge theory is believed to be invariant under certain
massive deformations which drastically simplify the theory. By using
a further deformation which corresponds to a $U(1)$ rotation on
${\cal O}(-p)$, it was argued in \cite{Vafa:2004qa} that the theory
localizes to $U(1)$-invariant modes and reduces to an effective
gauge theory on $\Sigma_g$. One could expect that the gauge theory
is still fully topological from the two-dimensional point of view.
In \cite{Vafa:2004qa} it was shown that the non-triviality of the
fibration ${\cal O}(-p)$ generates an extra term in the effective
two-dimensional action which basically contains all the information
about the four-dimensional structure. It is given by \beq
S_p=-\frac{p}{2g_s}\,\int_{\Sigma_g}{\rm Tr}\,\Phi^2~K  \ , \eeq
where \beq \Phi(z)=\oint_{S^1_{z, |u|=\infty}}A \eeq parameterizes
the holonomy of the gauge field $A$ around a circle at infinity in
the fiber over the point $z\in\Sigma_g$, with $u$ a local complex
coordinate of ${\cal O}(-p)$. The relevant two-dimensional action
becomes
\begin{equation}
S_{{\rm YM}_2}=\frac{1}{g_s}\,\int_{\Sigma_g}{\rm Tr}\bigl(\Phi\,
F\bigr)+\frac{\theta}{g_s}\,\int_{\Sigma_g}{\rm Tr}\,\Phi~K
-\frac{p}{2g_s}\,\int_{\Sigma_g}{\rm Tr}\,\Phi^2 K.\label{aqd}
\end{equation}
This is just the action of two-dimensional Yang-Mills
theory on the Riemann surface $\Sigma_g$. In the case that
$\Sigma_g$ is the torus $T^2$ ($g=1$), ordinary two-dimensional
Yang-Mills theory provides the correct instanton counting on ${\cal
O}(-p)\to T^2$~\cite{Vafa:2004qa}.

However, in the general case there is an important subtlety. The
two-dimensional path-integral should take into account the new
degree of freedom $\Phi$, which is periodic due to its origin as
the holonomy of the gauge field at infinity. It can be shown
\cite{Aganagic:2004js} that periodicity affects the path integral
measure in a well-defined way. As a consequence, the emerging
theory has a natural interpretation as a peculiar $q$-deformation
of two-dimensional Yang-Mills theory. This deformation has already
been introduced in \cite{Klimcik:1999kg} with different
motivations. The exact form of the partition function obtained by
performing the path integral with the action (\ref{aqd}) and
periodic measure can be derived~\cite{Aganagic:2004js} in a spirit
similar to~\cite{Blau:1993tv} or in a combinatorial approach
resembling that of \cite{Witten:1992xu}. The final result is
\begin{equation}
Z_{{\cal N}=4}=Z^{q}_{\rm YM}=\sum_{R}{\rm
dim}_q(R)^{2-2g}~q^{\frac{p}{2}\,C_2(R)}~\e^{\ii\theta\,
C_1(R)}\label{pqd}
\end{equation}
This is to be compared with the partition function of ordinary
Yang-Mills theory on $\Sigma_g$ given by the Migdal expansion
\cite{Migdal:1975zg} \begin{equation} Z_{\rm YM}=\sum_{R}{\rm
dim}(R)^{2-2g}~\e^{\frac{g^2A}{2}\,C_2(R)}~\e^{\ii\theta\,
C_1(R)} \ ,\label{migdal}
\end{equation}
where $R$ runs through the unitary irreducible representations of the gauge group
$U(N)$, ${\rm dim}(R)$ is its dimension, and
$C_1(R)$ and $C_2(R)$ are respectively its first and second
Casimir invariants. The dimensionless combination $g^2A$ of the Yang-Mills
coupling constant $g^2$ and area $A$ of $\Sigma_g$ is the effective coupling of the theory, as
dictated by invariance under area-preserving diffeomorphisms.
With $R_i$ labelling the lengths of the rows in the Young tableau corresponding
to the irreducible representation $R$, the main effect of the deformation is to turn the
ordinary dimension of the $R$ given by
\begin{equation}
{\rm dim}(R)=\prod_{1\leq i<j\leq N}\frac{R_i-R_j+j-i}{j-i}
\end{equation}
into the {\it quantum} dimension
\begin{equation}
{\rm
dim}_q(R)=\prod_{1\leq i<j\leq N}\frac{\bigl[R_i-R_j+j-i\bigr]_q}{\bigl[j-i\bigr]_q}
=\prod_{1\leq i<j\leq N}\frac{\bigl[q^{(R_i-R_j+j-i)/2}-q^{-(R_i-R_j+j-i)/2}\bigr]}
{\bigl[q^{(j-i)/2}-q^{-(j-i)/2}\bigr]} \ ,
\end{equation}
where the deformation parameter $q$ is related to the coupling $g_s$
through \beq q=\e^{-g_s} \ . \eeq Clearly as $g_s\to 0$ the quantum
dimension goes smoothly into the classical one. The effective
dimensionless coupling of the gauge theory is $g^2A=g_sp$.
The deformation arises only for $g\neq1$ and the
conclusions of \cite{Vafa:2004qa} still hold.

Eq.~(\ref{pqd}) provides the solution of the difficult
problem of instanton counting, as it represents
the answer on all four-manifolds $C_4$. The direct comparison with eq.
(\ref{N4}) seems at first sight puzzling, because the expansion of
the $q$-deformed gauge theory is in $\e^{-g_s}$ and not in
$\e^{-{1}/{g_s}}$. This implies that a modular transformation is required in
eq.~(\ref{pqd}) and we will discuss this point in Sect.~3.

\subsection{Large $N$ limit and topological strings}

Let us come back now to the conjecture. Eq.~(\ref{pqd}) gives, in
principle, the exact expression for the black hole partition function $Z_{\rm BH}$. In order to establish
the connection with the perturbative topological string partition
function on $X$, we need to take the limit of large charges and
consider the large $N$ limit of the $q$-deformed
Yang-Mills theory. It is important, first of all, to set the
relation between the parameters of the gauge theory describing the
black hole and the data of the closed topological string theory.
According to the conjecture, the moduli of the Calabi-Yau manifold
are fixed by the black hole attractor mechanism. The real parts of
the projective coordinates ($X^0$, $X^1$) on Calabi-Yau moduli
space are connected to the magnetic charges (D6 and D4-branes)
while their imaginary parts are the chemical potentials
$\phi_0$ and $\phi_2$ for the D0 and D2-brane charges. Here
D6-branes are absent and we have $N$ D4-branes. Their charges
should be measured in terms of electric units of D2-branes
wrapping $\Sigma_g$. By evaluating the intersection number of
$\Sigma_g$ and the four-cycle $C_4$ on which the D4-branes are wrapped, it can be
shown that the relevant charge is $p+2g-2$. The projective moduli
for the closed topological string are therefore
\begin{equation}
X^0=\ii\frac{\phi_0}{\pi}\ , \,\,\,\,\,\,
X^1=(p+2g-2)N-\ii\frac{\phi_2}{\pi} \ ,
\end{equation}
which from
eq.~(\ref{chem}) implies
\begin{equation}
X^0=\frac{4\pi\ii}{g_s}  \ , \,\,\,\,\,\,
X^1=(p+2g-2)N-2\ii\frac{\theta}{g_s} \ .
\end{equation}
The K\"ahler modulus $t_s$ of the base $\Sigma_g$ is given by
$t_s=2\pi \ii X^1/X^0$. The closed topological string emerging from
the large $N$ limit is thereby expected to possess the modulus
\begin{equation}
t_s=(p+2g-2)\,\frac{Ng_s}{2}-\ii\theta \ . \end{equation} This is
the first very non-trivial feature that should be reproduced by the
large $N$ limit.

The second non-trivial point is that a square modulus
structure should emerge. This is not completely unexpected due to
the relation with the large $N$ limit of ordinary Yang-Mills
theory. In the Gross-Taylor expansion~\cite{Gross:1993hu} two types
of representations contribute to the limit, called the chiral
representations (with much less than $N$ Young tableaux boxes) and the antichiral
representations (with order $N$ boxes). The partition function is
almost factorized into two copies, apart from the contribution of
some geometrical structures called orientation-reversing
tubes that are required to complete the description. According to the same
logic, one could wonder whether $Z^q_{\rm YM}$ factorizes as well. The
choice of relevant representations is dictated by the Casimir
dependence of the partition function, which is unchanged by
the deformation. We expect
\begin{equation}
Z^q_{\rm YM}\simeq Z^{q,+}_{\rm YM}\,Z^{q,-}_{\rm YM} \ .
\end{equation}
Moreover, one would expect that the chiral $q$-deformed Yang-Mills
partition function $Z^{q,+}_{\rm YM}$ could be written as a
holomorphic function of $t_s$ and identified with the topological
string amplitude $Z_{\rm top}(g_s,t_s)$ on $X$. In
\cite{Aganagic:2004js} these expectations have been confirmed, but
with some important subtleties.

For genus $g>1$ one finds~\cite{Aganagic:2004js}
\begin{equation}
Z^q_{\rm YM}(\Sigma_{g})=\sum_{l=-\infty}^{\infty}~\sum_{\hat
R_1,\dots, \, \hat R_{2g-2}} Z^{q{\rm YM},+}_{ \hat R_1,\dots, \,
\hat R_{2g-2}}(t_s+p\,g_s l)\, Z^{q{\rm YM},-}_{\hat R_1,\dots, \,
\hat R_{2g-2}}(\,\bar{t}_s-p\,g_s l) \ ,
\end{equation}
where $\hat R_{i}$ are irreducible representations of $SU(N)$ and
the chiral block $Z^{q{\rm YM},+}_{ \hat R_1,\dots, \, \hat
R_{2g-2}}(t_s)$ agrees exactly with the perturbative topological
string amplitude on $X$ \cite{Bryan:2004iq} with $2g-2$ stacks of
D-branes inserted in the fiber. It depends explicitly on the
choice of $2g-2$ arbitrary Young tableaux which correspond to the
boundary degrees of freedom of the fiber D-branes. When all the
Young tableaux are taken to be trivial, one recovers the expected
closed topological string partition function. The chiral and
anti-chiral parts are sewn along the D-branes and summed over
them. The extra integer $l$ which shifts the K\"ahler modulus is
not the naive expectation which would take only the $l=0$ term and
neglect the fiber D-brane contributions.

For $g=0$, the case studied extensively in this paper, the result
is slightly different and given by
\begin{equation}
Z^q_{\rm YM}(S^2)=\sum_{l=-\infty}^{\infty}~\sum_{ \hat R_1, \,
\hat R_{2}}Z^{q{\rm YM},+}_{ \hat R_1, \, \hat R_2}(t_s+p\,g_s
l)\, Z^{q{\rm YM},-}_{ \hat R_1, \, \hat R_2}(\,\bar{t}_s-p\,g_s
l)
\end{equation}
with \beq Z^{q{\rm YM},-}_{ \hat R_1, \, \hat
R_2}(\,\bar{t}_s\,)=(-1)^{| \hat R_1|+| \hat R_2|}\, Z^{q{\rm
YM},+}_{ \hat R_1^\top, \, \hat R_2^\top}(\,\bar{t}_s\,) \ , \eeq
where $| \hat R|$ is the total number of boxes of the Young
tableau of the representation $ \hat R$. The genus zero case is
also special because it admits a standard description in terms of
toric geometry. The fibration $X={\cal O}(p-2)\oplus {\cal
O}(-p)\to\proj^1$ is in fact a toric manifold \cite{CGPS} and the
partition function can be written in terms of the topological
vertex  $C_{ \hat R_1 \hat R_2 \hat R_3}(q)$
\cite{Aganagic:2003db} as
\begin{equation}
Z^{q{\rm YM},+}_{ \hat R_1, \, \hat R_2}(t_s)=Z_0~ q^{{k_{ \hat
R_1}}/{2}}~\e^{-\frac{t_s(| \hat R_1|+| \hat R_2|)}{p-2}}~
\sum_{\hat R}\e^{-t_s| \hat R|}~q^{{(p-1)k_{ \hat R_1}}/{2}}~C_{0
\hat R_1 \hat R^\top}(q)\,C_{0 \hat R \hat R_2}(q) \ ,
\end{equation}
where $k_{\hat R}$ is related to the Young tableaux labels through
$k_{\hat R}=\sum_i \hat R_i( \hat R_i-2 \, i+1)$ and $Z_0$
represents the contribution from constant maps. This is the
partition function of the topological A-model on $X$ with
non-compact D-branes inserted at two of the four lines in the web
diagram. The D-branes are placed at a well-defined ``distance"
$t_s/(p-2)$ from the Riemann surface, thereby introducing another
geometrical parameter.

We thus observe an apparent discrepancy between the prediction of
\cite{Ooguri:2004zv} that $Z_{\rm BH}=|Z_{\rm top}|^2$ and the
explicit computation leading to \beq Z_{\rm
BH}=\sum_{b,l}\,\bigl|Z^{(b,l)}_{\rm top}\bigr|^2 \ . \eeq We have
indicated with the index $b$ the sum over chiral blocks with branes
inserted. The extra sum over the integer $l$ originates from the
$U(1)$ degrees of freedom contained in the original gauge group
$U(N)$, which has been interpreted as a sum over Ramond-Ramond
fluxes through the Riemann surface \cite{Vafa:2004qa}. The sum over
the fiber D-branes seems instead related to the fact that the
Calabi-Yau is non-compact and has more moduli coming from the
non-compact directions \cite{Aganagic:2005dh}. We will not enter
into this subtle aspect of the comparison except for noticing that,
at least for $g=0$, the sum over the ``external" branes in the
complete partition function enters on the same footing as the sum
over the topological string amplitude constituents. This ``external"
sum is weighted with a different K\"ahler parameter
$\hat{t}=t_s/(p-2)$, and the partition function therefore
effectively depends on two parameters. The observation above
suggests that $\hat{t}$ could have an interpretation as a true
K\"ahler modulus. This follows from a different definition of the
chiral gauge theory which is directly connected to the ordinary
Yang-Mills one and leads to a closed topological string theory by
itself  \cite{CGPS}.

The partition function $Z^{q{\rm YM},+}_{\hat R_1,\dots, \, \hat
R_{|2g-2|}}(t_s)$ describes a truly open topological string theory
with the worldsheets ending on the $|2g-2|$ stacks of D-branes
that are geometrically represented by boundaries wrapping on the
non-contractible cycles of the Calabi-Yau threefold. Nevertheless
the boundaries are glued together, and the holomorphic and
anti-holomorphic worldsheets match up on the D-branes. Worldsheets
contributing to the total partition function are still closed,
except that they are piecewise holomorphic or anti-holomorphic. A
natural speculation is that boundary branes should be the analogue
of the orientation-reversing tubes (and possibly even of the
$\Omega$-points) appearing in the Gross-Taylor string description
of ordinary large $N$ Yang-Mills theory.

The whole picture therefore seems very convincing and pointing
towards a beautiful confirmation of the conjecture presented in
\cite{Ooguri:2004zv}. It also suggests a natural embedding of
two-dimensional Gross-Taylor string theory into a six-dimensional
topological string theory, providing a new understanding of the QCD$_2$
string. However, there is
a point that has been overlooked which could
have interesting ramifications. In taking the large $N$ limit it is
possible to encounter phase transitions. The prototype of this
kind of phenomenon was discovered long ago
\cite{Gross:1980he,Wadia:1980cp} in the one-plaquette model
of lattice gauge theory.

A well-known large $N$ phase transition is the Douglas-Kazakov
transition~\cite{Douglas:1993ii}. It concerns Yang-Mills theory on
the sphere, a close relative of the relevant black hole ensembles
discussed earlier. A strong-coupling phase, in which the theory
admits the Gross-Taylor string description with its
chiral-antichiral behaviour, is separated by a weak-coupling phase
with gaussian field theoretical behavior. Two-dimensional Yang-Mills
theory on $S^2$ is equivalent to a string theory only above a
certain critical value of the effective t'Hooft coupling constant
$\lambda=Ng^2A$ given by \begin{equation}\lambda_c=\pi^2 \
.\end{equation} Instanton configurations induce the phase transition
to the strong-coupling regime \cite{Gross:1994mr}. On the other
hand, the entropy associated to certain classes of branched covering
maps seems responsible for the divergence of the string perturbation
series above the critical point
\cite{Taylor:1994zm,Crescimanno:1994eg}. It is natural to expect
that, if $q$-deformed large $N$ Yang-Mills theory is related to the
undeformed one, some of these features could find a place in the
black holes/topological string scenario. Note that the partition
function $Z_{BH}$ describing the black-hole physics at large charges
is { not} the topological string partition function, but rather its
square modulus summed over a complicated set of boundary conditions.
The final result is thus { not} an analytic function of the K\"ahler
parameters, preventing in principle its analytical continuation
below certain points. We will explore this possibility in the
subsequent sections.

\section{$\mbf q$-deformed Yang-Mills theory on \boldmath$S^2$\unboldmath}

In this section we will study the gauge theory structure of $q$-deformed
Yang-Mills theory on $S^2$ at finite~$N$. Even without reference
to black hole physics, it is worthwhile studying this peculiar
theory which connects instanton counting in four dimensions, topological
strings in six dimensions and, as we explain momentarily, Chern-Simons theory
in three dimensions.

\subsection{General properties and the undeformed limit}

Recall the definition of the partition function of
$q$-deformed Yang-Mills theory on a genus $g$ Riemann surface $\Sigma_g$ given by
\begin{equation}
Z^{q}_{\rm YM}=\sum_{R}{\rm
dim}_q(R)^{2-2g}~q^{\frac{p}{2}\,C_2(R)}~\e^{\ii\theta\,
C_1(R)} \ .\label{pqdbis}
\end{equation}
In deriving eq.~(\ref{pqdbis}) there are
normalization ambiguities coming in part from a choice of
regularization which corresponds \cite{Witten:1991we} to additions of
terms of the form
\begin{equation}
\alpha\,\int_{\Sigma_g}R+\beta\,\int_{\Sigma_g}K=\alpha\,\chi(\Sigma_g)+\beta'
\end{equation}
to the action~(\ref{aqd}), where $\chi(\Sigma_g)=2-2g$ is the Euler
characteristic of $\Sigma_g$. The constants appearing here have been
fixed in \cite{Aganagic:2004js} by requiring consistency, in the
large $N$ limit, with topological string theory. They also have a
different overall normalization in eq.~(\ref{pqd}). In their case
the quantum dimension is multiplied by the quantity
\begin{equation}
S_{00}=\prod_{1\leq i<j\leq N}\bigl[q^{(j-i)/2}-q^{-(j-i)/2}\bigr] \ .
\label{0block}
\end{equation}
At finite $N$ this term is only a function of $g_s$ and $N$, and thus does not
affect the main properties of the theory. Instead, it will assume
an important role at large $N$
in reproducing the correct contributions from constant maps to the
topological string amplitudes.

We will momentarily ignore the
regularization ambiguities and write the partition function with the factor
$S_{00}$, so that
\begin{equation}
Z^{q}_{\rm YM}=\sum_{R}(S_{0R})^{2-2g}~q^{\frac{p}{2}\,C_2(R)}~\e^{\ii\theta\,
C_1(R)} \ .\label{pqd1}
\end{equation}
In this form one recognizes the building blocks of $U(N)$
Chern-Simons theory, \begin{equation}{\rm
dim}_q(R)=\frac{S_{0R}}{S_{00}} \ , \end{equation} where $S_{RP}$
is the modular S-matrix of the $U(N)$ WZW model on $\Sigma_g$ at level $k$,
with the important feature that the level $k$ is not an integer.
In fact, $k$ is a purely imaginary number. At
$p=0=\theta$ eq.~(\ref{pqd1}) formally coincides with the partition
function of Chern-Simons theory on the three-manifold $S^1\times \Sigma_g$. The
crucial difference is that we do not have here any truncation on
the sum over the representations, which would usually be associated with the periodicity
induced by integer-valued $k$. At more general values of $p$, it was
claimed in \cite{Aganagic:2004js} that eq.~(\ref{pqd1}) should correspond to
Chern-Simons theory on a circle bundle over $\Sigma_g$. A strong
suggestion along these lines comes from observing that this Seifert manifold is the
boundary the non-compact four-cycle $C_4$ wrapped by the
D4-branes. The partition function of $q$-deformed Yang-Mills theory should be viewed
as providing a definition of Chern-Simons theory at non-integer
values of $k$ through the coupling $g_s=2\pi/(k+N)$. We will substantiate this argument
in genus 0 by showing that part of the periodicity survives
in the sum over representations $R$, allowing for the explicit determination of
the relevant Chern-Simons gauge theory.

The other theory connected to eq. (\ref{pqdbis}) is of
course ordinary Yang-Mills theory. As we saw in the
previous section, as $g_s\to 0$ the quantum dimension goes smoothly
into the ordinary one. In order to recover
the undeformed partition function (\ref{migdal}), we have also to send
$p\to\infty$ with
\begin{equation}
g_sp=a=g^2A
\end{equation}
fixed. This simple observation has some far reaching consequences.
Because in this particular limit ordinary Yang-Mills theory is
reproduced, we can look at the $q$-deformed theory as a peculiar
$a/p$ expansion and hope that some interesting characteristic
survives coming down from $p=\infty$. We also expect that, in the
large $N$-limit, the Gross-Taylor string theory emerges from the
closed topological string theory on $X$. From the geometrical point
of view, the limit $p\to\infty$ should be understood as some kind of
limit in the Calabi-Yau manifold. Transition functions on fibers
become highly singular and the geometry is expected to collapse
effectively into a two dimensional one, reproducing eventually the
topological sigma-model proposed in \cite{Cordes:1994sd}. From the
three-dimensional perspective, this limit was shown
in~\cite{Beasley:2005vf} to reduce the Chern-Simons partition
function on the pertinent Seifert fibration over $\Sigma_g$ to that
of ordinary Yang-Mills theory on $\Sigma_g$. From a practical point
of view, taking this double-scaling limit and obtaining the
well-known QCD$_2$ results will be a useful consistency check in our
computations.

In order to perform concrete calculations, it is better to express
the partition function directly in terms of Young tableaux
variables. The irreducible representations $R$ of the $U(N)$ gauge group
are easily constructed from
irreducible representations $\hat{R}$ of $SU(N)$ by assigning the
$U(1)$ charge $Q$. Its values are determined accordingly to the
isomorphism $U(N)=SU(N)\times U(1)/\mathbb{Z}_N$ as
\begin{equation}
R=(\hat{R},Q) \ , \,\,\,\,\,\, Q=n+r\,N \ ,
\end{equation}
where $n$ is the number of Young tableau boxes contained in $\hat{R}$ and $r$ is any
integer. The first and second Casimir invariants of $R$ can be written as
\begin{eqnarray}
C_1(R)&=&Q \ , \nonumber\\
C_2(R)&=&C_2(\hat{R})+\frac{Q^2}{N} \ ,\nonumber\\
\end{eqnarray}
where the quadratic Casimir of the $SU(N)$ representation $\hat R$ is given by
\begin{equation}
C_2(\hat{R})=n\,N+\sum_{i=1}^N
\hat{n}_i(\hat{n}_i+1-2i)-\frac{n^2}{N}
\end{equation}
with $\hat{n}_i$ the lengths of the rows of the Young tableaux
corresponding to $\hat{R}$ obeying the constraints
$\hat{n}_1\geq\hat{n}_2\geq\dots\geq\hat{n}_N \geq 0$ and
$\sum_{i=1}^N\hat{n}_i=N$. Although the above parameterization is
well-suited for deriving the string representation at large $N$, to explore
the modular properties of the theory
we will find it useful to switch to the set of integers $n_i$ defined by
\begin{equation}
n_i=\hat{n}_i+r-i \ .
\end{equation}
The range of $n_i$ is $+\infty> n_1>n_2>\dots>n_N>-\infty$. In this parameterization, the
partition function of the $q$-deformed gauge theory on $S^2$ assumes
the simple form \beq
Z_{\rm YM}^{q}(g_s,p)=\frac{1}{N!}\, \sum_{n_i\in \mathbb{Z}} \e^{-\frac{g_s
p}{2}\, {( n_1^2+\cdots+n_N^2)}+\ii\theta(n_1+\cdots+n_N)}\,\prod_{1\le
i<j\le N} \sinh^2\left(\mbox{$\frac{g_s}{2}$}\,(n_i-n_j)\right) \ .
\label{Cas}\eeq In eq.~(\ref{Cas}) we have used the total symmetry
of the summand in $n_i$ to remove their ordering, and
the fact that the quantum dimension vanishes for coincident $n_i$ to
extend the sum over all of $\mathbb{Z}^N$.
We have disregarded some overall factor that combines
with the renormalization ambiguities. As we have remarked in the
previous section, the comparison with ${\cal N}=4$ gauge theory on
$C_4$ is not natural when the partition function is written in
series in $\e^{-g_s}$. A modular transformation of (\ref{Cas})
is required and, as a result, some important properties will be
unveiled.

\subsection{Instanton expansion}

The geometrical meaning of the $q$-deformed theory on $S^2$ and its
relation with Chern-Simons theory on a particular Seifert manifold
becomes more transparent when we consider the dual description in
terms of instantons of the gauge theory, provided by a modular
transformation of the series (\ref{Cas}). This is accomplished by
means of a Poisson resummation. It is also an efficient way to
investigate the behaviour of the theory at weak coupling $g_s$, as
we shall explain later. We begin by taking the Fourier-transform of
the function \beq \mathcal{F}(x_1,\dots,x_N)=\e^{-\frac{g_s p}{4}\,
\sum_{i=1}^N x_i^2}\,\prod_{1\le i<j\le N}
\sinh\left(\mbox{$\frac{g_s}{2}$}\,(x_i-x_j)\right) \eeq using the
technique of orthogonal polynomials
\cite{Tierz:2002jj,deHaro:2004id,deHaro:2005rz,Marino:2002fk}.
Shifting the variables of integration as $x_i\to
y_i+\frac{(1-N)}{p}+\frac{4\pi \ii s_i}{g_s p}$, we have \bea
\mathcal{S}(s_1,\dots,s_N)&=&\int^\infty_{-\infty} \dd x_1\cdots \dd
x_N~
\e^{2\pi \ii \sum_{i=1}^N s_i x_i}\,\mathcal{F}(x_1,\dots,x_N)\nonumber\\
&=&C_N~\e^{\sum_{i=1}^N \frac{{\left( {g_s}\,\left( N -1\right)  - 4
\pi \ii s_i \right) }^2}{4\,{g_s}\,p}}
\,\int^\infty_{-\infty}\dd y_1\cdots \dd y_N ~\e^{-\frac{g_s
p}{4}\, \sum_{i=1}^N y_i^2}\nonumber\\&&\times\,\prod_{1\le i<j\le N}
\left( \e^{g_s
 y_i+\frac{4\pi \ii s_i}{ p}}-\e^{
 g_s y_j+\frac{4\pi \ii s_j}{ p}
 }\right)
\eea where $C_N={2^{-\frac{N(N-1)}{2}}}
~\e^{-g_s\,\frac{N(1-N)^2}{2p}} $. We set $\e^{g_s
y_i}=\e^{-{2 g_s}/{p}}\, z_i$ to reduce the integral to the
standard form \bea
{\cal S}(s_1,\dots,s_N)&=&D_N ~\e^{\sum_{i=1}^N \frac{{\left(
{g_s}\,\left( N -1\right)  -  4 \pi \ii s_i \right)
}^2}{4\,{g_s}\,p}} \,\int^{\infty}_{0}\!\! {\dd z_1\cdots \dd z_N}
~\e^{-\frac{ p}{4 g_s}\, \sum_{i=1}^N
\log(z_i)^2}\nonumber\\&&\times\,\prod_{1\le i<j\le N}\left(z_i
~\e^{\frac{4\pi \ii s_i}{ p}}-z_j ~\e^{\frac{4\pi \ii s_j}{ p}
 }\right)
\label{pupo1}\eea where $D_N=C_N g_s^{-N} ~\e^{-{g_s N^2}/{ p}}$.

The factor $\e^{-\frac{ p}{4 g_s}\,  \log(z)^2}$ is the natural measure for
the Stieltjes-Wigert polynomials $W_k(z;q)$ \cite{SW} with the
parameter $q$ chosen to be $q=\e^{{2
g_s}/{p}}$. Some useful details and properties of these polynomials are given in Appendix~A.
We can expand the
Vandermonde determinant in terms of these polynomials as \bea
&&\prod_{1\le i<j\le N} \left(z_i ~\e^{\frac{4\pi \ii
s_i}{ p}}-z_j ~\e^{\frac{4\pi \ii s_j}{ p}
 }\right)\nonumber\\&&\qquad=~
\sum_{\pi\in S_N} (-1)^{|\pi|}\, W_{0}(z_{\pi(1)} ~\e^{{4\pi \ii s_{\pi(1)}}/{
p}},q)\cdots W_{N-1}(z_{\pi(N)} ~\e^{{4\pi \ii s_{\pi(N)}}/{
p}},q) \ . \label{pipo}\eea
Substituting the representation
(\ref{pipo}) into eq.~(\ref{pupo1}), we can perform exactly the
integral  in a closed form (see Appendix A.2) to get
\bea
 {\cal S}(s_1,\dots,s_N)&=&
F_N ~\e^{\sum_{i=1}^N \frac{{\left( {g_s}\,\left( N -1\right)  -
4 \pi \ii s_i \right) }^2}{4\,{g_s}\,p}} \nonumber\\&&
\times\,\sum_{\pi\in S_N} (-1)^{|\pi|}\, \hat S_{0}( -q^{(1-1)/2}
~\e^{{4\pi \ii s_{\pi(1)}}/{ p}})\cdots \hat S_{N-1}(-q^{(1-N)/2}
~\e^{{4\pi \ii s_{\pi(N)}}/{ p}}) \ , \nonumber\\ && \label{piro}
\eea where $\hat S_k(z)$ are proportional to the Szeg\`o
polynomials (see Appendix A) and \beq F_N={{\left(\frac{4\pi}{g_s
p}\right) }^{\frac{N}{2}}~\e^{-\frac{{g_s} \left( N-2 \right)
\,\left(  N-1 \right) \,N}{6\,p}}~2^{-\frac{\left(  N-1\right)
\,N}{2}}} \ . \eeq Recalling the definition of the Vandermonde
determinant, eq. (\ref{piro}) can be rewritten as \beq
\mathcal{S}(s_1,\dots,s_N)=F_N ~\e^{\sum_{i=1}^N \frac{{\left(
{g_s}\,\left( N -1\right)  -  4 \pi \ii s_i \right)
}^2}{4\,{g_s}\,p}} \prod_{1\le i< j\le N}\left(\e^{\frac{4\pi \ii
s_{i}}{ p}}-\e^{\frac{4\pi \ii s_{j}}{ p}}\right) \ . \eeq The
final step in the Poisson resummation employs a well-known
property of the convolution product under Fourier transformation
to get \bea \label{cipow} Z_{\rm
YM}^{q,\mathrm{inst}}(s_1,\dots,s_N)&=&\int^\infty_{-\infty} \dd
x_1\cdots \dd x_N ~\e^{2\pi \ii \sum_{i=1}^N s_i x_i}\,
\mathcal{F}(x_1,\dots,x_N)^2\nonumber\\
&=&\e^{-\frac{2\pi^2 }{g_s p}\,\sum_{i=1}^N s_i^2}
~{w}^{\mathrm{inst}}_q(s_1,\dots,s_N)
\eea where \bea \label{cipow2}
{w}_q^{\mathrm{inst}}(s_1,\dots,s_N)&=&\frac{1}{2}\,\left(\frac{2\pi}{g_s
p}\right)^N ~\e^{-\frac{{g_s}\left(  N^3-N \right) }{6
p}}\,\int^\infty_{-\infty} \dd z_1\cdots \dd z_N ~\e^{-\frac{2\pi^2}{g_s p}\,
\sum_{i=1}^N  z_i^2}\nonumber\\
&&\times\,\prod_{1\le i< j\le N}
  \left[ \cos \left(\mbox{$\frac{2\,\pi \,\left( {s_i} - {s_j} \right)
        }{p}$}\right) - \cos
  \left(\mbox{$\frac{2\,\pi \,\left( {z_i} - {z_j} \right)
      }{p}$}\right) \right] \ .
\eea

The total partition function thereby turns out to be \beq
Z_{\rm YM}^q(g_s,p)=\frac{1}{N!}\, \sum_{s_i\in \mathbb{Z}}\e^{-\frac{2\pi^2
}{g_s p}\,\sum_{i=1}^N (s_i-\theta)^2}
~{w}^{\mathrm{inst}}_q(s_1,\dots,s_N) \ . \label{pqdi}\eeq
Note that the coupling parameter $\theta$ simply appears in the
exponential factor as a constant
shift in the integers $s_i$,
as ${w}^{\mathrm{inst}}_q(s_1,\dots,s_N)$ depends only on their
differences. We shall call this expression the instanton
expansion of $q$-deformed Yang-Mills theory on $S^2$. This terminology
mimicks that of the undeformed theory and will be justified presently.

The partition function of ordinary two-dimensional Yang-Mills theory
can be computed exactly via a nonabelian generalization of the
Duistermaat-Heckman theorem~\cite{Witten:1992xu}. It is given by a sum over
contributions localized at the classical solutions of the theory. For
finite $N$ the $U(N)$ path integral is given by a
sum over unstable instantons where each instanton contribution is
given by a finite, but non-trivial, perturbative expansion. By
``instantons'' we mean solutions of the classical Yang-Mills field
equations
\beq {\dd_A}{}^*F=\dd{}^*F+\ii[A,{}^*F]=0  \label{Fcovconst}  \eeq
which are not gauge transformations of the trivial solution
$A=0$. This equation implies that the scalar field $f={}^*F$ is
covariantly constant and may therefore be regarded as a constant element of
the Lie algebra of the $U(N)$ gauge group. The background curvature
breaks the $U(N)$ gauge symmetry to the subgroup of $U(N)$ which
commutes with $f$. By gauge invariance we may assume that $f$ is an
$N\times N$ real diagonal matrix. Its
collection of eigenvalues has multiplicities $N_k$ with
$\sum_kN_k=N$. In other words, the solutions of (\ref{Fcovconst}) are
labelled by partitions of the rank $N$ of the gauge theory, and for a
given partition $\{N_k\}$ the symmetry breaking is
$U(N)\to\prod_kU(N_k)$. If $E\to\Sigma_g$ is the corresponding
principal $U(N)$-bundle of the gauge theory on $\Sigma_g$, then near
each critical point it admits a decomposition $E=\bigoplus_kE_k$ into
$U(N_k)$ sub-bundles $E_k\to\Sigma_g$.

On the sphere $S^2$, the most general solution (up to
gauge transformations) is given by
\beq \bigl(A(z)\bigr)_{ij}=\delta_{ij}~A^{(m_i)}(z) \label{monopole}   \eeq
where $A^{(m_i)}(z)$ is the Dirac monopole potential of magnetic
charge $m_i$. The bundle splitting is described by taking $E_k=({\cal
  L}^{\otimes m_k})^{\oplus N_k}$, where ${\cal L}\to S^2$ is the
monopole line bundle (or equivalently the canonical line bundle over
$\proj^1$) which is classified by the Hopf fibration $S^3\to S^2$.
The Yang-Mills action evaluated  on such an instanton is given by
\beq S_{\mathrm{\rm inst}}=\frac{2\pi^2}{g^2A}\,\sum_{i=1}^N m_i^2 \
.\eeq Poisson resummation exactly provides the representation of
ordinary Yang-Mills theory on $S^2$ in terms of instantons
\cite{Minahan:1993tp,Gross:1994mr}. Looking closer at
eq.~(\ref{pqdi}) we recognize a similar structure emerging. We
observe the expected exponential of the ``classical action" (at
$\theta=0$) $\e^{-\frac{2\pi^2  }{g_s p}\,\sum_{i=1}^N s_i^2}$ and
the fluctuations ${w}_q^{\mathrm{inst}}(s_1,\cdots,s_N)$ which
smoothly reduce to the undeformed ones in the double scaling limit.
The instanton representation is also useful to control the
asymptotic behaviour of the partition function as $g_s\to 0$. In
this limit, only the zero-instanton sector survives, the others
being exponentially suppressed (for fixed $p$). In ordinary
two-dimensional Yang-Mills theory the limiting partition function
computes the volume of the moduli space of flat connections on the
underlying Riemann surface and also intersection pairings on this
moduli space \cite{Witten:1992xu}. For the case of $S^2$ the moduli
space consists of a single point and the limit is trivial. Instead,
in the $q$-deformed case we expect non-trivial geometrical
structures emerge~\cite{Beasley:2005vf}.

The arguments based on localization follow from the Yang-Mills action
(\ref{aqd}) in undeformed case wherein the scalar field $\Phi$ is
non-compact. Varying the gauge field $A$ in this action gives the
equations of motion $\dd_A\phi=0$, implying as above that the classical
solutions for $\Phi$ may be taken to be constant, diagonal real
$N\times N$ matrices. Varying $\Phi$ gives the equation (at
$\theta=0$) $\Phi=f$, which thus implies the Yang-Mills equation
(\ref{Fcovconst}) for the gauge field. However, this argument breaks
down in the case that $\Phi$ is a compact $U(N)$ scalar
field. Heuristically, one should add to the non-compact scalar field
$f$ a sum over all image charges which render the classical solution
$\Phi=f$ effectively compact. This means that, in addition to the sum
over partitions of $N$ which specify the given bundle splitting, there
is an additional integer sum in each bundle component $E_k$. However,
even this heuristic argument is sloppy, because the effect of the
deformation is encoded in the
 measure for the scalar field.

We can gain some insight into the structure of the classical solutions
of $q$-deformed Yang-Mills theory via the following equivalent
reformulation in terms of generalized Yang-Mills theory~\cite{genYM2}. The aim
in this reformulation is to provide some insight on the classical
solutions of the theory and to clarify the use of the term
instanton. Let us start from the partition function expressed in the
form (\ref{Cas}). The quantum dimension of the  representations can be
exponentiated giving
\bea
Z_{\rm YM}^{q}(g_s,p)&=& \frac{1}{N!}\, \left(  \frac{g_s}{2} \right)^2 \,
 \sum_{n_i\in \mathbb{Z}} \e^{-\frac{g_s
p}{2}\, {( n_1^2+\cdots+n_N^2)}+\ii\theta(n_1+\cdots+n_N)}\, \prod_{1\le
i<j\le N} \left( n_i - n_j  \right)^2 \cr && \times\,\exp\left[
-\mbox{$\sum\limits_{i,j=1}^N\log\left(\frac{g_s(n_i-n_j)}{2\sinh
\left(\frac{g_s}2
\, (n_i -n_j)\right)}\right)$}\right] \ .
\eea
The function that appears in the last exponential factor can rewritten
using the expansion
\beq
\log\left(\frac{x}{\sinh(x)}\right) = \sum_{k=1}^\infty (-1)^k\,
\frac{\zeta(2k)}{k }\,\left(\frac{x}{\pi}\right)^{2k}
\eeq
to get
\bea
&&\sum_{i,j=1}^N\log\left[\mbox{$\frac{g_s(n_i-n_j)}{2\sinh(\frac{g_s}2\,
 (n_i -n_j))}$}\right]\nonumber\\
&& \qquad=~ \sum_{k=1}^\infty\frac{\zeta(2k)}{(2\pi)^{2k}\,k }\,g_s^{2k}\,
\sum_{\ell=0}^{2k}(-1)^{k+\ell}\, {2k\choose\ell}\,
\left(\sum_{i=1}^N n_i^\ell\right)\,
\left(\sum_{j=1}^N n_j^{2k-\ell}\right) \ .
\eea
If we now introduce the diagonal matrix $\phi = g_s~{\rm diag}\left(
  n_1, \dots, n_N \right) $, then we can finally rewrite (\ref{Cas}) as
\bea
Z_{\rm YM}^q (g_s,p)
 = \frac{1}{N!}\,\left(\frac{g_s}{2}\right)^{2}\, \sum_{n_i\in\mathbb{Z}}
 \left( n_i - n_j  \right)^2   ~\e^{
{V(\phi)-\frac{g_s p}{2}\,
  \mathrm{Tr}(\phi^2)+\ii{\theta}\,\mathrm{Tr}(\phi)} } \ .
\eea
In this representation the only difference from the usual partition
function of the undeformed theory is in the apparence of an auxilliary
potential for the matrix $\phi$ given by
\beq
V(\phi)=-\sum_{k=1}^\infty\frac{\zeta(2k)}{(2\pi)^{2k}\,k }\,
\sum_{\ell=0}^{2k}(-1)^{\ell+k}\, {2k\choose\ell}\,
\mathrm{Tr}(\phi^\ell)\,\mathrm{Tr}(\phi^{2k-\ell}) \ .
\eeq

These formal manipulations suggest that the partition function of the
$q$-deformed theory can be derived from a generalized (but undeformed)
Yang-Mills theory with action
\bea
S_{{\rm YM}_2}^{\rm gen}&=&\frac1{g_s}\,\int_{S^2} \mathrm{Tr}(\Phi \,F)-
\frac{p}{2 g_s}\int_{S^2}
\mathrm{Tr}(\Phi^2)~K+\frac{\theta}{g_s}
\int_{S^2} \mathrm{Tr}(\Phi)~K\nonumber\\
&&+\,\frac{1}{8\pi}\,\int_{S^2} V({\Phi})~ R \ , \label{genQCD}
\eea
where $R$ is the Ricci curvature two-form of a fixed metric on
$S^2$. This representation holds in fact for all genera. It assumes
that the path integral for the action (\ref{genQCD}) can be localized
onto a sum over $U(1)^N$ bundles for the gauge field, while the
integration over $\Phi$ is localized onto constant field
configurations $\phi$. The main difference between this representation and the
formulation of \cite{Aganagic:2004js} is that the scalar
 field $\Phi$ is not a periodic variable and its measure in the
path integral is the standard one. Moreover, the non-abelian
localization of the path integral is not altered by perturbing the
ordinary Yang-Mills action by deformations $V(\Phi)$ which depend only
on the scalar field. Thus the partition function will localize again
onto critical points, this time of the action functional
(\ref{genQCD}). By varying the action with respect
to the gauge field, one finds again that the field $\Phi$ is
covariantly constant. However, now the equation of motion for $\Phi$
itself involves a complicated non-linear relation between the scalar
field and the gauge field. Using gauge invariance to diagonalize
$\Phi$, by applying the covariant derivative to the equation of motion
for $\Phi$ we conclude also that $f={}^*F$ is covariantly constant and
again obeys the Yang-Mills equation (\ref{Fcovconst}). The solutions
for the gauge potential are therefore again of the form (\ref{monopole}). Thus
the critical points of the $q$-deformed theory (or equivalently of our
generalized QCD$_2$) behave exactly as the instantons of ordinary
two-dimensional Yang-Mills theory.

The effect of the deformation is now completely encoded in the
auxiliary potential for the field $\Phi$. However, its precise
dynamical role is not completely clear.
 To understand this problem, let us write down explicitly
 the equations of motion for the scalar field when the gauge group is
 $U(2)$. Diagonalizing the fields as described above to write them as
\beq
f = \left( \begin{array}{cc} f_1 & 0 \\ 0 & f_2   \end{array}
\right) \ , \qquad
\Phi = \left( \begin{array}{cc} \phi_1 & 0 \\ 0 & \phi_2   \end{array}
\right) \ ,
\eeq
the two equations of motion for $\Phi$ have the form
\bea
&& \frac{f_1}{g_s } - \frac{p}{g_s}\, \phi_1 - \frac{1}{\phi_1 -
  \phi_2}  + \frac{1}{2} \cot \left(\mbox{$ \frac{\phi_1-\phi_2}{2}$}  \right)
= 0 \ , \cr
&& \frac{f_2}{g_s } - \frac{p}{g_s} \,\phi_2 + \frac{1}{\phi_1 -
  \phi_2} - \frac{1}{2} \cot \left(\mbox{$ \frac{\phi_1-\phi_2}{2}$}  \right)
= 0 \ ,
\label{eomgenQCD}
\eea
where we have set the $\theta$-angle to zero. Due to the branch
structure of the cotangent function, these equations imply that for each
monopole configuration of the gauge fields there exists infinitely many
equivalent solutions for the scalar fields. This is likely to be
related to the image charges mentioned above, and agrees with the form
of the instanton expansion that we derive below.

We close this subsection by returning to the relation with the ${\cal
N}=4$ topological gauge theory. Eq.~(\ref{pqdi}) should be
equivalent to the formula obtained in \cite{Aganagic:2004js} for the modular
transformation of $Z^q_{\rm YM}$. They performed a weak check
for $p=1,2$, confronting the general structure of their
expressions with some results appearing in the literature
\cite{nakajima}. Unfortunately, there are very few results for the
Euler characteristic of the moduli space of instantons on
$C_4={\cal O}(-p)\to\proj^1$ for generic $p$. In confronting the two
theories, eq.~(\ref{pqdi}) should really reproduce the instanton
actions and their fluctuation determinants which are represented
here in terms of integrals. In the following, we shall try to
understand better the fluctuations around the instanton
$(s_1,\dots,s_N)$, i.e. the behaviour of the integral \beq
\int^\infty_{-\infty} \dd z_1\cdots \dd z_N ~\e^{-\frac{2\pi^2}{g_s p}
\,\sum_{i=1}^N z_i^2}\,\prod_{1\le i< j\le N} \left[
\cos \left(\mbox{$\frac{2\,\pi \,( {s_i} - {s_j}) }{p}$}\right)
- \cos \left(\mbox{$\frac{2\,\pi \,( {z_i} - {z_j})
}{p}$}\right) \right] \ . \eeq It can be localized onto a finite
interval by using the periodicity of the product appearing in the
integrand.

We now Poisson resum the generic series and
write it in terms of an elliptic theta-function as
\beq \sum_{n_i\in \mathbb{Z}} \e^{-\frac{2\pi^2 p}{g_s }\,
(z_i-n_i)^2}=\sqrt{\frac{{g_s}}
  {{}{2\,\pi\, p}}}\,\sum_{m_i\in \mathbb{Z}}\e^{-\frac{{g_s}  }{2 p}
\,{{m_i}}^2 + 2 \pi \ii {m_i}{z_i}}=
  \sqrt{\frac{{{g_s}}}
  {{}{2\,\pi \,p}}}\,\vartheta_3(\mbox{$\frac{{ \ii g_s}}
{2 \pi\, p }$} |z_i) \ .
\eeq Then the fluctuation  integral is given by \beq
\left(\frac{g_s p}{2\pi}\right)^{\frac{N}{2}}\,\int^1_{0}
\dd z_1\cdots \dd z_N~ \prod_{i=1}^N  \vartheta_3(\mbox{$\frac{{
\ii g_s}}{2 \pi\,
p }$} |z_i)\,\prod_{1\le i< j\le N}
  \left[ \cos \left(\mbox{$\frac{2\,\pi \,( {s_i} - {s_j}) }{p}$}\right) - \cos
  \bigl({2\,\pi \,\left( {z_i} - {z_j} \right) }\bigr) \right] \ .
\eeq The product can be equivalently written in terms
of Vandermonde determinants $\Delta(y_i)$ as \bea &&\prod_{1\le i< j\le N}
  \left[ \cos \left(\mbox{$\frac{2\,\pi \,( {s_i} - {s_j}) }{p}$}\right) - \cos
  \bigl({2\,\pi \,\left( {z_i} - {z_j} \right) }\bigr) \right]\nonumber\\
  &&\qquad =~\frac{\e^{-\frac{2\pi \ii }{p}\,(N-1) \sum_{l=1}^N s_l}}{2^N}
\,\Delta\left(\e^{2\pi \ii ( \frac{{s_i}}{p} - {z_i}) }\right)\,
\Delta\left(e^{2\pi \ii ( \frac{{s_i}}{p} + {z_i} ) }\right) \ .
\eea Thus the generic fluctuation term appears in the form \bea &&
\left(\frac{g_s p}{2\pi}\right)^{\frac{N}{2}}\,
\frac{\e^{-\frac{2\pi \ii}{p}\,(N-1) \sum_{l=1}^N s_l}}{2^N}
\,\int^1_{0}\dd z_1\cdots \dd z_N~ \prod_{i=1}^N
\vartheta_3(\mbox{$\frac{{ \ii g_s}}{2 \pi\, p }$} |z_i)
\nonumber\\ && \qquad\qquad \times ~ \Delta\left(\e^{2\pi \ii(
\frac{{s_i}}{p} - {z_i}) }\right)\, \Delta\left(\e^{2\pi \ii(
\frac{{s_i}}{p} + {z_i}) }\right) \ ,  \label{crys}\eea with the
Fourier transforms substituted by compact integrals and the
elliptic theta-function playing the role of the measure. In this
form, it strongly resembles the integrals appearing in
\cite{Okuda:2004mb,Okounkov:2003sp} where the B-model partition
function of open topological strings on the resolved conifold was
expressed in terms of a unitary matrix model.

\subsection{Relation to Chern-Simons theory on Lens spaces}

There is a very natural connection between the proposal of
\cite{Aganagic:2004js} and Chern-Simons theory on a Lens space. As
explained before, the instanton counting in the $\mathcal{N}=4$
 topologically twisted gauge theory living on the four-manifold
$C_4 = \mathcal{O}(-p) \to \mathbb{P}^1$ localizes onto the
two-dimensional deformed gauge theory on the base of this
holomorphic line bundle. The total space $C_4$ of the fibration is
an ALE space which asymptotes to the orbifold $\complex^2/\zed_p$,
where $\zed_p$ acts on $(z,w)\in\complex^2$ by
$(z,w)\mapsto\e^{2\pi\ii
  k/p}\,(z,w)$ for $k=0,1,\dots,p-1$~\cite{Lebrun}. More precisely, as
a scalar-flat K\"ahler four-manifold $C_4\cong
D(\mathcal{L}^{\otimes -p} )$ is the disk bundle of the dual of
the monopole line bundle over $S^2$ of magnetic charge $-p$. The
boundary of this manifold is a circle bundle over the sphere,
$\partial C_4\cong S(\mathcal{L}^{\otimes -p})$, which is
diffeomorphic to the Lens space  $L_p=S^3/\zed_p\cong
S(\mathcal{L}^{\otimes -p})$, where we regard
$S^3\subset\complex^2$ with the embedding $|z|^2+|w|^2=1$. Another
interesting perspective has been discussed in \cite{deHaro:2004uz}
where the partition function of ordinary Yang--Mills on a
cylinder, with trivial boundary conditions at the two ends of the
cylinder, was shown to be equivalent to the Chern--Simons
partition function on $S^3/\zed_p$ . In \cite{deHaro:2004wn}
 contact was made with the $q$--deformed theory on the sphere.

The relation to Chern-Simons gauge
theory on $\partial C_4 \simeq L_p$ now follows classically from the
action (\ref{4daction}).
An instanton excitation on a
generic four manifold with boundary is related to a Chern--Simons
theory on the boundary through
\beq \label{insttoCS}
\int_{{C}_4} \Tr F \wedge F =
\int_{{C}_4} \Tr~ \mathrm{d} \left( A \wedge
\mathrm{d} A + \mbox{$\frac{2}{3}$}\, A \wedge A \wedge A \right)=
\int_{\partial {C}_4} \Tr \left( A \wedge
\mathrm{d} A + \mbox{$\frac{2}{3}$}\, A \wedge A \wedge A \right) \ .
\eeq
Of course, this term only describes the pure topological sector of
the $q$-deformed gauge theory described by the BF lagrangian $\Tr(\Phi\,F)$,
which arises in the weak-coupling limit.
The remaining terms in the four-dimensional action, along with the massive deformation
$\Tr\,\Phi^2$, conspire to give a more complicated three-dimensional
gauge theory than just Chern-Simons theory.
According to \cite{Beasley:2005vf}, the three dimensional Chern-Simons
theory defined on a circle fibration over a Riemann surface is naturally
related to a two dimensional gauge theory on the base.
We will describe this relationship more explicitly later on in this section.
For now, we
will show how Chern-Simons
 theory on the Lens space $L_p$ emerges from the two dimensional point of view. We
will set $\theta=0$ as the $\theta$-angle is irrelevant for the present discussion.

Let us focus our attention on the partition function of the
$q$-deformed theory expressed in terms of instantons. A
remarkable property of the fluctuations is that they do not depend
really on the integers $s_i$ but only their values modulo $p$. In other words,
two instanton configurations $\vec{s}=(s_1,\dots,s_N)$ and
$\vec{s^\prime}=(s_1^\prime,\dots,s_N^\prime)$ that differ by
$p\cdot \vec{h}$ possess the same fluctuation factor, where $\vec{h}$ is any vector with $N$ integer
entries. The number of
independent fluctuations is then always finite and less than
$p^N$. It is natural to organize the partition function
by factorizing the independent fluctuations. This can be easily
achieved by writing each integer $s_i$ as \beq s_i= p\, \ell_i
+\hat s_i \eeq where $l_i\in\zed$
$\hat{s}_i\in\{0,1,\dots,p-1\}$. Then the instanton expansion of the
partition function $Z^q_{\rm YM}$ can be
written as \beq Z^q_{\rm YM}=\frac{1}{N!}\,\sum_{\hat
s_i=0}^{p-1}\,\sum_{\ell_i=-\infty}^\infty ~\e^{-\frac{2\pi^2  }{g_s
p}\,\sum_{i=1}^N (p\, \ell_i +\hat s_i)^2}~ {w}_q^{\mathrm{inst}}(\hat
s_1,\dots,\hat s_n) \ .\eeq
Two
fluctuations ${w}_q^{\mathrm{inst}}(\hat s_1,\dots,\hat s_n)$ that
differ only by a reordering of the $\hat s_i$ give again the same
contribution. From this observation we conclude that independent
fluctuations are completely characterized by the set of
non-negative integers $N_k$ which count the number of times the
integers $k\in\{0,\dots,p-1\}$ appears in the string $(\hat
s_1,\dots,\hat s_n)$. We have the obvious sum rule $\sum_k
N_k=N$.

The simplest way to eliminate the huge
degeneracy in the fluctuation factors is to reorder the
integers $\hat s_i$ and sum only over those configurations with
$\hat s_1\le \hat s_2\le \cdots \le \hat s_N$. In this way we arrive
at an elegant sum over ordered partitions $\{N_k\}$ of $N$ into
$p$ parts, and we obtain \bea
Z^q_{\rm YM}&=&\sum_{N_1,\dots,N_k\atop \sum_k
N_k=N}\frac{1}{\prod_{k} N_k!}\,\sum_{\ell_i\in \mathbb{Z}}
 \exp\left[-\frac{2\pi^2  }{g_s p}\,\left(\sum_{i=1}^{N_0} (p\, \ell_i )^2
+\sum_{i=N_0+1}^{N_0+N_1}(p\, \ell_i+1 )^2
 +\cdots\right.\right.\nonumber\\
 &&+ \left.\left.
 \sum_{i=N_0+ \cdots + N_{p-2}+1}^{N}(p\, \ell_i+p-1 )^2\right)\right]~{w}_q^{\mathrm{inst}}(\,
\underbrace{0,\dots,0}_{N_0},\dots,
 \underbrace{p-1,\dots,p-1}_{N_{p-1}}\,)\nonumber\\
&=&\sum_{N_1,\dots,N_k\atop \sum_k N_k=N}\frac{1}{\prod_{k}
N_k!}\,\prod_{k=0}^{p-1}\vartheta_3\left( \left.\mbox{$\frac{2\pi i p}{g_s}
$}\right|
 \mbox{$\frac{2\pi i k}{g_s}$}\right)^{N_k}\nonumber\\
 &&\times
\exp\left(-\frac{2\pi^2  }{g_s p}\,\sum_{m=0}^{p-1} N_m m^2\right)
~w_q^{\,\mathrm{inst}}(\underbrace{0,\dots,0}_{N_0},\dots,
 \underbrace{p-1,\dots,p-1}_{N_{p-1}}\,) \ .
\eea This is the central result of this section. We recognize in
the second line the partition function of $U(N)$ Chern-Simons
gauge theory on the Lens space $L_p$ in a non-trivial vacuum
given by~\cite{Marino:2002fk,Aganagic:2002wv} \beq Z^p_{\rm CS}\bigl(\{N_k\}\bigr)=\exp\left(-\frac{2\pi^2
}{g_s p}\,\sum_{m=0}^{p-1} N_m m^2\right)
~w_q^{\mathrm{inst}}(\,\underbrace{0,\dots,0}_{N_0},\dots,
 \underbrace{p-1,\dots,p-1}_{N_{p-1}}\,) \ .\label{LS}\eeq

The critical points of the
$U(N)$ Chern-Simons action on the Seifert manifold $L_p$ are flat connections which are
classified by the embeddings of the first fundamental group into
$U(N)$. Since the $\zed_p$-action on $S^3$ used to define $L_p$ is free,
one has $\pi_1(L_p)=\mathbb{Z}_p$. The cyclic generator $h$ of this fundamental group
can be taken to be any loop which is the projection of a
path on the universal cover $S^3\to L_p$ connecting
two points that are related by the $\zed_p$-action.
The critical points are therefore given by discrete $\mathbb{Z}_p$-valued
flat connections. They are easily described by
choosing $N$-component vectors with entries taking values in
$\mathbb{Z}_p$. Because the residual Weyl symmetry $S_N$ of the $U(N)$ gauge group permutes the
different components, the independent choices are in
correspondence with the partitions $\{N_k\}$.
The possible vacua of the gauge theory are in one-to-one correspondence with
the choices of flat connections. The full partition function of
Chern-Simons theory involves summing over all the flat
connections, and in fact the exact answer that can be obtained
from the relation with the WZW model \cite{Rozansky:1993zx,Rozansky:1994wv} gives such a sum.
Nevertheless, due to the fact that the flat connections here are
isolated points, it is not difficult to extract the particular
contribution of a given vacuum which coincides with eq.~(\ref{LS}).

The relation we have found should be understood as an analytical
continuation to imaginary values of the Chern-Simons coupling
constant $k$ by identifying \beq g_sp=\frac{2\pi \ii}{k+N} \ ,\eeq
in full agreement with the discussion of Sect. 3.1. With this
distinction in mind, we can write down the partition function of
$q$-deformed Yang-Mills theory in the suggestive form \beq
Z^q_{YM}= \sum_{\{N_k\}}\,\prod_{k=0}^{p-1}\frac{\theta_3\left(
\left.\frac{2\pi \ii p}{g_s} \right|
 \frac{2\pi \ii k}{g_s}\right)^{N_k}}{N_k!}~Z^p_{\rm CS}\bigl(\{N_p\}\bigr) \ .
\eeq The instanton contributions appear organized in definitive
way. Some of them, and in particular the ones having the lowest
classical action for each class in $\{N_k\}$, appear in the
analytic continuation of $Z^p_{\rm CS}(\{N_k\})$. The instanton
action, in this case, is exactly the Chern-Simons action evaluated
on its critical point (the flat connection). The other instantons
belonging to the class determined by $\{N_k\}$ are instead
contained in the theta-function contributions multiplying the
fundamental Chern-Simons partition function in the given
background. It is tempting to speculate that they are related to
the presence of the periodic measure in the path integral. A
simple indication of this comes from the limit $g_s\to 0$. We
expect that in this regime the effective periodicity goes to
$\infty$ and the theory ``decompactifies". In this limit
$\vartheta_3\to 1$ and the $q$-deformed partition function
coincides with the total Chern-Simons partition function on $L_p$
summed over all the non-trivial flat-connections.

The fact that Chern-Simons theory on Seifert manifolds localizes
around flat connections in the same way that Yang-Mills theory localizes around instantons has
been recently shown in \cite{Beasley:2005vf}. The
computation of quantum fluctuations around flat connections
resembles in many respects the calculation around non-trivial
instantons in Yang-Mills theory. This is in completely harmony with
what we find here and it opens up the possibility to understand even
the higher-instanton contributions as coming from classical
solutions of $q$-deformed Yang-Mills theory.
We will now describe more precisely how a flat connection
 on the three-manifold $L_p$ is related to a generic instanton on $S^2$.
 Three-dimensional flat connections can be obtained
 from two dimensional instantons by assigning a fixed holonomy along the generator $h$
of $\pi_1(L_p)$.
 The general correspondence on any Seifert fibration was proven in \cite{furuta} for the
global minima of the two-dimensional Yang-Mills action (the central
connections corresponding to the trivial partition having only one
component $N_1=N$). In the case of the Lens space $L_p$ the
correspondence can be rephrased in a particularly simple fashion as
follows.

Consider the monopole line bundle $\mathcal{L}$ over the sphere.
We can form the bundle $\mathcal{L}^{\otimes -p}$ endowed with the monopole connection $A^{(-p)}$
of magnetic charge $-p$.
We can
 describe a constant curvature instanton on $S^2$ with the two dimensional data $(E,A)$, where
$E$ is a principal $U(N)$-bundle over $S^2$ with first Chern number $q\in\{0,1,\dots,N-1\}$ and $A$ is a connection
on $E$ of curvature $F_A = {\mathbf 1}_N \,\frac{q}{N}\, F_{A^{(-p)}}$.
The two dimensional connections are in one-to-one correspondence
 with flat connections on the circle bundle $S(\mathcal{L}^{\otimes -p})=L_p$ with gauge group
 $SU(N)$ and fixed holonomy $\e^{-2 \pi \ii {q}/{N}}\,\mbf 1_N$ in the center
 of $SU(N)$ around the fiber.  The three dimensional flat connection can be
 explicitly constructed in terms of the two dimensional constant curvature
 connection as follows. Let $\pi : S(\mathcal{L}^{\otimes -p}) \to S^2 $ be the
 bundle projection. The pull-back under this projection naturally lifts to three
 dimensions the two dimensional data $(E,A)$. Moreover, it can be used to define
 the trivial bundle $\pi^* \mathcal{L}^{\otimes -p} \cong S(\mathcal{L}^{\otimes -p}) \times \mathbb{C} $
 with connection $a = \frac{1}{N}\, \pi^* A_0$, to which we have the freedom to add
 any trivial connection.  With these data we can construct the bundle $ E'= \pi^*E
 \otimes \pi^* \mathcal{L}^{\otimes-pq}$ over $L_p$ whose curvature vanishes by construction
 The structure group of $E'$ is $SU(N)$, because the determinant line bundle
 $\det E' = \det \pi^*E \otimes (\pi^* \mathcal{L})^{\otimes  -pqN}$ is endowed with a
 vanishing connection and the holonomy around the fiber is in the center of $SU(N)$
 since by construction $\pi^* A_0$ has trivial holonomy around the fiber of $L_p\to S^2$.

The above construction can be easily generalized to generic instantons corresponding to
 two dimensional connections which are not proportional to the identity $\mbf 1_N$.
 Let us suppose that the instanton is specified by a partition $\{N_k\}$ of
 $N$ which corresponds to the gauge symmetry breaking $U(N) \to
 \prod_k U(N_k)$. In the vicinity of such an instanton the data $(E,A)$
 decompose as $E = \bigoplus_k E_k$ and  $A = \bigoplus_k A^{(m_k)}$.
 The argument above can then be applied to each sub-bundle $E_k$ with its central Yang-Mills
 connection $A^{(m_k)}$ to lift the most general reducible connection in two dimensions to a three
 dimensional Chern-Simons critical point. The main difference now is that
 the holonomy around the circle fiber is not in general an element of the
 center of $SU(N)$ but will respect the symmetry breaking pattern. The
holonomy is given by a block diagonal matrix whose entries are all of the form
 $e^{- 2 \pi \ii{q_k}/{N_k}}$ where $N_k$ is the rank of the corresponding subgroup
 and $0 \le q_k < N_k$.

A flat connection on a circle bundle over a
 Riemann surface with a holonomy around the fiber non-trivial but still
 proportional to the identity, that arises as the pull-back from a central
 $U(N)$-connection on the base manifold, can be seen as the pull-back of a {\it flat}
 two dimensional connection on a non-trivial bundle whose structure group is $U(N)/U(1)\cong
SU(N)/\zed_N$. In other words, a constant curvature connection,
 which can be locally seen as arising from an element of the center of the structure group,
 becomes trivial when we divide the gauge group $U(N)$ by its center $U(1)$.
 However, a flat connection on a nontrivial
$U(N)/U(1)$-bundle over the base Riemann surface can be described as a flat
 connection on a {\it trivial} $SU(N)$-bundle with given monodromy in $\zed_N$ around
 an arbitrary point on the base. Thus the correspondence between two
 dimensional instantons and Chern-Simons critical points can be equivalently
 rephrased as a correspondence between three dimensional flat connections and
 two dimensional flat connections with non-trivial monodromy around an arbitrary point.
This description explains the coincidence of the instanton expansion above in the weak-coupling limit,
which effectively picks out the zero-instanton contributions, and the sum over vacua
of the Chern-Simons partition function on $L_p$. The argument does not generalize to higher-instantons
which instead give non-trivial contributions to the theta-functions appearing. The full effect
of the $q$-deformation is captured by these functions.

We close this section by returning to the relation with
topological string theory. The partition function of Chern-Simons
theory on $L_p=S^3/\mathbb{Z}_p$ in the background \beq
u\bigl(\{N_k\}\bigr)=(\,\underbrace{0,\dots,0}_{N_0},\dots,
\underbrace{p-1,\dots,p-1}_{N_{p-1}}\,)\eeq is the perturbative
open topological string theory on the total space of the cotangent
bundle $T^*L_p$ of the Lens space, with $N_0,N_1,\dots,N_{p-1}$
topological D3-branes wrapped around the non-trivial cycles of
$L_p$. The partition function $Z^p_{\rm CS}(\{N_k\})$ has been
obtained in \cite{Aganagic:2002wv} from a matrix-model computation
associated to the topological B-model on the mirror of $L_p$. The
large $N$ limit and the closed topological string theories
emerging from the relevant geometric transition have been
discussed in
\cite{Aganagic:2002wv,Halmagyi:2003ze,Halmagyi:2003mm}, assuming a
fixed configuration of branes. The appearance of $Z^p_{\rm
CS}(\{N_p\})$ in $q$-deformed Yang-Mills theory could suggest a
topological open string underlying the theory and one could wonder
if the geometry of $X$, in the large $N$ limit, can be understood
as a result of a geometric transition. There are a couple of basic
differences which should be taken into account. Firstly, the $N$
D3-branes are not fixed but the $q$-deformed theory sums over all
possible wrappings. Secondly, the theta-function contributions do
not appear in the perturbative topological string amplitudes.

\section{Instanton driven large $\mbf N$ phase transition}

We are ready now to discuss the large $N$ limit of $q$-deformed
Yang-Mills theory on the sphere. We have expressed the partition
function as a sum over classical solutions in a way that the
limit $g_s\to 0$ is quite transparent and well related to the
quantum field theoretical degrees of freedom. Let us see the fate
of the instanton contributions at large $N$.

\subsection{Instanton contributions at large $N$}

We start by defining the relevant parameters to be held fixed as
$N\to\infty$ by
\begin{equation} t=g_sN \ ,\,\,\,\,\,\,a=g_s p N=p\,t \ .
\end{equation}
Henceforth we will set $\theta=0$. In terms of these variables,
the partition function assumes the form
\beq Z^q_{\rm YM}=\frac{1}{N!}\, \sum_{s_i\in
\mathbb{Z}}\e^{-\frac{2\pi^2N }{a}\,\sum_{i=1}^N s_i^2}
~{w}^{\mathrm{inst}}_q(s_i;N,a,t) \ . \label{pqdiN}\eeq We immediately
recognize that all the non-trivial instanton contributions are
nonperturbative in the $\frac1N$ expansion, being naively
exponentially suppressed, suggesting that the theory could reduce
in this limit to the zero-instanton sector. In order for
this possibility to be correct, one should control the fluctuation
factors. Let us recall what happens in ordinary Yang-Mills theory. In that case the
corrections due to the contribution of instantons to the
free energy were calculated in \cite{Gross:1994mr}. They find that while in
the weak-coupling phase this contribution is exponentially small,
it blows up as the phase transition point is approached. The
transition occurs when the entropy of instantons starts dominating
over their Boltzmann weight $\e^{-S_{\rm inst}}$. The order of
perturbation theory required to compute the fluctuations around
the instantons grows with $N$, and as $N\to\infty$ one gets
contributions from all orders which compete with the exponential
suppression. The density of instantons then goes from 0 at weak
coupling to $\infty$ at strong coupling.

In principle, the $q$-deformed theory on $S^2$ could
experience the same fate. In eq.~(\ref{pqdiN}) the Boltzmann
weights are the same as in the undeformed case, and only the structure of the
fluctuations is changed by the deformation. As written in
eq.~(\ref{crys}), the instanton factors exhibit an ``entropic"
behavior because in the large $N$ limit one has $\theta_3\to 1$, and the
divergence appears thanks to the huge number of (compact)
integrals to be performed.

One way to detect the presence of a phase transition
at $N\to\infty$ is to look for a region in the parameter
space where the one-instanton contribution dominates the
zero-instanton sector \cite{Gross:1994mr}. In our case the ratio of the two
contributions is given by
\bea
\e^{F_0}&=&\frac{\mathcal{Z}^{1-{\rm inst}}}{\mathcal{Z}^{0-{\rm inst}}}\nonumber\\
&=& \frac{{ \,\int \prod\limits_{i=1}^N \dd z_i ~\e^{-\frac{8\pi^2
N}{a}\, \sum_{i=1}^N  z_i^2}\, \prod\limits_{j=2}^N
  \scriptstyle{\left[ \sin^2
  \left(\frac{2\pi\, t  }{a}\,\left( {z_1} - {z_j} \right)\right) -
\sin^2\left(\frac{\pi \,t }{a}\right) \right]}}
\,\prod\limits_{i< j\atop i\ge2}
\scriptstyle{\sin^2\left(\frac{2\pi\, t }{a}\,\left( {z_i}
- {z_j} \right)\right)}
  }{{~\e^{\frac{2\pi^2 N
}{a}}\,\int
  \prod\limits_{i=1}^N \dd z_i  ~\e^{-\frac{8\pi^2 N}{a}\,
\sum_{i=1}^N  z_i^2}\, \prod\limits_{1\le i< j\le
N}\scriptstyle{\sin^2\left(\frac{2\pi\, t}a\,\left( {z_i} -
{z_j} \right) \right)}}}
   \ , \nonumber\\ &&
\eea where we have dropped an irrelevant normalization factor.
 We shall employ the saddle-point technique to compute the
above ratio. Consider the integral \beq {\cal Z}^{0-{\rm
inst}}=2\int
  \prod_{i=1}^N \dd z_i  ~\e^{-\frac{8\pi^2 N}{a}\,
\sum_{i=1}^N  z_i^2} \,\prod_{1\le i< j\le
N}\sin^2\left(\mbox{$\frac{2\pi\, t\left( {z_i} - {z_j} \right)}
{a}$}\right) \ . \eeq In the large $N$ limit, it is dominated by the
solutions of the saddle-point equation \beq z_i=\frac{t}{4\pi\,
N}\,\sum_{j\ne i}\cot\left(\mbox{$\frac{2\pi\, t\left( {z_i} - {z_j}
\right)}{a}$}\right) \label{discrete}\eeq
Introduce the density \beq
\rho(z)=\frac{1}{N}\,\sum_{i=1}^N \delta(z-z_i) \eeq to rewrite eq.
(\ref{discrete}) as \beq z=\frac{t}{4\pi}\, \int \dd w~\rho(w)\,
\cot\left(\mbox{$\frac{2\pi \,t\left( {z} - {w} \right)}{a}$}\right) \ . \eeq

A convenient way to analyse this integral equation is to notice that
under the change of variables $t\mapsto \ii \hat t$ it maps to \beq
\frac{8\pi^2}{a}\,z=\frac{2\pi\, \hat t}{ a}\, \int \dd w~ \rho(w)\,
\coth\left(\mbox{$\frac{2\pi\, \hat t\left( {z} - {w}
\right)}{a}$}\right)\label{cippino5} \ . \eeq Fortunately, we do not
have to solve eq. (\ref{cippino5}) directly, because we can map this
equation to the one already solved in~\cite{Marino:2004eq} in the
context of topological strings and matrix models.
Borrowing the solution we get
 \beq
\rho(z)= \frac{4}{\hat t}\,\arctan\left(\mbox{$\sqrt{
\frac{\e^{{\hat t^2}/{a}}}{\cosh^2(\frac{2\pi\,\hat \,t\,
z}{a})}-1}$}~\right) \ . \eeq Now we analytically continue $\hat
t\to -\ii t$, giving the final result \beq \rho(z)=\frac{4}{
t}\,\mathrm{arctanh}\left(\mbox{$\sqrt{1- \frac{\e^{-{
t^2}/{a}}}{\cos^2(\frac{2\pi\, t\, z}{a})}}$}~\right)=\frac{4}{
t}\,\mathrm{arccosh}\left( \e^{{ t^2}/{2a}}\,
\cos\left(\mbox{$\frac{2\pi t z}{a}$}\right)\right) \ . \eeq The
support of the density $\rho(z)$ is easily determined by imposing
the condition $\cos^2(\frac{2\pi\, t \,z}{a})- {\e^{-{ t^2}/{a}}}\ge
0$, giving \beq\label{rhosupp} |z|<\frac{a}{2\pi
\,t}\,\arccos(\e^{-{ t^2}/{2a}}) \ . \eeq

In the
large $N$ limit, the ratio $\e^{F_0}$ is completely determined
by this distribution as\bea
&&\e^{F_0}\label{eFlargeN}\\
&&\quad=~\,\int \dd z~ \exp\left(\mbox{$ -\frac{8\pi^2 N\,
z^2}{a}-\frac{2\pi^2 N }{a}$}+N \,\int\dd w~
\log\left[\mbox{$\frac{\sin^2\left(\frac{\pi\, t}{a} \right)
-\sin^2\left(\frac{2\pi \,t(z-
w)}{a}\right)}{\sin^2\left(\frac{2\pi\, t (z-w)}{a}\right)
}$}\right]\,\rho(w)\!\! \right) \ .\nonumber \eea We can safely
assume that the integral over $z$ is peaked around $z=0$ in the
large $N$ limit. Then eq.~(\ref{eFlargeN}) reduces to \beq \e^{F_0}=
\, \exp\left( -\frac{2\pi^2 N }{a}+{N } \,\int \dd w~
\log\left[\mbox{$\frac{\sin\left(\frac{2\pi\, t(1/2+w)}{a}
\right)\sin\left(\frac{2\pi\, t(1/2-w)}{a} \right)
}{\sin^2\left(\frac{2\pi t\, w}{a}\right)}$}\right]\,\rho({ w})
\right) \ . \label{ratio}\eeq This integral can be computed in
closed form but we prefer to follow a different route. The basic
quantity to consider is \beq \label{cipu} G(x)=\int \dd w~ \rho({
w})\, \log\left|\sin\left(\mbox{$\frac{2\pi\, t
(x-w)}{a}$}\right)\right|-\frac{4 \pi^2}{a} \,x^2 \ . \eeq The
expression (\ref{ratio}) can then be written in terms of $G(x)$ as
\beq \label{fippo} \e^{F_0}=\, \exp\left[ N
\bigl(G(\mbox{$\frac12$})+G(-\mbox{$\frac12$})-2
G(0)\bigr)\right]=\, \exp\left[ 2N \bigl(G(\mbox{$\frac12$})-
G(0)\bigr)\right] \ , \eeq where we have used the reflection
symmetry $G(x)=G(-x)$.

The function (\ref{cipu}) has an interesting behaviour.
Both $G(x)$ and $G'(x)$
are continuous even when $x$ reaches one of
the endpoints of the support of the density $\rho$ which vanishes as $\sqrt x$ there.
The integral (\ref{cipu}) is thus
convergent also at $x=\pm\,\frac{a}{2\pi\, t}\, \arccos(\e^{-{
t^2}/{2a}})$. Moreover, $G(x)$ is constant when $x$ lies in the support region (\ref{rhosupp})
of the density.
By taking the derivative with respect to $x$ we obtain \beq
\label{cipponi} G^\prime(x)=\frac{2\pi\, t}{a}\,\int \dd w~ \rho({
w})\,\cot\left(\mbox{$\frac{2\pi\, t (x-w)}{a}$}\right)-\frac{8 \pi^2}{a} \,x=0 \eeq
due to the saddle-point equation. For $x$ outside of the support region (\ref{rhosupp}), the saddle-point equation
no longer holds and we cannot conclude that $G(x)$ is constant.
In order to understand
what happens in this instance it is sufficient
compute the second derivative
\beq
G^{\prime\prime}(x)=-\left(\frac{2\pi\, t}{a}\right)^2\,\int \dd w~
\frac{\rho({ w})}{\sin^2\left(\mbox{$\frac{2\pi t (x-w)}{a}$}\right)}-\frac{8
\pi^2}{a} \ , \eeq
which is easily seen to be negative in the interval under consideration.
This means that $G^\prime(x)$ is a
continuous decreasing function in the interval $|x|\ge
\frac{a}{2\pi \,t}\,\arccos(\e^{-{ t^2}/{2a}})$. Since
$G^\prime(x)$ vanishes at the endpoints of the interval, it must
be negative. $G(x)$ is therefore a decreasing function for $
|x|\ge \frac{a}{2\pi \,t}\,\arccos(\e^{-{ t^2}/
{2a}})$.

We conclude
that if $x=\frac12> \frac{a}{2\pi\, t}\,\arccos(e^{-{
t^2}/{2a}})$ then
\beq
G(0)=G\left(\mbox{$\frac{a}{2\pi\, t}\,\arccos(\e^{-{
t^2}/{2a}})$}\right)>G(\mbox{$\frac12$})  \ ,
\eeq
which implies that the zero-instanton contribution is exponentially larger
than the one-instanton contribution. Instead, if $x=\frac12\le
\frac{a}{2\pi\, t}\,\arccos(\e^{-{ t^2}/{2a}})$ then \beq
G(0)=G(\mbox{$\frac12$}) \eeq and from (\ref{fippo}) we see that
the two contributions are of the same order.
We conclude that a critical curve exists and is given by
\beq \frac{1}{2}=\frac{a}{2\pi \,t}\,\arccos(\e^{-{
t^2}/{2a}}) \ ,\label{curvtrans} \eeq which separates two different regimes
of the theory. In terms of the original parameters $(t,p)$ it can
be written in the form \beq t=p\,\log \left({\sec
(\mbox{$\frac{\pi }{p}$})}^2\right) \ .\label{trans} \eeq In particular, we see that
a critical value $t_c=p\,\log ({\sec (\frac{\pi }{p})}^2)$ is present
at fixed $p$, with the instanton contributions to the free energy being
exponentially suppressed for $t<t_c$.

\subsection{Resurrecting the resolved conifold}

In ordinary Yang-Mills on the sphere the analogous computations
single out a critical value for the effective 't Hooft coupling
$\lambda_c=\pi^2$. The statistical weight of instantons grows
large enough above $\lambda_c$ to make them the favorable
configurations. The first natural check of our result
(\ref{trans}) is to recover $\lambda_c$ in the limit $p\to\infty$.
Multiplying eq.~(\ref{trans}) by $p$ and recalling that the
ordinary gauge theory is reached by sending $p\to\infty$ with $N
\,g_sp=a$ fixed, we obtain $\lambda_c=\pi^2$ immediately.

The presence of $\lambda_c$ separates the strong-coupling phase
wherein instantons are not suppressed and a non-trivial master field
is generated, from a weak-coupling phase governed by a gaussian
master field. From the string theoretical point of view,
the Gross-Taylor string expansion accurately
describes the strong-coupling phase, while in weak-coupling phase stringy
degrees of freedom do not appear. This suggests that, in a certain
sense, strings emerge from instantons.

In the present case, for $t<t_c$, our result seems to
support again the conclusion that the theory truncates to
the zero-instanton sector in the large $N$-limit. The novelty
this time is that the weak-coupling phase still admits a string
description, but not the expected one! Let us consider the large
$N$ limit of the zero-instanton sector. Inserting
${w}^{\mathrm{inst}}_q(0,\dots,0)$ into the instanton expansion
we obtain
\beq {\cal Z}^{0-{\rm inst}}=\frac{1}{N!}\,\left(\frac{2\pi}{g_s p}\right)^N
~\e^{-\frac{{g_s}(  N^3-N ) }{6 p}}\,\int
  \prod_{i=1}^N \dd z_i  ~\e^{-\frac{8\pi^2 N}{a}\,
\sum_{i=1}^N  z_i^2} \,\prod_{1\le i< j\le
N}\sin^2\left(\mbox{$\frac{2\pi\, t( {z_i} - {z_j} )} {a}$}\right) \
. \label{CSM}\eeq As expected, the partition function (\ref{CSM})
coincides with the partition function $Z_{\rm CS}^p$ of Chern-Simons
theory on the Lens space $L_p$ in the background of the trivial
flat-connection with partition $\{N_k\}=(0,\dots,0)$. According
to~\cite{Marino:2002fk,Aganagic:2002wv}, eq.~(\ref{CSM}) is its
matrix-model representation. The large $N$ limit in the trivial
vacuum is much easier to handle than in the general case. It can be
explicitly performed by using, for example, the orthogonal
polynomial technique explained in \cite{Marino:2004eq} or it can be
obtained directly from the knowledge of the analogous computation
for $S^3$ \cite{Gopakumar:1998ki} by simply identifying the
parameters. The partition function of the closed topological string
theory on the resolved conifold ${\cal O}(1)\oplus{\cal O}(-1)\to
\proj^1$ emerges in the limit
 as an effect of the
geometric transition. The only subtle point is that the string
coupling constant, appearing in the closed string expansion, is
renormalized as \beq \hat{g}_s=\frac{g_s}{p} \ . \eeq This is
expected from the relation with Chern-Simons theory, since the
K\"ahler modulus is $\hat{t}=\hat{g}_s N$. This conclusion for the
weak-coupling phase of $q$-deformed theory should be true at all
order in $1/N$, the perturbative contribution to this expansion
coming solely from the zero-instanton sector. In order to check this
explicitly, an analysis in terms of discretized orthogonal
polynomials, as done in (\cite{Gross:1994mr}) in the ordinary case,
should be performed.

From the point of view of the conjecture \cite{Ooguri:2004zv}, we have the
rather surprising result that below a certain value of the K\"ahler
modulus the string theory associated to $q$-deformed Yang-Mills
theory does not seem to be the correct one. In particular, the
background geometry does not depend on the parameter $p$. The
Gopakumar-Vafa invariants, representing the BPS content of the
theory, are the same as the ones of the resolved conifold.

\begin{figure}[htb]
\begin{center}
\epsfxsize=3.5 in \epsfysize=2 in \epsfbox{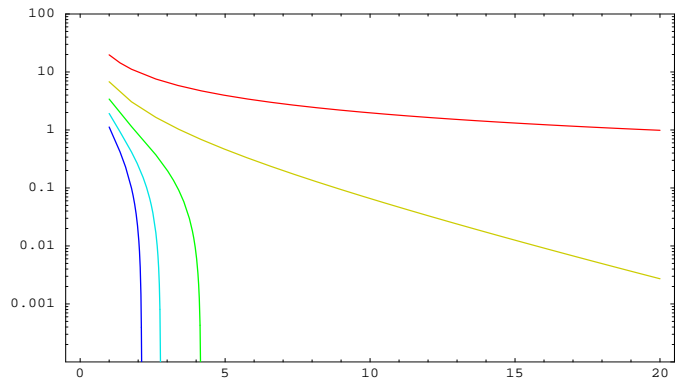}
\end{center}
\caption{\label{PLOT} Plotting $F_0$ for $p=1,2,3,4,5$ as a function
of $t$}
\end{figure}

One should expect that above the transition point $t>t_c$ the
non-perturbative contributions, associated to the non-trivial flat
connections, will generate the desired geometry. Looking closer at
the transition curve (\ref{curvtrans}), we realize that the cases
$p=1$ and $p=2$ are exceptional from this point of view. The
instanton contribution dominates when \beq
\frac{\pi}{p}\geq\arccos(\e^{-{ t}/{2p}}) \ .\label{curvtrans2} \eeq
For $p=1$ there are no real solutions for $t$ while the case $p=2$
experiences the transition at $t_c=\infty$. This suggests that for
$p\leq 2$ the theory always remains in the weak-coupling phase. This
conclusion is not affected by the presence of the parameter
$\theta$. One can easily check that the introduction of a
$\theta$-angle does not change the density, controlling the
zero-instanton large $N$ limit, nor the the ratio between the
nonperturbative and perturbative contributions. In the case $p=1$
the correct geometry is nevertheless obtained, but at this level we
do not see any chiral-antichiral factorization, nor any fiber
D-brane contributions. It is instructive to see explicitly how the
behaviour changes above the value $p=2$ by plotting the integral
defining $F_0$ in eq.~(\ref{ratio}) (Fig.~\ref{PLOT}).
In the subsequent sections we will recover and extend these results
by turning to a powerful saddle-point analysis, originally applied
to ordinary Yang-Mills theory in~\cite{Douglas:1993ii}.

\section{Saddle point equation}

An explicit result for the leading order (planar) contribution to
the free energy of ordinary Yang-Mills theory on the sphere was
obtained in \cite{Douglas:1993ii}. For large area it fits nicely with
the interpretation in terms of branched coverings that arises in the
Gross-Taylor expansion,
down to the phase transition point at $\lambda_c=\pi^2$
where the string series is divergent. We will now perform similar
computations for $q$-deformed Yang-Mills theory on $S^2$.

\subsection{Deformed Douglas-Kazakov phase transition}

Recall that the partition function of the
$q$-deformed gauge theory on $S^2$ is given by \beq
{Z}_{\rm YM}^{q}(g_s,p)= \sum_{n_1, \dots, n_N\in
\mathbb{Z}\atop n_i-n_j \ge i-j \ \mathrm{for}\ i\ge j}
\e^{-\frac{g_s p}{2}\, {( n_1^2+\cdots+n_N^2)}}\,\prod_{1\le i<j\le N}
\sinh^2\left(\mbox{$\frac{g_s}{2}(n_i-n_j)$}\right) \ . \eeq The constraint on
the sums keeps track of the meaning of the integers $n_i$ in
terms of Young tableaux labels and highest weights. As in the
previous section, we shall introduce the 't Hooft parameter $t=g_s
N$ which is kept fixed as $N\to\infty$, and the additional
parameter $a=g_s N \,p=t\, p$. In terms of these new variables, the
partition function at $N\to\infty$ is given by \beq
\label{ZYMqnewvars} {Z}_{\rm YM}^{q}(t,a)=\exp\bigl(N^2 F(a,t)\bigr)=
\lim_{N\to  \infty}\,\sum_{n_1, \dots, n_N\in
\mathbb{Z}\atop n_i-n_j \ge i-j \ \mathrm{for}\ i\ge j} \e^{-N^2
S_{\rm eff}(\vec{n})} \eeq where \beq
S_{\rm eff}(\vec{n})=-\frac{1}{N^2}\,\sum_{i\ne
j}\log\left[\sinh\left(\mbox{$\frac{t}{2}\,\frac{|n_i-n_j|}{N}$}\right)\right]+
\frac{a}{2 N}\,\sum_{i=1}^N\left(\frac{n_i}{N}\right)^2  \ . \eeq The limit
can be explicitly evaluated by means of saddle-point techniques
usually employed in matrix models.

We introduce the variable $x_i=i/N$ and the function $n(x)$ such that
\beq n(x_i)=\frac{n_i}{N} \ . \eeq In the large $N$ limit, $x_i$
becomes a continuous  variable $x\in[0,1]$ and
$S_{\rm eff}$ can be written as an integral \beq \label{pippino1}
S_{\rm eff}=-\int_0^1 \dd x ~\int^1_0
\dd y~\log\left[\sinh\left(\mbox{$\frac{t}{2}$}\,\bigl|n(x)-n(y)\bigr|\right)\right]+
\frac{a}{2 }\,\int_0^1 \dd x ~ n(x)^2  \ . \eeq In the continuum limit the constraint on the series
(\ref{ZYMqnewvars}) becomes \beq \label{pippino2} n(x)-n(y)\ge
x-y \ \ \ \ \ \mathrm{for}\ \ \ \ x\ge y  \ .\eeq From
eq.~(\ref{pippino2}) it follows that $n(x)$ is a monotonic function, and so we can consider
its inverse $x(n)$. In
eq.~(\ref{pippino1}) we perform the change of variable $x\mapsto x(n)$
to get \beq
\label{pippino3} S_{\rm eff}=-\int_{n(0)}^{n(1)}\dd n~
\int_{n(0)}^{n(1)}\dd n^\prime ~\frac{\partial x(n)}{\partial n}\,
 \frac{\partial y(n^\prime)}{\partial n^\prime}\,\log\left[
\sinh\left(\mbox{$\frac{t}{2}\,|n-n^\prime|$}\right)\right]+
\frac{a}{2 }\,\int_{n(0)}^{n(1)}\dd n ~ \frac{\partial
x(n)}{\partial n} \ n^2 \ .\eeq To simplify notation, we
introduce the function \beq \rho(n)=\frac{\partial x(n)}{\partial
n} \eeq and denote the interval $[n(0),n(1)]$ by $C$.
Then \beq \label{pippino4} S_{\rm eff}=-\int_{C} \dd w~
\int_{C} \dd w^\prime \rho(w)\,\rho(w^\prime\,)\,
 \log\left[\sinh\left(\mbox{$\frac{t}{2}$}\,|w-w^\prime|\right)\right]+
\frac{a}{2 }\,\int_C \dd w ~\rho(w) \ w^2 \ . \eeq From eq.~(\ref{pippino2}) we
may translate the original constraint on the series in (\ref{ZYMqnewvars})
in terms of the function $\rho$ as \beq
\rho(n)=\lim_{n^\prime\to n}\,\frac{x(n^\prime)-x(n)}{n^\prime-n}\le
1 \ . \label{bound1}\eeq

The constraint (\ref{bound1}) is of fundamental importance in the
calculation of the free energy, and in the underformed
case~\cite{Douglas:1993ii} it leads to non-trivial consequences such
as the large $N$ phase transition and the existence of the
strong-coupling phase. In the deformed case this same constraint
holds as it follows from the fact that we started from a discrete
summation over Young tableaux. The density $\rho(x)$ is positive
since the function $x(n)$ is monotonic. Furthermore, one has \beq
\int_C \dd n ~\rho(n)=x\bigl(n(1)\bigr)-x\bigl(n(0) \bigr)=1-0=1 \ .
\eeq The distribution $\rho(x)$ in eq.~(\ref{pippino4}) can be
determined by requiring that it minimizes the action. This implies
that it satisfies the saddle-point equation \beq \label{cippino5a}
\frac{a}{2}\, z=\frac{t}{2}\,\int_C \dd w ~\rho(w)\,
\coth\left(\mbox{$ \frac{t}{2}$} \,(z-w)\right) \ . \eeq This
equation is a deformation of the usual Douglas-Kazakov equation that
governs ordinary QCD$_2$ on the sphere and it is of the same type of
eq.~(\ref{cippino5}) which controls the zero-instanton sector. The
ordinary gauge theory is recovered when $t\to 0$ while $a$ is kept
fixed according to the double-scaling limit. Eq.~(\ref{cippino5a})
then reduces to the usual saddle-point equation for the gaussian
one-matrix model. The one-cut solution of eq.~(\ref{cippino5a}) is
given in \cite{Marino:2004eq} where the large $N$ limit of the
Chern-Simons matrix model \cite{Marino:2004eq,Tierz:2002jj} was
considered.

Borrowing again the solution, the distribution $\rho(z)$ is given by
\beq \rho(z)= \frac{a
}{\pi\, t }\, \arctan\left(\mbox{$\sqrt{\frac{ \e^{t^2/a}}{\cosh^2(\frac{t
\,z}{2})}-1}$}~\right) \eeq with the symmetric support \beq
\label{ioppi} z\in\bigl[-\mbox{$\frac{2}{t}$}\, \mathrm{arccosh}(\e^{-t^2/2
a})~,~\mbox{$\frac{2}{t}$}\, \mathrm{arccosh}(\e^{-t^2/2 a}) \bigr] \ . \eeq It
can be considered as the $q$-deformation of the well-known Wigner
semi-circle distribution. We now come to the crucial point. In
order for $\rho(z)$ to be an acceptable solution, it has to
satisfy the bound in eq.~(\ref{bound1}), i.e. it must be bounded from above by~$1$.
We recall that this constraint originates from the structure of the discrete sum
defining the partition function and it
defines the regime in which the theory is well-approximated
by a matrix model (up to nonperturbative $\frac1N$
corrections). To verify
this constraint, we simply notice that the function $\rho(z)$ attains
its absolute maximum at $z=0$. In terms of the variables $t$ and $p$,
this produces the
inequality\beq
\arctan\left(\sqrt{{ \e^{t/p}}-1}~\right)\le \frac{\pi}{p} \ .
\label{pippino6}\eeq Since the $-\frac\pi2\leq\arctan(z)\leq\frac\pi2$,
this condition is always satisfied for
$p=1$ or $p=2$. The situation changes for $p>2$. Since $\tan(z)$
is a monotonically increasing function for $z\in[-\frac\pi2,\frac\pi2]$, the inequality
(\ref{pippino6}) can be equivalently written as $\sqrt{{
\e^{t/p}}-1}\le \tan\frac{\pi}{p}$ which implies that \beq t\le
p\,\log\left(\sec^2\left(\mbox{$\frac{\pi}{p}$}\right)\right)  \ . \eeq This means
that our solution breaks down when the '{t}~Hooft coupling $t$ reaches the
critical value $t_c$ found from the instanton computation in
the previous section. This result reappears here from a different
and more general point of view. Again, note that the cases $p=1$
and $p=2$ are special because then the one-cut solution is always valid. The
relevant closed topological string theory is generated through a
geometric transition, with the master field producing strings living on
the resolved conifold.

The breakdown of the one-cut solution parallels exactly what
happens in ordinary two-dimensional Yang-Mills theory, where it signals the
appearance of a phase transition. In the saddle-point approach it
is possible to go further and to find a solution describing the
strong-coupling phase. To understand the nature of the phase
transition and to describe the stringy geometry of the strong-coupling
regime, we have to determine the distribution $\rho(z)$ that
governs the system at $t>t_c$. Before proceeding with this analysis, we will
compute the free energy in the weak-coupling phase.

\subsection{Resolved conifold from the one-cut solution}

The free energy in the weak-coupling phase is determined in
terms of the gaussian potential and density $\rho(x)$ through \bee
\label{eq1} \mathcal{F}( t,a)=-\int_{-b}^b\dd x~
\int_{-b}^b\dd y\ \rho(x)\, \rho(y)\, \log\left |\sinh\left(\mbox{$\frac{
t}{2}$}\,(x-y)\right) \right|+\frac{a}{2}\,\int_{-b}^b \dd x~ x^2\, \rho(x) \ ,
\eeq where the interval $C=[-b,b]$, $b=\frac2t\,{\rm arccosh}(\e^{t^2/2a})$
is the support of $\rho(x)$ given in
eq.~(\ref{ioppi}). Instead of working with eq.~(\ref{eq1}), it is simpler
to compute its derivative with respect to $a$. In fact, when the
derivative acts on one of the limits of the integrals it produces
the distribution $\rho(x)$ evaluated at $x=b$ or $x=-b$, where it
vanishes. Moreover, when the derivative is applied directly to
$\rho(x)$ the different contributions cancel because of the saddle-point
equation. The only remaining contribution is \beq
\frac{\partial\mathcal{F}( t,a)}{\partial a}=\frac{1}{2}\,\int_{-b}^b
\dd x~x^2\, \rho(x) \ , \eeq as in the usual one-matrix model with a
gaussian potential. There is, however, an important difference from the
standard case. Normally, derivatives of
the free energy are easily related to the Laurent expansion of the
resolvent which solves the Riemann-Hilbert problem associated to the
saddle-point equation. Here the presence of a deformed kernel (or
of a non-polynomial potential in the matrix-model variables)
means that no such simple relation exists, and the derivative of the
free energy has to be computed by brute force techniques. This characteristic
will complicate computations in the next section when we study two-cut solution. In the
single-cut case, the final result is well-known \cite{Marino:2004eq} but
it is instructive to perform the integral explicitly.

We have to evaluate the integral \beq \label{eq5} \int^{b}_{-b}
\dd x~ x^2\, \rho(x)=\frac{a}{\pi\, t}\int^{b}_{-b} \dd x~ x^2\,
\arctan\left(\mbox{$\sqrt{\frac{\e^{{{ t}^2}/{a}}}{\cosh^2( \frac{ t\, x
}{2})}-1}$}~\right) \ . \eeq
Changing variables $x=
(\log( y)-\frac{{t}^2}a)/{t}$ we have \bea \int^b_{-b}
\dd x~x^2\,\rho(x)&=& \frac{a}{\pi\,
{t}^4}\,\int^{\infty}_{-\infty}\frac{\dd y}{y}~ \left(\log
(y)-\mbox{$\frac{{t}^2}a$}\right)^2 \,\arctan\left(\mbox{${\frac{\sqrt{4 y-(1+y
~\e^{-{t}^2/a})^2}}{1+y ~\e^{-{t}^2/a}}}$}~\right)\nonumber\\\label{eq6}
&&\times\, \Theta\left[4\,y -
{\left( 1 + {y}~{\e^{-{t^2 }/{a}}} \right) }^2\right]
\eea
where $\Theta(y)$ is the step function. Starting from the integral
representation
\beq\rho\bigl(y(x);a,t\bigr)=
\int_{0}^1 \frac{\dd \xi}{\pi\, {t}^2 }~ \frac{\e^{-{{t}^2\,\xi
}/{a}}\,\Theta\left[4\,y - {\left( 1 + {y}{~\e^{-{{t}^2\,\xi
}/{a}}} \right) }^2\right]
 }{{\sqrt{4\,y - {\left( 1 + {y}{~\e^{-{{t}^2\,\xi }/{a}}} \right)
 }^2}}}
\eeq
we may rewrite eq.~(\ref{eq6}) as
\beq \int^{b}_{-b}\dd x~ x^2
\,\rho(x) =\int_{0}^1 \frac{\dd \xi}{\pi\, {t}^2 }~
\int_{y_1(\xi)}^{y_2(\xi)} \dd y~\left(\log
(y)-\mbox{$\frac{{t}^2}a$}\right)^2\,
\frac{\e^{-{{t}^2\,\xi }/{a}} }{{\sqrt{4\,y -
{\left( 1 + {y}{~\e^{-{{t}^2\,\xi }/{a}}} \right) }^2}}}  \ , \eeq
where $y_1$ and $y_2$ are the zeroes of the polynomial in the variable
$y$ which is the
argument of the step function $\Theta$ in eq.~(\ref{eq6}),
so that $[y_1(\xi),y_2(\xi)]$ is the support of $\Theta$.

After another change of variable \bee y=-\e^{{{t}^2
\xi}/{a}} + 2~\e^{{2\,{t}^2\xi}/{a}} + 2\cos(\theta)\,
{\sqrt{-\e^{{3\,{t}^2\xi}/{a}} + \e^{{4\,{t}^2\xi}/{a}}}}
\eeq the integral reduces to
\bea \int^b_{-b}\dd x~x^2\,\rho(x)&=&\int_0^1 \frac{\dd\xi}{a^2\pi\,
  {t}^2}~ \int_0^\pi\dd\theta
~\Bigl( t^2\left( 2\xi -1 \right)\Bigr.\nonumber\\&&\left.  +\, a\,\log \left[2 - \e^{-
{t^2\,\xi }/{a}  } +
        2\,{\sqrt{1 - \e^{-{t^2\,\xi }/{a}  }}}\,\cos \theta \right]
    \right)^2
\nonumber\\
&=&\frac{t^2}{3 a^2} +\frac{2}{t^2}\, \int_0^1 \dd\xi~
\textrm{Li}_2\left(1 - \e^{-{t^2\,\xi }/{a}}\right) \ , \eea
where ${\rm Li}_n(x)$ is the polylogarithm function of order $n$.
The remaining integral can be evaluated by integration
by parts and all the remaining
integrals are easily computed by employing the Mellin representation
of the polylogarithm function to get\bea \int_{-b}^b\dd x~x^2\,\rho(x)&=&
\frac{t^2}{3 a^2} +\frac{2}{t^2}~
\textrm{Li}_2\left(1 - \e^{- {t^2 }/{a}}\right)-
\frac{4a}{t^4}\,\zeta(3)-\frac{2}{a}\,\log \left(1 - \e^{-{t^2}/{a} }\right)
\nonumber\\ &&+\,
  \frac{4}{t^2}~\mathrm{{Li}}_2\left(\e^{-{t^2}/{a}  }\right) +
  \frac{4a}{t^4}~\mathrm{Li}_3\left(\e^{-{t^2}/{a} }\right) \ .
\eea Next we use the polylogarithm identity
\beq
-\mbox{$\frac{{\pi }^2}{6}$} - {x}\,\log (1 - \e^{-x}) +
\mathrm{Li}_2(\e^{-x}) + \mathrm{Li}_2(1 - \e^{-x})=0
\eeq
to simplify the integral finally to \beq \int_{-b}^b\dd x~x^2\,\rho(x)=\frac{t^2}{3 a^2}
+\frac{\pi^2}{3 t^2} - \frac{4a}{t^4}\,\zeta(3) +
  \frac{2}{t^2}~\mathrm{{Li}}_2(\e^{-{t^2}/{a}  }) +  \frac{4a}{t^4}~\mathrm{Li}_3(\e^{-{t^2}/{a} }) \ .
\eeq This expression is easily integrated to give the free energy
\beq {\cal F}(t,a)=-\frac{t^2}{6 a} +\frac{\pi^2 a}{6 t^2} -
\frac{a^2}{t^4}\,\zeta(3) +  \frac{a^2}{t^4}~\mathrm{Li}_3(\e^{-
{t^2}/{a} })+c(t) \ , \eeq which as expected coincides with the
genus 0 free-energy of closed topological string theory on the
resolved conifold. The $a$-independent function $c(t)$ can be easily
determined by looking at the asymptotic expansion in $t$. In the
limit $t\to 0$, $p\to \infty$ the free-energy of ordinary Yang-Mills
theory in the weak-coupling phase is easily recovered. We have also
to remark that the free energy depends only on the combination
$t^2/a$ (or $t_s/(p(p-2))$ in string variable). This dependence is
in contrast with the one expected from the geometry  of the
Calabi-Yau $X={\mathcal O}(p+2g-2)\oplus{\mathcal
O}(-p)~\longrightarrow~\Sigma_g $. This picture, in fact, contains
as a necessary ingredient two independent moduli $e^{-t_s}$ and
$e^{-\frac{2 t_s}{p-2}}$.

\section{Two-cut solution and the strong-coupling phase}

For $p>2$, the one-cut solution constructed earlier breaks down when
the 't Hooft coupling constant $t$ reaches the critical value $t_c$.
To capture the behaviour of the theory above this threshold, we have
to look for an ansatz that keeps track of the physical bound
(\ref{bound1}) satisfied by the distribution $\rho(x)$. We expect
that the free energy obtained in this regime will reproduce the
black hole partition function described in Sect.~2 in terms of
topological strings on $X={\cal O}(-p)\oplus{\cal O}(p-2)\to
{\mathbb P}^1$, with its chiral-antichiral dynamics and fiber
D-brane contributions.

\subsection{The two-cut solution}

\begin{figure}[htb]
\begin{center}
\epsfxsize=3.0 in \epsfysize=1.5 in \epsfbox{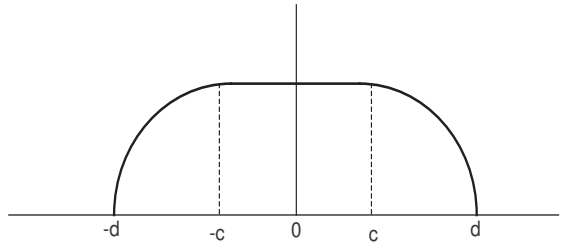}
\end{center}
\caption{\label{dbcut} The double cut ansatz for the distribution
$\rho(z)$.}
\end{figure}

For $t>t_c$ we still have to solve the saddle point equation, but in
the presence of the boundary condition (\ref{bound1}). The new
feature which may arise is that a finite fraction of Young tableaux
variables $n_1,\dots,n_N$ condense at the boundary of the inequality
by respecting the parity symmetry of the problem, \beq
n_{k+1}=n_{k+2}=\cdots=n_{N-k}=0 \ , \eeq while all others are
non-zero. This observation translates into a simple choice for the
profile of $\rho(z)$ as depicted in Fig. \ref{dbcut}.
 In order to
respect the bound, the distribution function  is chosen constant and
equal to $1$ everywhere in the interval $[-c,c]$.  In the intervals
$[-d,-c]$ and $[c,d]$ its form is instead dynamically determined by
the saddle-point equation. Since the potential is symmetric under
parity $z\to-z$, so is the function $\rho(z)$. The choice of
symmetric $\rho$ has a deeper meaning at the level of $q$-deformed
Yang-Mills theory. It hides the assumption that the two-cut solution
is dominated by the sector with vanishing $U(1)$ charge. The
asymmetry of the distribution can be related to this charge. This
has been verified explicitly in the context of ordinary
two-dimensional QCD in
 \cite{Minahan:1993tp,Alimohammadi:1997wc}.

Let us denote the set $[-d,-c]\cup [c,d]$ by $\mathcal{{R}}$ and the
restriction of the distribution function to this set by $\tilde
\rho=\rho|_{\cal R}$. Then the saddle-point equation \beq
\frac{a}{2}\, z=\frac{t}{2}\,\int_{-d}^d \dd w~ \rho(w)\, \coth
\left(\mbox{$\frac{t}{2}$}\,(z-w)\right) \eeq
 can be rewritten as
\beq \label{cipow3}\frac{a}{2}\, z-\log\left
|\frac{\sinh\left(\frac{t}{2}\,(z+c)\right)}{\sinh\left(\frac{t}{2}\,(z-c)\right)}\right|=
\frac{t}{2}\,\int_\mathcal{R}\dd w~ {\tilde{\rho}}(w)\, \coth
\left(\mbox{$\frac{t}{2}$}\,(z-w)\right) \ . \eeq It can be cast in
a more standard form if we perform the change of variables
$y=\e^{t\, z}~\e^{t^2/a}$ and $u=\e^{t\, w}~\e^{t^2/a} $. The
saddle-point equation can then be written as \beq \frac{a}{2 t^2 }
\,\frac{\log(y)}{y}-\frac{1}{t\, y}\,\log \left(\mbox{$\frac{\e^{
c\,t-t^2/a}\, y-1}
   {\e^{-c\,t-t^2/a} \,y -1}$}\right)
= \int_\mathcal{R} {\dd u}~ \frac{\hat\rho(u)}{y-u} \ , \eeq  where
we have defined \beq \hat \rho(u)=\frac{\tilde\rho\bigl(\log(u
~\e^{-t^2/a})/t\bigr)}{u\, t}\ . \eeq The support $\mathcal{R}$ in
terms of these new variables is given by \beq {\cal R}=\bigl[ \
\underbrace{\e^{-t\, d+t^2/a}}_{\e^{d_-}}\ \,,\,\underbrace{\e^{-t\,
c+t^2/a}}_{\e^{c_-}}\ \bigr]~\cup~\bigl[ \ \underbrace{\e^{t
\,c+t^2/a}}_{\e^{c_+}}\ \,,\, \underbrace{\, \e^{t\,
d+t^2/a}}_{\e^{d_+}}\ \bigr] \ . \eeq

The original potential has been modified by a new term that depends
explicitly on the endpoints of the support of the distribution. This
apparently mild modification of the original one-cut problem will
lead to a number of consequences. The original symmetry of our
ansatz seems to have completely disappeared when the saddle-point
equation is written in this way. This equation actually exhibits a
more subtle symmetry. Let us consider the transformation \beq
\label{symmegh} y\longmapsto~\frac{\e^{{2 t^2}/{a}}}{y} \ \eeq which
is the version of the original parity symmetry $z\mapsto -z$ in the
new variables. It is straightforward to check that the saddle point
equation and the region of integration are unaltered by
(\ref{symmegh}). This invariance will be very important in the
following, since it will govern many of the mysterious cancelations
that will occur. In particular, it will play an important role in
solving the apparent puzzle of having more equations than unknowns.
At the level of the distribution function $\hat\rho(y)$ and of the
corresponding resolvent $\omega(z)$, it implies the two useful
functional relations \beq \hat \rho(y)=\frac{\e^{2
t^2/a}}{y^2}\,\hat\rho\left(\mbox{$\frac{\e^{{2 t^2}/{a
}}}{y}$}\right) \eeq
 and
\beq \omega(z)=\int_{\mathcal{R}} \dd w~\frac{\hat\rho(w)}{w-z}=
\frac{1}{z}\,\int_{\mathcal{R}} \dd  y~ \frac{y \,\hat
\rho(y)}{\frac{\e^{{2 t^2}/{a }}}{z}-y}=-\frac{1-2c}{z}
-\frac{\e^{{2 t^2}/{a }}}{z^2}\,\int_{\mathcal{R}} \dd y ~\frac{
\hat \rho(y)}{y-\frac{\e^{{2 t^2}/{a }}}{z}} \ , \eeq which imply
that \beq \omega(z)+\frac{\e^{{2 t^2}/{a }}}{z^2}\,
\omega\left(\mbox{$\frac{\e^{{2 t^2}/{a }}}{z}$}\right)=-\frac{1-2
c}{z} \ . \label{omegafunrel}\eeq

\begin{figure}[htb]
\begin{center}
\epsfxsize=4 in \epsfysize=2 in \epsfbox{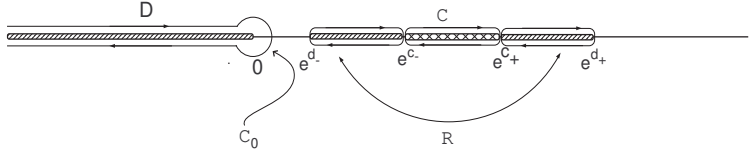}
\end{center}
\caption{\label{dbcut1} The cut structure of the distribution
$\rho(z)$.}
\end{figure}

The functional relation (\ref{omegafunrel}) states that the
behaviour of the resolvent $\omega(z)$ at large $z$ is determined by
its behaviour about the origin. The only new information at
$z\to\infty$ which is not present in the expansion around $z=0$ is
contained in the first term of the Laurent expansion and is given by
\beq \label{behav1} \omega(z)\simeq -\frac{1-2 c}{z} \ \ \ \ \ \
\mathrm{as}\ \ \ z\to \infty \ . \eeq It is exactly this term that
carries the dynamical information. Imposing this condition will lead
to two equations that determine the endpoints of the support
intervals. It may seem puzzling that we are trying to fix four
variables ($c_\pm, d_\pm$) with only two equations, but by
definition one has \beq c_++c_-=\frac{2 t^2}{a}  \ , \ \ \ \  \ \
d_-+d_+=\frac{2 t^2}{a} \ . \eeq If we choose the cuts of the square
root and of the logarithms as indicated in Fig. \ref{dbcut1},
then we can write the solution of the integral equation
(\ref{cipow3}) as a contour integral around the cuts
$[\e^{d_-},\e^{c_-}]$ and $[\e^{d_+},\e^{c_+}]$ as \bea
\omega(z)&=&-\frac{1}{2\pi \ii}\,\oint_\mathcal{R} \dd w ~
 \frac{\frac{a}{2 t^2 }
\,\frac{\log(w)}{w}-\frac{1}{t\, w}\,\log \left(\frac{\e^{-c_-}w -1}
   {\e^{-c_+}w -1
   }\right)}{z-w}\nonumber\\ && \times\,
\sqrt{\frac{(z-\e^{c_-})(z-\e^{c_+})(z-\e^{d_+})(z-\e^{d_-})}{(w-\e^{c_-})
(w-\e^{c_+})(w-\e^{d_+})(w-\e^{d_-})}} \ . \eea Since this integral
decays as $\frac1{w^3}$ at infinity, we can deform the contour of
integration so that it encircles the cuts of the two logarithms.
This costs an additional contribution coming from the pole at $w=z$
and we find \bea \omega(z)&=& -\frac{a}{2t^2
z}\,\underbrace{\int_{-\infty}^{0} \frac{\dd w}{z-w}~
\sqrt{\frac{(z-\e^{c_-})(z-\e^{c_+})(z-\e^{d_+})(z-\e^{d_-})}{(w-\e^{c_-})
(w-\e^{c_+})(w-\e^{d_+})(w-\e^{d_-})}}}_{\mathcal{I}}
\nonumber\\
&&-\,\frac{a}{2t^2 z}\,\underbrace{\int_{-\infty}^{-\epsilon}
\frac{\dd w}{w}~
\sqrt{\frac{(z-\e^{c_-})(z-\e^{c_+})(z-\e^{d_+})(z-\e^{d_-})}
{(w-\e^{c_-})(w-\e^{c_+})
(w-\e^{d_+})(w-\e^{d_-})}}}_{\mathcal{C}_0}
\nonumber\\
&&+\,\underbrace{\frac{1}{t}\,\int_{\e^{c_-}}^{\e^{c_+}} \frac{\dd
w}{w(z-w)}~
\sqrt{\frac{(z-\e^{c_-})(z-\e^{c_+})(z-\e^{d_+})(z-\e^{d_-})}{
(w-\e^{c_-})(\e^{c_+}-w)(\e^{d_+}-w)(w-\e^{d_-})}}}_{\mathcal{H}}\nonumber\\
&&\underbrace{-\,\frac{a}{2 t^2 }\,
\frac{\log(z)}{z}}_{\mathcal{I}}\,+\, \underbrace{
\frac{a~\e^{-{2\,t^2}/{a}}\,\log (\epsilon
)}{2\,t^2\,z}\,\sqrt{{(z-\e^{c_-})(z-\e^{c_+})(z-\e^{d_+})
(z-\e^{d_-})}}}_{\mathcal{C}_0}\nonumber\\
&&+\,\frac{1}{t\, z}\,\log \left(\mbox{$\frac{\e^{-c_-}\, z -1}
   {\e^{-c_+}\,z -1 }$}\right) \ .
\label{omegazints}\eea

In order to determine the endpoints of the intervals, we have to
impose the boundary condition (\ref{behav1}). Recovering the
expansion around $z=\infty$ is quite subtle and uninformative, and
we describe it in detail in Appendix B. We also show there that the
spurious cut due to the presence of the logarithm disappears and a
nice expansion to all orders can be given. The symbol $\mathcal{I}$
above represents the first integral together with the $\log(z)$
term, while $\mathcal{C}_0$ is the combination of the second
integral and of the $\log(\epsilon)$ term. Both combinations are
divided by the factor
$\sqrt{{(z-\e^{c_-})(z-\e^{c_+})(z-\e^{d_+})(z-\e^{d_-})}}$, and
similarly for the integral denoted $\mathcal{H}$. For $\mathcal{I}$,
we have the expansion \beq
\mathcal{I}=\frac{{\sf{B}}_0}{z}-\frac{{\sf{B}}_1}{z^2}+\cdots \eeq
and  for the integral $\mathcal{H}$ we can write \beq
\mathcal{H}=\frac{\mathcal{H}_{-1}}{z}+\frac{\mathcal{H}_0}{z^2}+\frac{\mathcal{H}_{-1}}{z^3}+
\cdots \ . \eeq The definitions of  the integrals ${\sf{B}}_i$,
$\mathcal{C}_0$ and $\mathcal{H}_i$ are given in Appendix B.
Combining these expansions, we can write down a symbolic series for
the resolvent up to order $\frac1z$ as
 \bea
\omega(z)&\simeq&-\frac{a}{2 t^2}\,\left[z\,\mathcal{
C}_0-\mathcal{C}_0 S_1+\frac{S_2}{z}\,\mathcal{C}_0+{{\sf{B}}}_0-
\frac{1}{z}\,\left({\sf{B}}_1+{\sf{B}}_0 S_1\right)\right]\\
&&+\,\frac{1}{t}\,\left[\mathcal{H}_{-1} z+\mathcal{H}_0 -S_1
\mathcal{H}_{-1}+\frac{1}{z}\,\left(S_2 \mathcal{H}_{-1}-S_1
\mathcal{H}_0+ \mathcal{H}_1\right)\right]+\frac{c_+-c_-}{t\,
z}\nonumber \ , \eea where the coefficients $S_i$ are defined
through the asymptotic expansion \beq
\frac{\sqrt{{(z-\e^{c_-})(z-\e^{c_+})(z-\e^{d_+})(z-\e^{d_-})}}}{z}=z-{S_1}+
\frac{S_2}{z}+\cdots \ . \eeq

Imposing the boundary condition (\ref{behav1}) thereby produces
three equations \bea
(z)&:\ \ \ \ \ &\frac{a}{2 t}\, \mathcal{C}_0-\mathcal{H}_{-1}=0 \ , \nonumber\\
(z^0)&:\ \ \ \ \ &\frac{a}{2 t}\, \left(-S_1
\mathcal{C}_0+{\sf{B}}_0\right)-\mathcal{H}_0+S_1 \mathcal{H}_{-1}=0
\ , \\
(z^{-1})&:\ \ \ \ \ &-\frac{a}{2
t}\,\left({S_2}\mathcal{C}_0-{\sf{B}}_1-{\sf{B}}_0 S_1\right)+S_2
\mathcal{H}_{-1}-S_1 \mathcal{H}_0+ \mathcal{H}_1+{c_+-c_-}{ }=-{t
(1-2 c)}{}\nonumber \eea which combine to give \bea
\frac{a}{2 t}\, \mathcal{C}_0-\mathcal{H}_{-1}&=&0  \ , \nonumber\\
\frac{a}{2 t}\, {\sf{B}}_0-\mathcal{H}_0 &=&0 \ , \nonumber\\
\frac{a}{2 t}\,{\sf{B}}_1+ \mathcal{H}_1+2 t \,c&=&-{t (1-2 c)}  \ .
\label{3eqs2un}\eea There are three equations in two unknowns, but
they are not independent. As shown in Appendix B, the symmetry
(\ref{symmegh}) implies the two further relations \beq
\left({\sf{B}}_1+\mbox{$\frac{2 t^2}{a}$}\right)=-\e^{{2
t^2}/{a}}\,\mathcal{C}_0 \ , \ \ \ \ \ \ \ \mathcal{H}_1=\e^{{2
t^2}/{a}}\,\mathcal{H}_{-1} \ . \eeq Using these relations it is
easy to show that the third equation in eq.~(\ref{3eqs2un}) implies
the first one. It suffices therefore consider only the first two
equations, whose explicit integral forms are given respectively by
\bea &&\frac{a}{2 t}\,\int_0^{\infty} \dd w ~\log(w)\,\frac{\dd}{\dd
w}\left(\frac{1}{
\sqrt{(w+\e^{c_-})(w+\e^{c_+})(w+\e^{d_+})(w+\e^{d_-})}}\right)\nonumber\\
&&\qquad\qquad=~\int_{\e^{c_-}}^{\e^{c_+}} \frac{\dd
w}{w}~\frac{1}{\sqrt{(w-\e^{c_-})(\e^{c_+}-w)(\e^{d_+}-w)(w-\e^{d_-})}}
\label{DK1st} \ , \\ &&\frac{a}{2 t}\,\int_0^{\infty} \dd w~
\frac{1}{
\sqrt{(w+\e^{c_-})(w+\e^{c_+})(w+\e^{d_+})(w+\e^{d_-})}}\nonumber\\
&&\qquad\qquad=~\int_{\e^{c_-}}^{\e^{c_+}} \frac{\dd
w}{\sqrt{(w-\e^{c_-})(\e^{c_+}-w)(\e^{d_+}-w)(w-\e^{d_-})}} \ .
\label{DK2nd}\eea

We expect our two-cut solution to follow closely what happens in
ordinary Yang-Mills theory at large $N$. When  the two cuts  merge
($c=0$), we should recover the one-cut values for the endpoints at
the phase transition point. The equations above for the endpoints at
$c=0$ reduce to \bea \pi\, \frac{\e^{{t\,( d - \frac{4\,t}{a})
}/{2}}}{\e^{d\,t}-1}&=& -\frac{a\,\log
\left(4~\e^{{2\,t^2}/{a}}\right)}{2~\e^{{t^2}/{a}}\,t} +
  \frac{a\,\log \left(\e^{-d\,t   + {t^2}/{a}} + \e^{d\,t +{t^2}/{a}}+2~\e^{{t^2}/{a}}
\right)}{2~\e^{\frac{t^2}{a}}\,t}\nonumber\\&& +\,\frac{a~\e^{{t\,(
d - \frac{4\,t}{a} ) }/{2}}\,\arcsin \bigl(\tanh
(\frac{d\,t}{2})\bigr)}{t - \e^{d\,t}\,t} \eea and \beq \pi\,
\frac{\e^{{t\,( d - \frac{2\,t}{a}) }/{2}}}{ \e^{d\,t}-1}
=\frac{a}{2 t}\,\frac{2~\e^{{t\,( d - \frac{2\,t}{a} )
}/{2}}\,\arcsin \bigl(\tanh (\frac{d\,t}{2})\bigr)}{ \e^{d\,t}-1} \
, \eeq where we have explicitly performed the integrals. Some simple
algebra shows immediately that these  two equations are  equivalent
to
 \beq
d=\frac{2}{t}\,\mathrm{arccosh}\left(\e^{{t^2}/{2a}}\right) \ \ \ \
\  \mathrm{and }\ \ \ \ t=-p\,\log\cos^2\left(\mbox{$\frac{\pi
}{p}$}\right) \ . \eeq These are  exactly the conditions that
determine endpoint $b$ and the critical coupling constant in the
one-cut solution.

\subsection{The Douglas-Kazakov equations}

We want to now understand how our equations are connected to the
Douglas-Kazakov equations that arise in the large $N$ limit of
ordinary Yang-Mills theory on $S^2$. We can proceed easily once the
integrals are performed in terms of elliptic functions (see
Appendix~B). The equation~(\ref{DK2nd}) reads \beq \label{cip12}
\frac{a}{2 t}\, {\bigl(F(\nu_\infty,k)-F(\nu_0,k)\bigr)}=\frac{a}{2
t} \,{F}\left(\arcsin \left(\tanh \bigl(\mbox{$\frac{t\left( c + d
\right) }{2}$}\bigr)\right),{k}\right) ={ K\left(k\right)} \eeq
where \beq \label{defk}
k=\sqrt{\frac{(\e^{c_+}-\e^{c_-})(\e^{d_+}-\e^{d_-})}{(\e^{d_+}-\e^{c_-})
(\e^{c_+}-\e^{d_-})}} \eeq and \bea \sin
(\nu_0)=\sqrt{\frac{(\e^{d_+}-\e^{c_-})~\e^{d_-}}{(\e^{d_+}-\e^{d_-})~
\e^{c_-}}} \ , \ \ \ \ \ \ \ \sin
(\nu_\infty)=\sqrt{\frac{\e^{d_+}-\e^{c_-}}{\e^{d_+}-\e^{d_-}}} \ .
\eea The symbols $K(k)=F(\pi/2,k)$ and $F(\nu,k)$ denote
respectively the complete and incomplete elliptic integrals of the
first kind. The intermediate equality in eq.~(\ref{cip12}) uses a
well-known summation formula for elliptic integrals
\cite{matematica1}. In the limit $t\to0$, the small angle behaviour
of $F$ reduces eq.~(\ref{cip12}) to \beq \label{cip121}
\mbox{$\frac{a}{4}$}\,(c+d) =K(\hat k) \eeq where $\hat k=2\,
\sqrt{c\, d}/(c+d)=2\, \sqrt{\ell}/(1+ \ell)$ and $\ell=c/d$ is the
Douglas-Kazakov modulus. From the transformation property \beq
K\left(\mbox{$\frac{2\, \sqrt{\ell}}{1+\ell}$}\right)=(1+\ell)
K(\ell)=\frac{c+d}{d}\, K(\ell) \eeq we recover \beq
d=\mbox{$\frac{4}{a}$}\, K(\ell) \eeq which is the first
Douglas-Kazakov equation.

Let us now turn to eq.~(\ref{DK1st}). The situation here is more
involved since we have to deal with (incomplete and complete)
elliptic integrals of the third kind $\Pi(\nu,n,k)$ and
$\Pi(n,k)=\Pi(\pi/2,n,k)$, which do not arise in ordinary
two-dimensional Yang-Mills theory. Evaluating all the integrals, we
find \bea &&\frac{2~\e^{-{2
t^2}/{a}}}{\sqrt{(\e^{d_+}-\e^{c_-})(\e^{c_+}-\e^{d_-})}}\,\left[\left(\e^{c_-}-\e^{d_-}\right)
\Pi\left(\mbox{$\frac{\e^{c_+}-\e^{c_-}}{
\e^{c_+}-\e^{d_-}}$},k\right)+\e^{d_-}\, K(k)\right]\nonumber\\
&&\quad=~-\frac{a}{2 t}~\e^{-{2 t^2}/{a}}\,
\left[\frac{2}{\sqrt{(\e^{c_+}-\e^{d_-})(\e^{d_+}-\e^{c_-})}}\,\left\{\left(\e^{c_-}-\e^{d_-}\right)
\left(\Pi\bigl(\nu_{0},\mbox{$\frac{\e^{c_+}-\e^{c_-}}{\e^{c_+}-\e^{d_-}}$},k
\bigr)\right.\right.\right.\nonumber\\
&&\quad\qquad-\left.\left.\Pi\bigl(\nu_{\infty},\mbox{$\frac{\e^{c_+}-\e^{c_-}}{\e^{c_+}-
\e^{d_-}}$},k\bigr)\right)-\e^{d_-}\, \bigl(F(\nu_{\infty},k)-
F\left(\nu_{0},k\right)\bigr)\right\}\nonumber\\
&&\quad\qquad-\Biggl.\log\left(\mbox{$\frac{1}
  {4}$}\,( \e^{d_-} + \e^{c_-} ) \,( \e^{-d_+-d_-}+ \e^{-c_- -d_-
})\right)\Biggr] \ . \eea With the help of eq.~(\ref{DK2nd}) we
obtain \bea \label{pippo}
\Pi\left(\mbox{$\frac{\e^{c_+}-\e^{c_-}}{\e^{c_+}-\e^{d_-}},k$}\right)&=&
\frac{a}{2 t}\,\left[\Pi\left(
\nu_\infty,\mbox{$\frac{\e^{c_+}-\e^{c_-}}{\e^{c_+}-\e^{d_-}}$},k\right)-
\Pi\left(
\nu_0,\mbox{$\frac{\e^{c_+}-\e^{c_-}}{\e^{c_+}-\e^{d_-}}$},k\right)
\right]\nonumber\\
&&+\,\frac{a}{4t}\,\frac{{\sqrt{(\e^{d_+}-\e^{c_-})
(\e^{c_+}-\e^{d_-})}}}{\e^{c_-}-\e^{d_-}}\nonumber\\ &&
\times\,\log\left(\mbox{$\frac{1}
  {4}$}\,( \e^{d_-} + \e^{c_-} ) ( \e^{-d_+-d_-}+ \e^{-c_- -d_-  }
)\right) \ . \eea To reach ordinary two-dimensional QCD, we expand
around $t=0$ with $a$ fixed. At lowest order $t^0$ we find
 \beq \Pi\left(\mbox{$\frac{2 c}{c+d},\frac{2\, \sqrt{c\,
d}}{c+d}$}\,\right)={a}\, \frac{d\,\left( c + d \right) }{4\,\left(
d -c\right) } \ . \eeq Since \beq n=\mbox{$\frac{2
c}{c+d}$}=1-\sqrt{1-k^2} \eeq the complete elliptic integral of the
third kind reduces to one of the first kind due to the identity \beq
\Pi\left(\mbox{$\frac{2 c}{c+d},\frac{2\, \sqrt{c
\,d}}{c+d}$}\,\right)= \frac{d}{d-c}\, K\left(\mbox{$\frac{2\,
\sqrt{c\, d}}{c+d}$}\,\right) \ . \eeq We can therefore write the
lowest order equation as  \beq K\left(\mbox{$\frac{2\, \sqrt{c
\,,d}}{c+d}$}\, \right)=a \,\frac{c+d}{4} \ , \eeq which coincides
with eq. (\ref{cip12}) and hence the first Douglas-Kazakov equation.

To obtain the second equation, we expand eq.~(\ref{pippo}) to second
order in $t$ to get \beq \frac{\left( c + d \right)^2 }{4\,\left( c
- d \right) } \,
    \left[  \,
       {E}\left(\mbox{$\frac{2\, \sqrt{c\, d}}{c+d}$}\,\right) -
      {K}\left(\mbox{$\frac{2\, \sqrt{c \,d}}{c+d}$}\,\right)
 \right]=-\frac{a\, c\,d\,\left( c + d \right)  }{8\,\left( c - d
   \right) }+\frac{\left( c + d \right)
 \,\left( 4 - a\,( c^2 + d^2)  \right) }{16\,\left( c - d \right)
 }
\eeq where $E(k)$ is the complete elliptic integral of the second
kind. Simplifying we get \beq \left( c + d \right)
       {E}\left(\mbox{$\frac{2 \,\sqrt{c\, d}}{c+d}$}\,\right) =1 \ ,
\eeq which finally yields the second Douglas-Kazakov equation after
employing the modular property \beq
      (c+d) {E}\left(\mbox{$\frac{2\, \sqrt{c\, d}}{c+d}$}\,\right)
=d\,\left[ 2{E}\left(\mbox{$\frac{c}{d}$}\right)-
\left(1-\mbox{$\frac{c^2}{d^2}$}\right)K\left(
\mbox{$\frac{c}{d}$}\right)\right] \ . \eeq One can verify that
expanding eq. (\ref{cip12}) to the next order in $t$ does not
produce any new conditions because its expansion does not contain
terms linear in $t$. The fact that the second equation appears only
at second order in $t$ should come as no surprise, as we have summed
the equations with a weight depending on $t$ in order to have the
simplest possible expressions. This procedure mixes the various
orders of the expansion.

\subsection{Saddle-point solution and the transition
curve}

It is instructive to show that the double cut equations are solvable
only for $p>2$, in agreement with the result that the phase
transition exists only in this region. This can be achieved in two
ways. From the summation formula \cite{matematica1} we can easily
show that \beq F(\nu_0,k)+F(\nu_\infty,k)=K(k) \ . \eeq Then
eq.~(\ref{cip12}) can be written in the form \beq
\left(p-2\right)F(\nu_\infty,k)=\left(p+2\right)F(\nu_0,k) \eeq
where we have used $a= p\, t$. Since \beq \frac{\partial
F(\nu,k)}{\partial\nu}=\frac{1}{\sqrt{1-k^2 \sin^2(\nu)}}>0 \eeq the
function $F$ is real and monotonic  for any $k$. Since $F(0,k)=0$,
it is positive for any $\nu>0$ and $0\le k<1$.  Since  both $\nu_0$
and $\nu_\infty$ are positive, $F(\nu_\infty,k)$ and $F(\nu_0,k)$
are also positive. The equation thereby admits solutions if and only
if \beq p>2 \ . \eeq

Alternatively, eq. (\ref{cip12})  can be solved by determining
$(c+d)$ in terms of the modular parameter $k$. We have \beq
\label{fava}
c+d=\frac{2}{t}\,\mathrm{arctanh}\left(\mathrm{sn}\bigl(\mbox{$\frac{2
t \,K(k)}{a}$},k\bigr)\right)=
\frac{2}{t}\,\mathrm{arctanh}\left(\mathrm{sn}\bigl(\mbox{$\frac{2
K(k)}{p}$},k\bigr)\right) \eeq where $\mathrm{sn}(x,k)$  is the
elliptic sine function of modular parameter $k$. From this
representation one can explicitly check that no solution exists  for
$p=1$ and $p=2$. For $p=1$ we have $\tanh(\frac{(c+d)t}{2})=0$
implying $c=d=0$, since both $c$ and $d$ are non-negative. For $p=2$
we find $\tanh(\frac{(c+d)t}{2})=1$ implying $t(c+d)=\infty$, and so
in this case we can obtain finite values for $c$ and $d$  only if
$t\to \infty$.

\begin{figure}[htb]
\label{cut1}
\begin{center}
\epsfxsize=4.0 in \epsfysize=2.4 in \epsfbox{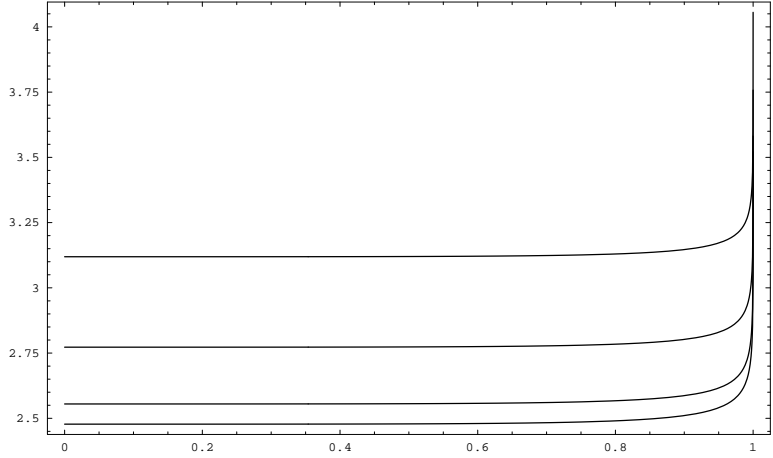}
\end{center}
\caption{\label{saddle} The right-hand side of the saddle-point
  equation times $p$ for $p=3,4,7,20$.}
\end{figure}

The appearance of elliptic functions suggests that the natural
unknown parameter for our equations are not the endpoints $(c,d)$ of
the support interval, but rather the modulus $k$~
\cite{Douglas:1993ii}. We can eliminate all explicit dependence on
$(c,d)$ through the condition (\ref{fava}) and \beq \label{fava1}
\tanh\left(\mbox{$\frac{(d-c)t}{2}$}\right)=
k^\prime\,\frac{\sn(x,k)}{\dn(x,k)}=-\cn\left(\mbox{$\frac{p+2}{p}$}
\,K(k),k\right) \eeq where $x=2 K(k)/p$,  $k^\prime=\sqrt{1-k^2}$ is
the complementary modulus and $\mathrm{dn}(x,k)$ is the elliptic
amplitude. For future convenience we shall also introduce the
parameter ${\hat x=(p+2)K(k)/p}$.  This second relation is the
translation of the definition (\ref{defk}).  Then we can write down
a table of translations for our parameters as \beq
n=\frac{\e^{c_+}-\e^{c_-}}{\e^{c_+}-\e^{d_-}}= k^2
\,\sn^2\left(\mbox{$\frac{\hat x}{2}$},k\right) \eeq and \beq
\sin(\nu_{\infty})=\frac{\sqrt{n}}{k}=\sn\left(\mbox{$\frac{\hat
      x}{2}$},k\right) \ , \ \ \
\sin(\nu_0)=\frac{\cos(\nu_\infty)}{\sqrt{1-k^2
\sin^2(\nu_\infty)}}= \frac{{\cn\left(\mbox{$\frac{\hat
x}{2}$},k\right)}}{ {\dn \left(\mbox{$\frac{\hat x}{2}$},k\right)}}
\ . \eeq With this table of translations, eq.~(\ref{pippo})  can be
written as \bea \label{pirippo} \!\!\!\!\!\!\!\frac{t}{4}&=&
\frac{p}{4}\,\log\left(\mbox{$\frac{\dn(x,k)}{\cn^2(x,k)}$}
\right)+\frac{p}{2}\,\frac{k^\prime\,\cn(x,k)}{1+\sn(x,k)}\,
\left[\Pi\left( \nu_\infty,n,k\right)-  \Pi\left(
\nu_0,n,k\right)-\mbox{$\frac{2}{p}$}\,  \Pi\left(n,k\right) \right]
\ . \eea The graphical behaviour of the function on the right-hand
side of eq.~(\ref{pirippo}) is depicted in Fig.~\ref{saddle}, and
one can check that it is a monotonically increasing function as $k$
runs from $0$ to $1$.
\medskip
Thus in order to have a solution, $\frac t4$ must be greater than
the value of the right-hand side of eq. (\ref{pirippo}) at $k=0$.
For $k=0$, only the first term in eq.~(\ref{pirippo}) survives
giving the bound \beq \label{fp} \frac{t}{4}\ge
\frac{p}{4}\,\log\left(\mbox{$\frac{\dn\left(\frac{\pi}{p},0\right)}
{\cn^2\left(\frac{\pi}{p},0\right)}$}\right)=\frac{p}{4}\,
\log\left(\sec^2\left(\mbox{$\frac{\pi}{p}$}\right)\right)\equiv
\frac{t_c}{4} \ . \eeq In other words,  two-cut solutions exist only
when $t$ is above the transition curve determined by the one-cut
analysis. The procedure to determine the solutions is now clear.
Given $(t,p)$ one solves eq. (\ref{pirippo}) for $k$. Since the
right-hand side of eq.~(\ref{pirippo}) is monotonic, the solution is
unique. Then from eqs. (\ref{fava}) and (\ref{fava1}) we obtain
$(c,d)$.

Eq. (\ref{pirippo}) is very complicated and not promising for an
analytical treatment. The situation improves dramatically if we
change variable from $k$ to the modular parameter $q=\exp(-\pi\,
K^\prime(k)/K(k))$. Then the right-hand side of the equation reduces
to an elegant $q$-series given by \bea \label{nos1}
\frac{t}{4}&=&\frac{t_c}{4}-2 p \,\sum_{n=1}^\infty
\frac{(-1)^n}{n}\,\frac{q^{2 n}}{1-q^{2
n}}\,\sin^2\left(\mbox{$\frac{\pi \,n}{p}$}\right) \eea as shown in
Appendix C. This series can also be summed in terms of
theta-functions to get \beq
\frac{t}{4}=-\frac{p}{2}\,\left[\log\left(\cos\bigl(\mbox{$
\frac{\pi}{p}$}\bigr)\right)+4\sum_{n=1}^\infty
\frac{(-1)^n}{n}\,\frac{q^{2 n}}{1-q^{2 n}}\,
\sin^2\left(\mbox{$\frac{\pi\, n}{p}$}\right)\right]=-\frac{p}{2}\,
\log\left(\mbox{$\frac{\vartheta_2\left(\left.\frac{\pi}{p}
\right|q\right)}{\vartheta_2\left(\left.0\right|q\right)}$}\right) \
. \eeq In this form the connection with the DK equation determining
the modulus $q$ is very simple. Multiplying by $p$ and taking the
limit $p\to\infty$, at leading order we find \beq
\frac{a}{4}=\frac{\pi^2}{4}-2\pi^2 \,\sum_{n=1}^\infty
{(-1)^n}\,\frac{n\, \hat{q}^{2 n}}{1-\hat{q}^{2 n}}= \frac{{\pi }^2
}{4}\, \left( 1 + 8\,\hat{q}^2 - 8\,\hat{q}^4 + 32\,\hat{q}^6 -
      40\,\hat{q}^8 +\dots\right)
\eeq where $\hat q=\lim_{p\to\infty}\,q$. This is exactly the
Douglas-Kazakov equation, except that it is a series in $\hat q^2$
rather than in $\hat q$. This apparent discrepancy can be easily
resolved if we recall that our modulus $k$ and consequently $\hat
q$ are different from the Douglas-Kazakov moduli. One can show
that \beq \hat q=\exp\left(-\pi\,\mbox{$\frac{ K^\prime(\hat
k)}{K(\hat k)}$}\right)=\exp\left(-\mbox{$\frac{\pi}{2}\, \frac{
K^\prime\left(\ell\right)}{ K\left(\ell\right)}$}\right)\eeq and
so our modular parameter is the square root of the Douglas-Kazakov
modular parameter. We stress that the  knowledge of $q$ is
equivalent to knowledge of $k$. By employing  the $q$-expansions
of the elliptic trigonometric functions, the endpoints $(c,d)$ can
also be expressed in terms of $q$. Explicit formulas are given in
Appendix~C.

\subsection{The third-order phase transition}

Let us now investigate the behaviour of the theory near the
critical point. It is possible to derive a perturbative solution
of our equation just above the phase transition. At the critical
point $t_c$, we have $q=0$. From eqs. (\ref{fava}) and
(\ref{fava1}) we find that $d\pm
c=\frac{2}{t_c}\,\mathrm{arctanh}(\sin(\frac{\pi}{p}))$, which
implies that $c=0$ and \beq
d=\frac{2}{t_c}\,\mathrm{arctanh}\left(\sin\bigl(\mbox{$
\frac{\pi}{p}$}\bigr)\right)=\frac{2}{t_c}\,
\mathrm{arccosh}\left(\sec\bigl(\mbox{$\frac{\pi}{p}$}
\bigr)\right)=\frac{2}{t_c}\, \mathrm{arccosh}\left(\e^{{t^2_c}/{2
a_c}}\right) \ . \eeq At the critical point $t_c$ the two cuts
merge and form a single cut whose endpoints coincide with those of
the one-cut solution. To go further and see what happens in a
neighborhood around $t_c$ we have to solve eq.~(\ref{nos1}) as a
series in $(t-t_c)$. We assume that $q$ admits an expansion of the
form \beq q=(t-t_c)^\alpha\,\sum_{n=0}^\infty a_n (t-t_c)^n \ .
\eeq Substituting this expansion  into eq.~(\ref{nos1}), we can
iteratively solve for the coefficients $a_n$ and find \bea
q&=&{\sqrt{\hat t}}\,\left( \frac{\csc (\frac{\pi }{p})}
     {{\sqrt{8 p}}} +
    \frac{{\hat t}\,\cos (\frac{2\,\pi }{p})\,
       {\csc^3 (\frac{\pi }{p})}}{32\,{\sqrt{2}}\,
       p^{{3}/{2}}} +
    \frac{{{\hat t}}^2\,
       \left( -27 - 32\,\cos (\frac{2\,\pi }{p}) +
         5\,\cos (\frac{4\,\pi }{p}) \right) \,
       {\csc^5 (\frac{\pi }{p})}}{6144\,{\sqrt{2}}\,
       p^{{5}/{2}}}\right.\nonumber\\
       &&+\left.
    \frac{{{\hat t}}^3\,
       \left( -64 - 157\,\cos (\frac{2\,\pi }{p}) -
         64\,\cos (\frac{4\,\pi }{p}) +
         \cos (\frac{6\,\pi }{p}) \right) \,
       {\csc^7 (\frac{\pi }{p})}}{65536\,{\sqrt{2}}\,
       p^{{7}/{2}}} +O(\hat t^4)\right)
\eea where $\hat t=t-t_c$. From this expansion we can obtain all
information about the gauge theory in the strong-coupling phase
around the critical point.

To investigate the behaviour of the theory beyond the phase
transition, we need to understand what happens to the distribution
function $\rho$. Because our potential is non-polynomial, we have no
simple relations relating derivatives of the free energy to the
expansion of the resolvent, as in the standard matrix models or in
ordinary Yang-Mills theory. We have to resort therefore to a brute
force calculation. The first step is to compute the resolvent in a
closed form in terms of elliptic functions. Using the first
saddle-point equation (\ref{DK1st}) to simplify the expression
(\ref{omegazints}), we find \bea \omega(z)&=&-\frac{a}{2t^2z}\,
\int_{0}^{\infty} \frac{\dd w}{z+w}~
\sqrt{\frac{(z-\e^{c_-})(z-\e^{c_+})(z-\e^{d_+})(z-\e^{d_-})}
{(w+\e^{c_-})(w+\e^{c_+})(w+\e^{d_+})(w+\e^{d_-})}}
\nonumber\\
&&+\,\frac{1}{t \,z}\,\int_{\e^{c_-}}^{\e^{c_+}} \frac{\dd
w}{z-w}~\sqrt{
\frac{(z-\e^{c_-})(z-\e^{c_+})(z-\e^{d_+})(z-\e^{d_-})}
{(w-\e^{c_-})(\e^{c_+}-w)(\e^{d_+}-w)(w-\e^{d_-})}}
\nonumber\\
&&-\frac{a}{2 t^2 }\, \frac{\log(z)}{z} +\frac{1}{t\, z}\,\log
\left(\mbox{$\frac{\e^{-c_-}\,z -1}
   {\e^{-c_+}\,z -1 }$}\right) \\
&=&\frac{2\,\sqrt{(z-\e^{c_-})(z-\e^{c_+})(z-\e^{d_+})(z-\e^{d_-})}
\,(\e^{c_-}-\e^{d_-})}{t\, z(z-\e^{c_-})(z-\e^{d_-})\,
\sqrt{(\e^{d_+}-\e^{c_-})(\e^{c_+}-\e^{d_-})}}\nonumber\\
&&\times\,\left[
\Pi\left(\mbox{$\frac{(\e^{c_+}-\e^{c_-})(z-\e^{d_-})}
{(\e^{c_+}-\e^{d_-})(z-\e^{c_-})},k$}\right)+\frac{a}{2 t}\,
 \left\{\Pi\left(\mbox{$\nu_\infty,\frac{(\e^{d_+}-\e^{d_-})
(z-\e^{c_-})}{(\e^{d_+}-\e^{c_-})(z-\e^{d_-})},k$}\right)
\right.\right.\\\nonumber &&-\left.\left.
\Pi\left(\mbox{$\nu_0,\frac{(\e^{d_+}-\e^{d_-})(z-\e^{c_-})}
{(\e^{d_+}-\e^{c_-})(z-\e^{d_-})}$},k\right)\right\}- K(k)\right]-
\frac{a}{2 t^2 }\, \frac{\log(z)}{z}+\frac{1}{t\, z}\, \log
\left(\mbox{$\frac{\e^{-c_-}\,z -1}
   {\e^{-c_+}\,z -1 }$}\right) \ . \nonumber
\eea

From the discontinuities of this expression we can derive the
distribution of Young tableaux variables in the various intervals.
In the region $z\in[\e^{c_+},\e^{d_+}]$ we find \bea
\label{rhoelliptic} \rho&=&\frac{1}{2\pi
\ii}\,\bigl(\omega(z+\ii\epsilon)-
\omega(z-\ii\epsilon)\bigr)\nonumber\\ &=&
\frac{2(\e^{c_-}-\e^{d_-})}{\pi\, t\, z\,
\sqrt{(\e^{d_+}-\e^{c_-})(\e^{c_+}-\e^{d_-})}}\,
\sqrt{\frac{(z-\e^{c_+})(\e^{d_+}-z)}{(z-\e^{d_-})(z-\e^{c_-})}}
\nonumber\\ &&\times\,\left[
\Pi\left(\mbox{$\frac{(\e^{c_+}-\e^{c_-})(z-\e^{d_-})}
{(\e^{c_+}-\e^{d_-})(z-\e^{c_-})},k$}\right)+\frac{a}{2 t}\,
 \left\{\Pi\left(\nu_\infty,\mbox{$\frac{(\e^{d_+}-\e^{d_-})
(z-\e^{c_-})}{(\e^{d_+}-\e^{c_-})(z-\e^{d_-})}$},k\right)
\right.\right.\nonumber\\
&&\left.\left.-
\Pi\left(\nu_0,\mbox{$\frac{(\e^{d_+}-\e^{d_-})(z-\e^{c_-})}
{(\e^{d_+}-\e^{c_-})(z-\e^{d_-})}$},k\right)\right\}- K(k)\right] \
. \eea The explicit form of the distribution function in the dual
region $z\in[\e^{d_-},\e^{c_-}]$ is obtained by means of the
symmetry (\ref{symmegh}). The elliptic integrals of the third kind
appearing in the expression for $\rho$ have parameter $n$ (the
second entry of the function) in the interval $k^2<n<1$ when
$z\in[\e^{c_+},\e^{d_+}]$. They are thus of circular type~\cite{AS}.
In what follows we shall use the $q$-expansion derived in Appendix C
to determine the behaviour of the distribution $\rho(z)$ near the
critical point $t_c$.

One can write \bea \label{rhoexpansion}
\rho(z)&=&\frac{2}{\pi\,t\,
z}\Biggl(\frac{\pi}{2}-\frac{\pi}{2}
\,\Lambda(\eta,k)+\frac{p}{2}\,\arctan\left(\tan\bigl(\mbox{$
\frac{\pi}{p}$}\bigr)\,\tanh\bigl(2\beta_1\bigr)\right)-
{2\pi}\,\mu(\beta_1)
\Biggr. \nonumber\\
&&-\left.\sum_{s=1}^\infty \,\frac{p}{s}\,\frac{q^{4 s}}{1-q^{4
s}}\, \sin \left(\mbox{$\frac{2\pi\, s}{p}$}\right)\, \sinh\bigl(2
s\,\beta_1\bigr)\right) \eea where $\Lambda$ is the Heuman
lambda-function and $\mu$ is a ratio of theta-functions which is
given explicitly in Appendix~C. The definitions of the parameters
$\eta$ and  $\beta_1$ can also be found in Appendix~C. For the
moment their explicit forms are not important. At $t=t_c$, the
modular parameter $q$ vanishes and the only term surviving in the
expansion of $\rho$ is the one involving the arctangent function.
At the critical point the distribution is therefore given by \beq
\rho_c(z)=\frac{p}{\pi\,t\,
z}\,\arctan\left(\tan\bigl(\mbox{$\frac{\pi}{p}$}\bigr)\,
\tanh\bigl(2\beta^c_1(z)\bigr)\right) \ . \eeq In the original
variables it takes the form \beq \label{rhocritical}
\rho_c(x)=\frac{p}{\pi}\,\arctan\left(\tan\bigl(\mbox{$
\frac{\pi}{p}$}\bigr)\,\tanh\bigl(2\beta^c_1(\e^{ t_c
x+t_c/a_c^2}) \bigr)\right)=\frac{p}{\pi}\,\arctan\left(
\mbox{$\sqrt{\frac{\e^{{t^2}/{a}}}{\cosh^2(\frac{t\,x}{2})}-1
}$}~\right) \eeq and thus coincides with the one-cut distribution.
This result shows that the free energy and its first derivative
are continuous at the critical point. The second derivative of the
free energy can be reduced to computing \beq \frac{\partial^2{\cal
F}}{\partial a^2}\propto\frac{1}{2}\,\int_{\e^{c_+}}^{\e^{d_+}}
\dd z~\frac{\partial{\rho}(z)}{\partial a}\,
\left(\log(z)-\mbox{$\frac{t^2}a$}\right)^2+\frac{1}{2}\,
\int_{\e^{d_-}}^{\e^{c_-}} \dd z~\frac{\partial{\rho}(z)}{\partial
a}\,\left(\log(z)- \mbox{$\frac{t^2}a$}\right)^2 \eeq as all other
contributions vanish because of the boundary conditions on the
distribution and its symmetries. The derivatives are taken at
constant $t$. A tedious expansion of this quantity around $t=t_c$
using {\sl Mathematica} shows that it vanishes linearly in
$(a-a_c)$ and thus the phase transition is of third order.

\subsection{Evidence for the string picture}

The complete discussion of the string picture, namely the
identification of the  Calabi-Yau geometry and the appearance of the
perturbative  topological string  expansion, is out of the scope of
the present paper whose intent is to focus more on field theoretical
features. These string theoretic issues will be discussed in full
detail in a forthcoming paper~\cite{CGPS}. Nevertheless, we can
present some evidences for how the topological string expansion
emerges which also resolves an apparent puzzle. The topological
string perturbative expansion is naturally  organized as a double
series  in  two modular parameters $\e^{-t_{s}}$ and $\e^{-{2
t_{s}}/{(p-2)}}$, where $t_s$ is the K\"ahler modulus (related to
our $t$ by $t=\frac{2
  t_s}{p-2}$).  The appearance of this double dependence from our
equation is non-trivial and absolutely necessary for the string
interpretation. It is absent in the weak coupling region where
everything can be written as a series in a single modulus given by
$\e^{-{t}/{2 p}}=\e^{-{t_s}/{p(p-2)}}$.

We expect that the topological string theory will arise when $t$
is large. Thus we have to investigate the solution of our
saddle-point equation around $t=\infty$. In this region, $k$ and
consequently $q$ approach $1$. If we employ the standard
parametrization for the modular parameter $q=\e^{\pi \ii\tau}$,
then $\tau\to0$. When studying the behaviour of an elliptic
function as $\tau\to0$, the natural technique is to perform a
modular transformation. This procedure exchanges $\tau$ and
$-\frac1\tau$, and thus takes us back to the weak-coupling phase
where a perturbative solution can be attempted. Our saddle-point
equation then becomes \bea \label{nos12} \frac{t}{4}
&=&-\frac{p}{2}\, \log\left(\mbox{$\frac{\vartheta_2\left(\left.
\frac{\pi}{p}\right|q\right)}{\vartheta_2\left(\left.0\right|q\right)}$}
\right)= -\frac{\tilde \tau}{2 p}-\frac{p}{2}\,\log\left(\mbox{$
\frac{\vartheta_4\left(\left.\frac{\ii\tilde \tau}{p}\right|
\tilde q\right)}{\vartheta_4\left(\left.0\right|\tilde q\right)}$}
\right) \eea where we have defined $\tau=\ii K^\prime/K\equiv
\pi/\ii\tilde \tau$ and $\tilde q=\e^{\pi\ii\tilde \tau}$. At
leading order in the solution we have \beq \tilde \tau= -\frac{t
\,p}{2}=-t_{s}-\frac{2 t_{s}}{p-2} \ . \label{tau0}\eeq The
corrections coming from the theta-functions are exponentially
suppressed at this level.

To explore the subleading order, it is convenient to write our
equation as \beq \label{coppa} \tilde \tau=-t_{s}-\frac{2
t_{s}}{p-2}-p^2\, \log\left(\mbox{$
\frac{\vartheta_4\left(\left.\frac{\ii\tilde \tau}{p}\right| \tilde
q\right)}{\vartheta_4\left(\left.0\right|\tilde q\right)}$}\right)
\eeq and proceed iteratively. The first non-trivial order is
obtained by substituting back the zeroth-order solution (\ref{tau0})
into eq.~(\ref{coppa}) to get \beq \tilde \tau=-t_{s}-\frac{2
t_{s}}{p-2}+4 p^2\,
 \sum_{n=1}^\infty \,\frac{ \e^{-n \,t_{s}-\frac{2 n \,t_{s}}{p-2}}}{1-
 \e^{-2 n\, t_{s}-\frac{4 n\, t_{s}}{p-2}}}\,
\sinh^2\left(\mbox{$\frac{n\, t_{s}}{p-2}$}\right) \ , \eeq and so
on. Here we have used the standard expansion of the logarithm of a
ratio of two theta-function as a $q$-series. It is evident that the
solution for the modular parameter $\tilde \tau$ and thus the
partition function nicely organizes into a double expansion in the
two moduli $\e^{-t_{s}}$ and $\e^{-{2t_{s}}/{(p-2)}}$ as expected
from string theory.
\section{Conclusions}
In this paper we have discussed the existence of a large $N$ phase
transition in $q$-deformed Yang-Mills theory on the sphere which
occurs for integer parameters $p>2$ at the critical 't Hooft
coupling constant $t_c=p\,\log ({\sec (\frac{\pi }{p})}^2)$. The
transition appears to be of third order and to separate a
weak-coupling phase, where nonperturbative contributions (instantons
or better, in the present context, flat connections) are suppressed,
from a strong-coupling regime where they are enhanced. This occurs
in complete analogy with the undeformed case and we can consider the
new phase transition as a $q$-deformed version of the familiar
Douglas-Kazakov transition on the sphere. The two phenomena are in
fact smoothly connected. In ordinary two-dimensional Yang-Mills
theory, the strong-coupling region is described by the Gross-Taylor
string theory while the weak-coupling regime is essentially trivial.
Instead, in the $q$-deformed case both regimes seem to have a string
description due to the intimate relation with closed topological
string theory.

In the weak-coupling phase, which reduces the gauge theory to the
trivial flat connection sector of Chern-Simons theory on the Lens
space $L_p=S^3/\mathbb{Z}_p$, we found the familiar resolved
conifold geometry that is expected by the well-known geometric
transition. On the strong-coupling side, the theory should
reproduce the black hole partition function in terms of closed
topological string amplitudes on $X={\cal O}(p-2)\oplus {\cal
O}(-p)\to\mathbb{P}^1$. In particular, we expect the emergence of
chiral-antichiral dynamics responsible for the appearance of
$|Z_{\rm
  top}|^2$, along with fiber
D-brane contributions and a sum over Ramond-Ramond fluxes. Our
strong-coupling solution which is based on a symmetric ansatz
probably misses the non-trivial fluxes, because it only computes the
$Q=0$ sector of the $U(1)$ component of the gauge group, but it
should capture the correct Calabi-Yau geometry as well as the
D-brane contributions. These aspects will be the subjects of a
subsequent paper devoted to the comparison of the strong-coupling
regime with the topological string amplitudes and the black hole
partition function \cite{CGPS}. Here we concentrated our attention
on the construction of the strong-coupling solution based on a
two-cut ansatz. We proved that the relevant equations always admit a
unique solution for $p>2$, and by constructing the distribution
function above the critical point we found a third-order phase
transition. The most surprising result is the absence of phase
transitions for $p\leq 2$. In these cases the gauge theory always
remains in the weak-coupling phase and the strong-coupling equations
have no solutions.

It is tempting at this point to speculate about the meaning and
possible explanations of these results by resorting to the relation
with topological string amplitudes. At the same time it is natural
to wonder about their implications for the conjecture formulated in
\cite{Ooguri:2004zv}. The first observation is that the presence of
the phase transition signals from the strong-coupling side a
divergence of the string expansion of the large $N$ partition
function. In ordinary two-dimensional Yang-Mills theory on the
sphere, the Gross-Taylor series diverges at the critical coupling
$\lambda_c=\pi^2$ \cite{Crescimanno:1994eg} and the entropy of the
branched covering maps is considered to be responsible for the
critical behaviour. In analogy, we can raise the question about the
string degrees of freedom causing the divergence in the $q$-deformed
case. The obvious suspects are the fiber D-branes, for a number of
reasons. First of all, as we will show in \cite{CGPS}, their
contributions are related to the leading area-polynomial behaviour
of the Gross-Taylor series that is exactly the source of the
divergence. Furthermore, their appearance is intrinsically tied to
the non-compactness of the Calabi-Yau
manifold~\cite{Aganagic:2005dh} and summing over them could simulate
the thermodynamic limit which is necessary to drive the phase
transition. Another peculiarity about the presence of the fiber
D-branes is that the black hole partition function, as written in
\cite{Aganagic:2004js}, appears to depend on them in an intrinsic
non-holomorphic way. A different K\"ahler modulus is associated to
their contributions which  measures the ``distance'' of the fiber
D-branes from the sphere. While the series for the chiral and
antichiral partition functions are separately analytic, the series
defining the D-brane contributions to the full black hole partition
function depends only on the real part of the relevant K\"ahler
parameter which presumably prevents a suitable analytic
continuation.

The existence of the phase transition would establish the validity
of the conjecture of \cite{Ooguri:2004zv} only above the critical
coupling. It could be that the extension of our computation to
complex saddle-points and/or to the sum over all $U(1)$ sectors
smooths out the transition. If this is not the case, then we should
conclude that the exact black hole partition function is equivalent
to the related closed topological string theory only over a critical
area parameter  defining the geometry of the black hole itself. More
dramatically, for $p\leq 2$ the topological string description
should never be valid, a fact that is quite surprising. It is
therefore important to understand the effective degrees of freedom
dominating the black hole physics for large charges in the
weak-coupling phase and to understand the phase transition at the
gravitational level. A particulary intriguing possibility is that
the transition truly describes a topology change in the Calabi-Yau
background. This topology change could also fit in with the baby
universe splitting picture proposed on the torus in
\cite{Dijkgraaf:2005bp}. The nature of the phase transition itself
could lead to new insights into the relations between strings and
reduced models. An interesting third-order phase transition was
discovered some years ago \cite{Hoppe:1999xg}  in the apparently
unrelated context of a quantum mechanical model obtained as a
dimensional reduction of four-dimensional ${\cal N}=1$
supersymmetric Yang-Mills theory to a periodic light-cone time. In
spite of the completely different starting point, the equations
describing the phase transition there bear an impressive similarity
with ours in the strong-coupling phase. We believe that this
relation is worthy of further pursuit.

\bigskip
\noindent \textbf{Acknowledgements}\\
We thank F. Bonechi, F. Colomo, L. Cornalba, C. Imbimbo, A. Tanzini,
M. Tarlini, A. Torrielli for helpful discussions. L.G. would like to
thank K. Konishi for having invited him to present part of these
results at the miniworkshop "Dynamics of Supersymmetric Gauge
Theories", held in Pisa, 7 June 2005. He also acknowledges N. Dorey
and S. Benvenuti for very inspiring discussions there. M.C. and S.P.
thank H. Ooguri for  interesting conversations on these subjects at
the "Spring School on Superstring Theory and Related Topics" at ICTP
and at the "Summer School on Strings, Gravity and Cosmology" at the
Perimeter Institute. Moreover S.P. thanks  the organizers of the
"Convegno Informale di Fisica Teorica", held in Cortona, May 2005,
for giving her the chance to present preliminary results before
publication and M.C. thanks the organizers of the $12^{th}$ meeting
of the ``North British Mathematical Physics Seminar'', York, June
2005 for the invitation to present some of our results. L.G and D.S.
thank Heriot-Watt University for a week of kind hospitality and many
Edinburgh's pubs. The work of M.C. and R.J.S. was supported in part
by PPARC Grant PPA/G/S/2002/00478 and the EU-RTN Network Grant MRTN
-CT-2004-005104.

\newpage
\appendix

\addcontentsline{toc}{section}{Appendices}
\section{Stieltjes-Wigert and Szeg\`o polynomials}

This appendix is devoted to collecting some useful results about the
Stieltjes-Wigert and Szeg\`o polynomials.

\subsection{Definitions and general properties}

Define a deformation of integers, called $q$-numbers, by \beq
\label{A1} [n]_q=\frac{q^{n/2}-q^{-n/2}}{q^{1/2}-q^{-1/2}} \ . \eeq
For $q\to1$, $[n]_q=n+O(q-1)$. This is just one of the possible
definitions of $q$-numbers in the literature, but it is the one most
convenient for our purposes. By means of eq. (\ref{A1}) we can also
generalize all functions over the integers. In particular, we can
introduce the $q$-factorial \beq [n]_q!\equiv\prod_{k=1}^n
[k]_q=\frac{\prod\limits_{k=1}^n
(q^{-k/2}-q^{k/2})}{(q^{-1/2}-q^{1/2})^n}=
\frac{q^{-\frac{n(n-1)}{4}}}{(1-q)^n}\,{\prod_{k=1}^n
(1-q^{k})}{}\equiv \frac{q^{-\frac{n(n-1)}{4}}}{(1-q)^n}\,(q)_n \eeq
and the $q$-binomial coefficient \beq \left[{n \atop
k}\right]_q\equiv\frac{[n]_q!}{[n-k]_q!\,[k]_q!}=
\frac{q^{-\frac{n(n-1)}{4}}}{q^{-\frac{k(k-1)}{4}}q^{-\frac{(n-k)(n-k-1)}{4}}}
\,\frac{(q)_n }{(q)_k(q)_{n-k}}= q^{\frac{k\left( k-n \right)
}{2}}\,\frac{(q)_n }{(q)_k\,(q)_{n-k}} \ . \eeq

The Szeg\`{o} polynomials are then defined by \beq
S_n(x)=\sum_{k=0}^n \left[{n \atop k}\right]_q \,x^k \ . \eeq For
$q\to1$, this expression reduces to \beq S_n(x)~\overset{q\to
1}{=}~(1+x)^n+O(q-1) \ . \eeq For generic values of $q$, one has the
Euler identity \beq \sum_{k=0}^n \left[{n \atop k}\right]_q \,x^k=
\prod_{k = 0}^{ n-1}\left(1 + q^{k -\frac{n-1}{2}}\,x\right) \ .
\eeq

The Stieltjes-Wigert polynomials are defined by \beq W_n(x)=(-1)^n\,
q^{n^2+n/2}\, \sum_{k=0}^n \left[{n \atop k}\right]_q
\,q^{\frac{k(k-n)}{2}-k^2}\, \left(-q^{-{1}/{2}}\, x\right)^k \ .
\eeq The overall normalization is chosen so that these polynomials
are monic, \beq W_n(x)=x^n+\cdots \ . \eeq In our specific case the
deformation parameter is given by $q=\e^{-g_s}$, and one can show
that these polynomials are orthogonal with respect to the measure
\beq \dd\mu(x)=\e^{-\frac{\log(x)^2}{2 g_s}}~\frac{\dd x}{2\pi} \eeq
on the positive real half-line.

For this, consider the integral \bea
&&\int_0^\infty \dd\mu(x)~W_n(x)\, W_k(x)\nonumber\\
&&\qquad=~(-1)^{n+k}\, q^{n^2+n/2+k^2+k/2}\, \sum_{j=0}^n
\,\sum_{l=0}^k \left[{n \atop j}\right]_q\, \left[{k \atop
l}\right]_q\, q^{\frac{j(j-n)}{2}-j^2+ \frac{l(l-k)}{2}-l^2}\,
\left(-q^{-{1}/{2}} \right)^{j+l}\nonumber\\
&&\qquad\qquad\times\, \int_0^\infty \frac{\dd
x}{2\pi}~\e^{-\frac{\log(x)^2}{2 g_s}}\,
 x^{j+l}\nonumber\\
&& \qquad=~\sqrt{g_s}\,\frac{(-1)^{n+k}}{\sqrt{2\pi}}\,
 q^{n^2+k^2+\frac{n+k}{2}}\,
\sum_{j=0}^n\, \sum_{l=0}^k \left[{n \atop j}\right]_q\, \left[{k
\atop l}\right]_q \,(-1)^{j+l}\, q^{\frac{1 + j + l + 2\,j\,l - k\,l
- j\,n}{2}}
\nonumber\\
&&\qquad=~\sqrt{g_s}\,\frac{(-1)^{n+k}}{\sqrt{2\pi}}\,
 q^{n^2+k^2+\frac{n+k+1}{2}}\,
\sum_{j=0}^n (-1)^{j}\,\left[{n \atop j}\right]_q\,
q^{j(1-n)/2}\,\sum_{l=0}^k \left[{k \atop l}\right]_q\,  \left(-
q^{j+(1-k)/2}\right)^l \ . \nonumber\\ && \eea With the help of the
Euler identity we can compute the sum over $k$ to get \bea
\label{norma1}
&&\int_0^\infty \dd\mu(x)~W_n(x)\, W_k(x)\nonumber\\
&& \qquad =~\sqrt{g_s}\,\frac{(-1)^{n+k}}{\sqrt{2\pi}}\,
 q^{n^2+k^2+\frac{n+k+1}{2}}\,
\sum_{j=0}^n (-1)^{j}\,\left[{n \atop j}\right]_q\,
q^{j(1-n)/2}\,\prod_{l=0}^{k-1} \left(1-q^{l+j-k+1}\right) \ .
\nonumber\\ && \eea Since the computation is symmetric under the
exchange $k\leftrightarrow n$, we can assume $n\le k-1$ without loss
of  generality. This implies that $k-j-1\in\{0,1,\dots,k\}$ for
$j=0,1,\dots,n$. In other words, for all possible values of $j$ the
product $\prod_{l=0}^{k-1}\,(1-q^{l+j-k+1})$ always contains a term
that vanishes identically and so the whole product is zero. As a
consequence, the sum over $j$ vanishes as well.

The symmetry implies that this property extends to all $k\le n-1$.
The only remaining possibility is $k=n$. In that case eq.
(\ref{norma1}) reduces to \bea \label{norma2}
&&\int_0^\infty \dd\mu(x)~W_n(x)\, W_k(x)\nonumber\\
&&\qquad =~\sqrt{\frac{{g_s}}{{2\pi}}}\, q^{2 n^2+n+\frac{1}{2}}\,
\sum_{j=0}^n (-1)^{j}\,\left[{n \atop j}\right]_q\,
q^{j(1-n)/2}\,\prod_{l=0}^{n-1} \left(1-q^{l+j-n+1}\right) \ . \eea
Again  the product vanishes for $0\le j\le n-1$, because then $0\le
n-1-j\le n-1$. The only non-zero contribution comes from the $j=n$
term which yields \bea \label{norma3} \int_0^\infty
\dd\mu(x)~W_n(x)\, W_k(x)&=& \sqrt{\frac{{g_s}}{{2\pi}}}\, q^{2
n^2+n+\frac{1}{2}}\, (-1)^{n}\, q^{n(1-n)/2}\,\prod_{l=1}^{n}
\left(1-q^{l}\right)\nonumber\\ &=& \sqrt{\frac{{g_s}}{{2\pi}}}\,
q^{\frac{7}{4} (n^2+n)+\frac{1}{2}}  \prod_{l=1}^{n}
\left(q^{l/2}-q^{-l/2}\right)\nonumber\\
&=&\sqrt{\frac{g_s}{2\pi}}\,
 q^{\frac{7}{4}n(n+1)+\frac{1}{2}}\,
\left({q}^{1/2}-{q^{-1/2}}\right)^n\, [n]_q! \ . \eea

Summarizing, we have proven the orthogonality relation \bea
\int_0^\infty \dd\mu(x)~W_n(x)\, W_k(x)= \sqrt{\frac{g_s}{2\pi}}\,
q^{\frac{7}{4}n(n+1)+\frac{1}{2}}\,
\left({q}^{1/2}-{q^{-1/2}}\right)^n\, [n]_q!~\delta_{nk} \ . \eea In
standard matrix model notation this yields the orthogonal polynomial
normalization constants \beq h_n=\sqrt{\frac{g_s}{2\pi}}\,
q^{\frac{7}{4}n(n+1)+\frac{1}{2}}\,
\left({q}^{1/2}-{q^{-1/2}}\right)^n\, [n]_q! \ . \eeq These
polynomials also satisfy the three-term recurrence relation \beq
W_{n+1}(x)=\left(x-q^{n+\frac{1}{2}}\,(q^{n+1}+q^n-1) \right)\,
W_n(x)-q^{3n}\,(q^n-1)\, W_{n-1}(x) \ . \eeq

\subsection{Some useful integrals and Identities}

Let us now compute some useful integrals containing the
Stieltjes-Wigert  polynomials. Consider first \bea \int_0^\infty
\dd\mu(x)~W_n(a\, x) &=&\sqrt{\frac{g_s}{2\pi}}\,(-1)^{n}\,
 q^{n^2+(n+1)/2}\,S_n(-q^{\frac{1-n}{2}}\, a )=
\sqrt{\frac{g_s}{2\pi}}\,q^{\frac{(n+1)^2}{2}}\, \hat
S_n(-q^{\frac{1-n}{2}}\, a ) \nonumber \eea where $\hat
S_n(-q^{\frac{1-n}{2}}\, a )$ is obtained form
$S_n(-q^{\frac{1-n}{2}}\, a )$ by normalizing the coefficient of the
monomial of highest degree in $a$ to $1$. Next we compute \bea
&& \int_0^\infty \dd\mu(x)~\bigl(W_n(a\, x)\bigr)^2 \nonumber\\
&& \qquad=~\sqrt{\frac{{g_s}}{{2\pi}}}\, q^{2 n^2+n+1/2}\,
\sum_{j=0}^n \left(-a\, q^{(1-n)/2}\right)^{j}\,\left[{n \atop
j}\right]_q \,\sum_{l=0}^n
\left[{n \atop l}\right]_q \, \left(- q^{j+(1-n)/2}\,a\right)^l\nonumber\\
&& \qquad=~\sqrt{\frac{{g_s}}{{2\pi}}}\, q^{2 n^2+n+1/2}\,
\sum_{j=0}^n \left(-a\, q^{(1-n)/2}\right)^{j}\,\left[{n \atop
j}\right]_q\, S_n\left(- q^{j+(1-n)/2}\,a\right) \ .
\label{SWNaxint}\eea To simplify notation, we introduce the variable
$\hat a= q^{(1-n)/2}\, a$ so that eq.~(\ref{SWNaxint}) reduces to
\beq \int_0^\infty \dd\mu(x)~\bigl(W_n(a\, x)\bigr)^2=
\sqrt{\frac{{g_s}}{{2\pi}}}\, q^{2 n^2+n+1/2}\,\sum_{j=0}^n (-\hat a
)^{j}\,\left[{n \atop j}\right]_q\, S_n(- q^{j}\,\hat a ) \ . \eeq
\section{Elliptic integrals}

\subsection{Asymptotic expansions of elliptic integrals}

\renewcommand{\theequation}{\thesection.\arabic{equation}}

We want to compute the large $z$ asymptotic expansion of an integral
of the type \beq I=\int^\infty_0 \frac{\dd t}{t+z}~ f(t) \ , \eeq
where $f(t)$ decays polynomially for large $t$. In our specific case
the function $f(t)$ is given by \beq
f(t)=\frac{1}{\sqrt{(t+a)(t+b)(t+c)(t+d)}} \eeq and it admits the
large $t$ expansion \beq f(t)=\frac{1}{t^{2}} -\frac{1}{2\,t^3}\,(a
+ b + c + d) +
  {{O}\left(\mbox{$\frac{1}{t^4}$}\right)} \ .
\eeq The behaviour of $I$ for large $z$ can be evaluated by means of
the following theorem which is proven in~\cite{Wong}.

\begin{theorem}
Let $f(t)$  be a locally integrable function on $[0,\infty)$ and
$\{A_k\}$ a sequence of complex numbers. Suppose that $f(t)$ has a
large $t$ expansion for each $n=1,2,\dots$ of the form \beq
f(t)=\sum_{k=0}^{n-1} \frac{A_k}{t^{k+1}}+f_n(t) \eeq with
$f_n(t)={O}(t^{-n-1})$ as $t\to\infty$. Then for every $z,\rho>0$
and $n=1,2,\dots$ one has \bea \int_0^\infty \dd t ~
\frac{f(t)}{(t+z)^\rho}=\sum_{k=0}^{n-1} \frac{(-1)^k}{k!\,
z^{k+\rho}}\, (\rho)_k\, \left[A_k \bigl(\log(z)-\gamma-
\Psi(k+\rho)\bigr)+B_k\right]+R_n(z) \ , \eea where for
$k=0,1,2,\dots$ the coefficients $B_k$ are given by \beq
B_{k}=A_{k}\,\sum_{j=1}^k \frac{1}{j} +\lim_{T\to
  \infty}\,\biggl\{\int_0^T \dd t~t^k\, f(t) -\sum_{j=0}^{k-1}
A_j\, \frac{T^{k-j}}{k-j}-A_k\, \log(T)\biggr\}\equiv A_{k}\,
\sum_{j=1}^k \frac{1}{j} +{\sf B}_k \eeq and empty sums are
understood as zero. The remainder term is given by \bea
R_n(z)=(\rho)_n \,\int_0^\infty \dd t
~\frac{f_{n,n}(t)}{(t+z)^{n+\rho}} \eea where \beq
f_{n,n}(t)=\frac{(-1)^n}{(n-1)!}\,\int_t^\infty  \dd u~
 (u-t)^{n-1}\, f_n(u) \ .
\eeq The constant $\gamma$ is the Euler constant, $\Psi$ is the
digamma function and $(\rho)_k$ is the Pochhammer symbol.
\end{theorem}

This result simplifies drastically when $\rho=1$. In that case one
has \beq (1)_k=k! \ , \ \ \ \ \ \  \Psi(k+1)=-\gamma+\sum_{j=1}^k
\frac{1}{j} \ , \eeq and the asymptotic expansion takes the very
simple form \bea \int_0^\infty  \dd t~ \frac{f(t)}{t+z}=
\sum_{k=0}^{n-1} \frac{(-1)^k}{ z^{k+1}} \,\bigl[A_k\, \log(z)+{\sf
B}_k \bigr]+R_n(z) \ . \eea This result can also be expressed in the
more convenient form\footnote{Note that the coefficient of $\log(z)$
is the large $z$ asymptotic expansion of $-f(-z)$.} \beq
\int_0^\infty \dd t ~\frac{f(t)}{t+z}+f(-z)\, \log(z) =
\sum_{k=0}^{n-1} \frac{(-1)^k \,{\sf B}_k}{ z^{k+1}} +R_n(z) \ .
\eeq In other words, the combinations on the right-hand sides of
these equations admit an expansion in powers of $\frac1z$ alone at
$z\to\infty$. This ensures that our resolvent has only physical
cuts.

We now apply this result to evaluate the asymptotic series of \bea
\mathcal{I}&=& \frac{1}{z}\,\int^0_{-\infty} \frac{\dd
w}{z-w}~\sqrt{\frac{{(z-\e^{c_-})(z-\e^{c_+})(z-\e^{d_+})(z-\e^{d_-})}}
{{(w-\e^{c_-})(w-\e^{c_+})(w-\e^{d_+})(w-\e^{d_-})}}}+\frac{\log(z)}{z}
\nonumber\\
&=&\frac{1}{z}\,\int_0^{\infty} \frac{\dd w
}{z+w}~\sqrt{\frac{{(z-\e^{c_-})(z-\e^{c_+})
(z-\e^{d_+})(z-\e^{d_-})}}{{(w+\e^{c_-})(w+\e^{c_+})(w+\e^{d_+})
(w+\e^{d_-})}}}+\frac{\log(z)}{z}\nonumber\\
&=&\frac{\sqrt{(z-\e^{c_-})(z-\e^{c_+})(z-\e^{d_+})(z-\e^{d_-})}}{z}\,
\left(\frac{{\sf B}_0}{z}-\frac{{\sf B}_1}{z^2}+
\cdots\right)\nonumber\\
&=&{{\sf B}_0} -\frac{1}{z}\,\left({{\sf B}_1} + \frac{{{\sf B}_0}
}{2}\, \left( \e^{{c_-}} + \e^{{c_+}} + \e^{{d_-}} + \e^{{d_+}}
\right) \right) + {{O}\left(\mbox{$\frac{1}{z^2}$}\right)} \ . \eea
Since $A_0=0$, $A_1=1$ and $A_2=-\frac{1}{2}\,(a+b+c+d)$ in this
case, the coefficients of the expansion are easily given in terms of
elliptic integrals \bea {\sf B}_0&=&\int_0^\infty  \frac{\dd
t}{\sqrt{(t+\e^{c_-})
(t+\e^{c_+})(t+\e^{d_+})(t+\e^{d_-})}} \ , \\
{\sf B}_1&=&\lim_{T\to\infty}\,\biggl(\int_0^T \dd t~ \frac{t}
{\sqrt{(t+\e^{c_-})(t+\e^{c_+})(t+\e^{d_+})(t+\e^{d_-})}}-\log(T)\biggr)
\nonumber\\
&=&-\int^\infty_0 \dd t~ \log(t)\, \frac{\dd}{\dd t }\left(
\frac{t^2}{\sqrt{(t+\e^{c_-})(t+\e^{c_+})(t+\e^{d_+})(t+\e^{d_-})}}
\right)\nonumber\\ &=& \int^\infty_0 \dd t ~\log(t)\, \frac{\dd}{\dd
t }\bigl(t^2\, f(t) \bigr)\ . \eea

The other integral function whose asymptotic expansion plays an
important role is \bea \mathcal{H}&=&\int_{\e^{c_-}}^{\e^{c_+}}
\frac{\dd w}{w(z-w)}~
\sqrt{\frac{(z-\e^{c_-})(z-\e^{c_+})(z-\e^{d_+})(z-\e^{d_-})}
{(w-\e^{c_-})(\e^{c_+}-w)(\e^{d_+}-w)(w-\e^{d_-})}}\nonumber\\
&=&\frac{1}{z}\,\int_{\e^{c_-}}^{\e^{c_+}} \frac{\dd w}{w}~
\sqrt{\frac{(z-\e^{c_-})(z-\e^{c_+})(z-\e^{d_+})(z-\e^{d_-})}
{(w\-e^{c_-})(\e^{c_+}-w)(\e^{d_+}-w)(w-\e^{d_-})}}\nonumber\\
&&+\,\frac{1}{z^2}\,\int_{\e^{c_-}}^{\e^{c_+}} {\dd w}~
\sqrt{\frac{(z-\e^{c_-})(z-\e^{c_+})(z-\e^{d_+})(z-\e^{d_-})}
{(w-\e^{c_-})(\e^{c_+}-w)(\e^{d_+}-w)(w-\e^{d_-})}}\nonumber\\
&&+\,\frac{1}{z^3}\,\int_{\e^{c_-}}^{\e^{c_+}} {\dd w}~ w\,
\sqrt{\frac{(z-\e^{c_-})(z-\e^{c_+})(z-\e^{d_+})(z-\e^{d_-})}
{(w-\e^{c_-})(\e^{c_+}-w)(\e^{d_+}-w)(w-\e^{d_-})}
}+\cdots\nonumber\\
&=&{\mathcal{H}_{-1}}\,z + {\mathcal{H}_{0}}-\mbox{$\frac12$}\,
{\left( \e^{{c_-}} + \e^{{c_+}} + \e^{{d_-}} + \e^{{d_+}} \right)
 \,{\mathcal{H}_{-1}}} \nonumber\\
   &&+\,\frac{1}{z}\, \biggl[\mbox{$\frac{1}{2}$}\,
\biggl( \e^{{c_-} + {c_+}} + \e^{{c_-} + {d_-}} + \e^{{c_+} + {d_-}}
+
          \e^{{c_-} + {d_+}} + \e^{{c_+} + {d_+}} + \e^{{d_-} + {d_+}}
\biggr.\biggr.\nonumber\\ &&-\left.\left.\mbox{$
          \frac14$}\,{{\left( \e^{{c_-}} + \e^{{c_+}} + \e^{{d_-}} +
\e^{{d_+}} \right) }^2} \right) \,{\mathcal{H}_{-1}} -
\mbox{$\frac12$}\,{\left( \e^{{c_-}} +\e^{{c_+}} + \e^{{d_-}} +
\e^{{d_+}} \right) \,{\mathcal{H}_{0}}} +
{\mathcal{H}_{1}}\right]\nonumber\\ && +\,
  {{O}\left(\mbox{$\frac{1}{z^2}$}\right)}
\eea where the constants $\mathcal{H}_{n}$ are defined by \beq
\mathcal{H}_n=\int_{\e^{c_-}}^{\e^{c_+}} {\dd
w}~\frac{w^n}{\sqrt{(w-\e^{c_-})(\e^{c_+}-w)(\e^{d_+}-w)(w-\e^{d_-})}}
\ . \eeq The constant $\mathcal{H}_1$ is not independent of
$\mathcal{\mathcal{H}}_{-1}$. Using the change of variable $w\to
\e^{{2 t^2}/{a}}/w$ we find \beq \mathcal{H}_1= \e^{{2 t^2}/{a}}\,
\mathcal{H}_{-1} \ . \eeq

Finally, we consider the integral \bea
\hat{\mathcal{C}}&=&\frac{1}{z}\,\biggl(\int_{-\infty}^{-\epsilon}
\frac{\dd
w}{w\,\sqrt{(w-\e^{c_-})(w-\e^{c_+})(w-\e^{d_+})(w-\e^{d_-})}}-
{\e^{-{2\,t^2}/{a}}\,\log (\epsilon )}{}
\biggr)\nonumber\\
&=&-\frac{1}{z}\,\int^0_{-\infty} \dd w ~\log|w|\,\frac{\dd}{\dd w}
\left(\frac{1}{
\sqrt{(w-\e^{c_-})(w-\e^{c_+})(w-\e^{d_+})(w-\e^{d_-})}}\right)\nonumber\\
&=&\frac{1}{z}\,\int_0^{\infty} \dd w~ \log(w)\,\frac{\dd}{\dd w}
\left(\frac{1}{
\sqrt{(w+\e^{c_-})(w+\e^{c_+})(w+\e^{d_+})(w+\e^{d_-})}}\right)\nonumber\\
&=&\frac{\mathcal{C}_0}{z}=-\e^{-{2 t^2}/{a}}\,\frac{\left(\mbox{$
\frac{2t^2}a$}+{\sf B}_1\right)}{z} \ . \eea The last equality
relating $\mathcal{C}_0$ to ${\sf B}_1$ is obtained through the
usual change of variable $w\to \e^{{2 t^2}/{a}}/w$. The asymptotic
expansion of $\mathcal{C}$ is then given by\bea
\mathcal{C}&=&\sqrt{(z-\e^{c_-})(z-\e^{c_+})(z-\e^{d_-})(z-\e^{d_+})}
~\hat{\mathcal{C}}\nonumber\\
&=&z\,\mathcal{C}_0 -\frac{\mathcal{C}_0}{2}\,\left(\e^{{c_-}} +
\e^{{c_+}} +\e^{{d_-}} + \e^{{d_+}}\right) +
  \frac{1}{2\,z}\,\biggl(\e^{{c_-} + {c_+}} + \e^{{c_-} + {d_-}} +
\e^{{c_+} + {d_-}} + \e^{{c_-} + {d_+}} \biggr.\nonumber\\
     &&+\left.
     \e^{{c_+} + {d_+}} + \e^{{d_-} + {d_+}} -
     \mbox{$\frac{1}{4}$}\,{\left( \e^{{c_-}} + \e^{{c_+}} +
\e^{{d_-}} + \e^{{d_+}} \right) }^2\right)\,\mathcal{C}_0 +
     {{O}\left(\mbox{$\frac{1}{z}$}\right)}^2 \ .
\eea

\subsection{Evaluation of elliptic integrals }

We want to compute the elliptic integral \bea
\int_{\e^{c_-}}^{\e^{c_+}} \frac{\dd w}{\sqrt{(w-\e^{c_-})
(\e^{c_+}-w)(\e^{d_+}-w)(w-\e^{d_-})}} \ . \eea Its modulus is given
by \beq k=\sqrt{\frac{(\e^{c_+}-\e^{c_-})(\e^{d_+}-\e^{d_-})}
{(\e^{d_+}-\e^{c_-})(\e^{c_+}-\e^{d_-})}}=\sqrt{\frac{( \e^{{a
\,c}/{p}}-\e^{-{a \,c}/{p}}) (\e^{{a \,d}/{p}}-\e^{-{a\,
d}/{p}})}{(\e^{{a\, d}/{p}}- \e^{-{a \,c}/{p}})(\e^{{a
\,c}/{p}}-\e^{-{a\, d}/{p}})}}
 ~ \overset{\scriptsize{p\to \infty}}{=}~ 2\, \frac{\sqrt{c\, d}}{c+d}
\eeq and the result is \bea \int_{\e^{c_-}}^{\e^{c_+}} \frac{\dd
w}{\sqrt{(w-\e^{c_-})(\e^{c_+}-w)(\e^{d_+}-w)(w-\e^{d_-})}}= \frac{2
K\left(k\right)}{\sqrt{(\e^{d_+}-\e^{c_-})(\e^{c_+}-\e^{d_-})}} \ .
\nonumber\\ \eea With the same definitions, we have \bea
&&\int_{\e^{c_-}}^{\e^{c_+}} \frac{\dd w}{w}~
\frac{1}{\sqrt{(w-\e^{c_-})(\e^{c_+}-w)(\e^{d_+}-w)(w-\e^{d_-})}}
\nonumber\\ && \qquad =~ \e^{-{2
t^2}/{a}}\,\int_{\e^{c_-}}^{\e^{c_+}} {\dd w}~
\frac{w}{\sqrt{(w-\e^{c_-})(\e^{c_+}-w)(\e^{d_+}-w)(w-\e^{d_-})}}
\\ && \qquad =~
\frac{2 ~\e^{-{2 t^2}/{a}}}{\sqrt{(\e^{d_+}-\e^{c_-})
(\e^{c_+}-\e^{d_-})}}\,\left[(\e^{c_-}-\e^{d_-})\,\Pi\left(
\mbox{$\frac{\e^{c_+}-\e^{c_-}}{
\e^{c_+}-\e^{d_-}}$},k\right)+\e^{d_-}\, K(k)\right] \ . \nonumber
\eea

The next integral to be computed is \bea \int_0^{\infty} \dd w
~\frac{1}{ \sqrt{(w+\e^{c_-})(w+\e^{c_+})(w+\e^{d_+})(w+\e^{d_-})}}
=\frac{2\bigl(F(\nu_\infty,k)-F(\nu_0,k)\bigr)}
{\sqrt{(\e^{d_+}-\e^{c_-})(\e^{c_+}-\e^{d_-})}}\nonumber\\
\eea with \beq \!\!\!\!\!
\nu_0=\arcsin\left(\mbox{$\sqrt{\frac{(\e^{d_+}-\e^{c_-})
~\e^{d_-}}{(\e^{d_+}-\e^{d_-})~\e^{c_-}}}$}~\right)= \arcsin
\left(\mbox{${\sqrt{\frac{1}{2} + \frac{c}{2\,d}}}$}~ \right)-
{\sqrt{d^2 - {c^2}}}\,\left(\frac{t}{4}+ \frac{ c \,t^2}{48}\right)
+O(t^3) \eeq and \beq
\nu_\infty=\arcsin\left(\mbox{$\sqrt{\frac{\e^{d_+}-\e^{c_-}}
{\e^{d_+}-\e^{d_-}}}$}~\right)= \arcsin
\left(\mbox{${\sqrt{\frac{1}{2} + \frac{c}{2\,d}}}$}~
\right)+{\sqrt{d^2 - {c^2}}}\,\left(\frac{t}{4}- \frac{ c
\,t^2}{48}\right)+O(t^3) \ . \eeq In the limit $t\to0$ we thus have
\beq F(\nu_\infty,k)-F(\nu_0,k)=\mbox{$\frac{1}{2}$}\,(c+d)\,
t+O(t^2) \ . \eeq

Finally, consider the integral \beq {\cal C}_0=\int_0^{\infty} \dd w
~\log(w)\,\frac{\dd}{\dd w} \left(\frac{1}{
\sqrt{(w+\e^{c_-})(w+\e^{c_+})(w+\e^{d_+})(w+\e^{d_-})}}\right) \ .
\eeq Instead of evaluating this, we shall evaluate ${\sf B}_1$.
${\cal
  C}_0$ may then be recovered through the identity
$\mathcal{C}_0=-\e^{-{2 t^2}/{a}}\,{({\sf B}_1+2 t^2/a)}$. We have
\bea {\sf B}_1&=&-\int_0^{\infty} \dd w ~\log(w)\,\frac{\dd}{\dd
w}\, \left(\frac{w^2}{
\sqrt{(w+\e^{c_-})(w+\e^{c_+})(w+\e^{d_+})(w+\e^{d_-})}}
\right)\nonumber\\ &=& \lim_{\epsilon\to
  0}\,\biggl(\log(\epsilon)+\int_0^{{1}/{\epsilon}}
\dd w~\frac{w}{
\sqrt{(w+\e^{c_-})(w+\e^{c_+})(w+\e^{d_+})(w+\e^{d_-})}}\biggr)\\
&=& \lim_{\epsilon\to 0}\,\left[\log(\epsilon)+\frac{2}
{\sqrt{(\e^{c_+}-\e^{d_-})(\e^{d_+}-\e^{c_-})}}\left\{(\e^{c_-}-\e^{d_-})
\left(\Pi\left(\nu_{{1}/{\epsilon}},\mbox{$
\frac{\e^{d_+}-\e^{d_-}}{\e^{d_+}-\e^{c_-}}$},k\right)
\right.\right.\right.\nonumber\\
&&-\left.\left.\left.\Pi\left(\nu_{0},\mbox{$
\frac{\e^{d_+}-\e^{d_-}}{\e^{d_+}-\e^{c_-}}$},k\right)\right\}-\e^{c_-}\,
\bigl(F(\nu_{{1}/{\epsilon}},k)-
F\left(\nu_{0},k\right)\bigr)\right)\right] \nonumber \eea where
\bea \nu_{{1}/{\epsilon}}=\arcsin\left(\mbox{$
\sqrt{\frac{(\e^{d_+}-\e^{c_-})(1+\epsilon~
\e^{d_-})}{(\e^{d_+}-\e^{d_-})(1+\epsilon ~\e^{c_-})}}$}~\right) \ .
\eea The elliptic integrals of the third kind appearing above are
hyperbolic of type~II~\cite{AS}, since \bea
n=\frac{\e^{d_+}-\e^{d_-}}{
\e^{d_+}-\e^{c_-}}=\frac{1}{\sin^2(\nu_\infty)}>1 \ . \eea We want
to reduce them to hyperbolic integrals of  type I having $0<n< k$.
This is achieved through the identity \beq
\Pi(\nu,n,k)=-\Pi\left(\nu,\mbox{$\frac{k^2}{n}$},k\right)+F(\nu,k)+\frac{1}{2
p_1}\,\log\left(\mbox{$\frac{\Delta(\nu)+p_1\tan(\nu)}{\Delta(\nu)-p_1\tan(\nu)}
$}\right) \eeq with $\Delta(\nu)=\sqrt{1-k^2\sin^2(\nu)}$ and
$p_1=\sqrt{(n-1)(1-\frac{k^2}{n})}$. Then \bea
\Pi(\nu_{{1}/{\epsilon}},n,k)-\Pi(\nu_0,n,k)~&\overset{\epsilon\to
0}{=}&~\Pi\left(\nu_0,\mbox{$\frac{k^2}{n}$},k\right)-
\Pi\left(\nu_\infty,\mbox{$\frac{k^2}{n}$},k\right)+
F(\nu_\infty,k)-F(\nu_0,k)\nonumber\\
&&+\,\frac{1}{2
p_1}\,\log\left(\mbox{$\frac{\Delta(\nu_{{1}/{\epsilon}})+p_1
\tan(\nu_{{1}/{\epsilon}})}{\Delta(\nu_{{1}/{\epsilon}})-p_1
\tan(\nu_{{1}/{\epsilon}})}\,
\frac{\Delta(\nu_0)-p_1\tan(\nu_0)}{\Delta(\nu_0)+p_1\tan(\nu_0)}$}
\right) \ . \eea

The argument of the logarithm to first order in $\epsilon$ is \beq
\frac{\Delta(\nu_{{1}/{\epsilon}})+p_1
\tan(\nu_{{1}/{\epsilon}})}{\Delta(\nu_{{1}/{\epsilon}})-p_1
\tan(\nu_{{1}/{\epsilon}})}\,
\frac{\Delta(\nu_0)-p_1\tan(\nu_0)}{\Delta(\nu_0)+p_1\tan(\nu_0)}
=\frac{\e^{-t \left( c + d \right)+\frac{t^2 }{a}}}
  {4}\,\left( \e^{c\,t} + \e^{d\,t} \right) \,\left( 1 +
\e^{\left( c + d \right) \,t} \right)
  \,\epsilon \ .
\eeq Since \beq
\frac{2\left(\e^{c_-}-\e^{d_-}\right)}{\sqrt{(\e^{d_+}-\e^{c_-})
(\e^{c_+}-\e^{d_-})}}\,\frac{1}{2 p_1}=-1 \eeq the $\epsilon$
dependence in the argument of the logarithm cancels with the
subtraction appearing in the definition of the integral. We can
therefore safely drop it. The original integral is then equal to
\bea {\sf
B}_1&=&\frac{2}{\sqrt{(\e^{c_+}-\e^{d_-})(\e^{d_+}-\e^{c_-})}}\,
\left((\e^{c_-}-\e^{d_-})
\left\{\Pi\left(\nu_{0},\mbox{$\frac{\e^{c_+}-\e^{c_-}}
{\e^{c_+}-\e^{d_-}}$},k\right)\right.\right.\nonumber\\
&&-\left.\left.\Pi\left(\nu_{\infty},\mbox{$
\frac{\e^{c_+}-\e^{c_-}}{\e^{c_+}-\e^{d_-}}$},k\right)\right\}-\e^{d_-}\,
\bigl(F(\nu_{\infty},k)-
F\left(\nu_{0},k\right)\bigr)\right)\nonumber\\
&&-\log\left(\mbox{$\frac{\e^{-t \left( c + d \right)+{t^2 }/{a}}}
  {4}$}\,( \e^{c\,t} + \e^{d\,t}) \,( 1 + \e^{\left( c + d \right) \,t}
  )\right)
\eea and thus \bea \mathcal{C}_0 &=&-\e^{-{2 t^2}/{a}}\,\left[
\frac{2}{\sqrt{(\e^{c_+}-\e^{d_-})(\e^{d_+}-\e^{c_-})}}\left((\e^{c_-}-\e^{d_-})
\left\{\Pi\left(\nu_{0},\mbox{$\frac{\e^{c_+}-\e^{c_-}}{\e^{c_+}-\e^{d_-}}$}
,k\right)-\right.\right.\right.\nonumber\\
&&-\left.\left.\Pi\left(\nu_{\infty},\mbox{$\frac{\e^{c_+}-\e^{c_-}}
{\e^{c_+}-\e^{d_-}}$},k\right)\right\}-\e^{d_-}\,
\bigl(F(\nu_{\infty},k)-
F\left(\nu_{0},k\right)\bigr)\right)\nonumber\\
&&-\left.\log\left(\mbox{$\frac{1}
  {4}$}( \e^{d_-} + \e^{c_-} ) \,( \e^{-d_+-d_-}+ \e^{-c_- -d_-
})\right)\right] \ . \eea

\section{Series expansions}

\subsection{Saddle-point equation}

We will now derive eq.~(\ref{nos1}) from eq.~(\ref{pirippo}). It
will prove useful to introduce the modular parameter
$q=\exp\left(-\pi\, K(k^\prime)/K(k)\right)$. Let us consider the
 first term on the RHS of eq.~(\ref{pirippo}). By using the
 expansions
\begin{eqnarray}
\log\left(\cn\bigl(\mbox{$\frac{2 K(k)}{p}$}, \, k\bigr)\right)&=&
\log\left(\cos\bigl(\mbox{$\frac{\pi}{p}$}\bigr)\right)-
4\,\sum_{n=1}^\infty  \frac{1}{n}\,\frac{q^n}{1+(-1)^n\, q^n}\,
\sin^2\left(\mbox{$\frac{n\,\pi}{p}$}\right) \ , \nonumber\\
 \log\left(\dn\bigl(\mbox{$\frac{2
K(k)}{p}$}, \, k\bigr)\right)&=& -8\,\sum_{n=1}^\infty\frac{1}{2
n-1}\, \frac{q^{2n-1}}{1-q^{2(2n-1)}}\,
\sin^2\left(\mbox{$\frac{(2n-1) \, \pi}{p}$}\right) \ ,
\end{eqnarray}
it can be rewritten as
\begin{eqnarray}
\log\left(\mbox{$\frac{\dn\left(\frac{2 K(k)}{p},k\right)}
{\cn^2\left(\frac{2 K(k)}{p},k\right)}$}\right)&=&
-2\log\left(\cos\bigl(\mbox{$\frac{\pi}{p}$}\bigr)\right)+
4\,\sum_{n=1}^\infty  \frac{1}{n}\,\frac{q^{2 n}}{1+ q^{2 n}}\,
\sin^2\left(\mbox{$\frac{2 n\,\pi}{p}$}\right)\nonumber\\ &&+\,
8\,\sum_{n=1}^\infty  \frac{1}{2 n-1}\,\frac{q^{2(2 n-1)}} {1-
q^{2(2n-1)}}\sin^2\left(\mbox{$\frac{(2 n-1) \, \pi}{p}$}\right) \
.
\end{eqnarray}
To expand the elliptic integral of the third kind in the second
term on the RHS of eq.~(\ref{pirippo}) requires a bit more of
work. We introduce a set of auxiliary parameters given by
\begin{eqnarray}
\epsilon&=&\arcsin\left(\mbox{$\frac{\sqrt{n}}{k}$}~\right)
=\arcsin\left(\sn\bigl(\mbox{$\frac{\hat x}{2}$},k\bigr)\right) \ , \\
\beta&=&\frac{\pi}{2}\,\frac{F(\epsilon,k)}{K(k)}=\frac{\pi}{2}\,
\frac{F\left[ \, \arcsin\left(\sn\left(\frac{\hat
x}{2},k\right)\right),k \, \right]}{K(k)}= \frac{\pi}{2
K(k)}\,\frac{\hat x}{2}=\frac{\pi}{2}\, \frac{ p+2}{2p} \ ,
\\
v&=&\frac{\pi}{2}\,\frac{F(\nu,k)}{K(k)} \ , \\
\delta_1&=&\sqrt{\frac{n}{(1-n)(k^2-n)}}=
\frac{\sn\left(\frac{\hat x}{2},k\right)}{\dn\left(\frac{\hat
x}{2},k\right) \cn\left(\frac{\hat x}{2},k\right)} \ .
\end{eqnarray}
A generic elliptic integral of the third kind with $0\le n\le k^2$
can be represented as
\begin{equation} \label{Celliptic}
\Pi(\nu,n,k)=\delta_1\left[ \, -\mbox{$\frac{1}{2}\,\log\left(
\frac{\vartheta_4(v+\beta)}{\vartheta_4(v-\beta)}\right)$}+v\,
\frac{\vartheta_1^\prime(\beta)}{\vartheta_1(\beta)} \, \right] \
.
\end{equation}
In the following we will need the expansions in powers of $q$ of
the two terms in the RHS of eq.~(\ref{Celliptic})
\begin{eqnarray} \mbox{$\frac{1}{2}\,\log\left(\frac{\vartheta_4(v+\beta)}
{\vartheta_4(v-\beta)}\right)$} &=& 2\,\sum_{n=1}^\infty
\frac{q^n}{n \, (1-q^{2n})} \,\sin( 2 \, n \, v)\,\sin( 2 \, n \,
\beta) \ ,
\nonumber\\
\frac{\vartheta_1^\prime(\beta)}{\vartheta_1(\beta)} &=&
\cot(\beta)+4\sin( 2 \, \beta)\,
\sum_{n=1}^\infty\frac{q^{2n}}{1-2 \, \cos(2 \,
\beta)\,q^{2n}+q^{2n}} \ .
\end{eqnarray}
By using these expansions we can simplify the relevant combination
of elliptic integrals in eq.~(\ref{pirippo}) as
\begin{eqnarray} \label{Celliptic2}
\Pi(\nu_\infty,&n&,k)-\Pi(\nu_0,n,k)-\frac{2}{p}\Pi(n,k) \nonumber\\
&=&\delta_1\left[-2\sum_{n=1}^\infty\frac{q^n}{n \, (1-q^{2n})}
\left[\sin( 2 \, n \, v_\infty)-\sin(2 \, n \, v_0) -\frac{2}{p}
\sin{\pi \, n}\right]\sin( 2 \, n \, \beta) \right.
\nonumber\\
&& \left. + \, (v_\infty-v_0-\frac{\pi}{p})\left(
\cot(\beta)+4\sin( 2 \,
\beta)\sum_{n=1}^\infty\frac{q^{2n}}{1-2\cos(2 \,
\beta)q^{2n}+q^{2n}}\right)\right]
\nonumber\\
&=& -4 \, \delta_1\sum_{n=1}^\infty\frac{q^n}{n \, (1-q^{2n})}
\sin\left( \frac{\pi \, n}{p} \right)\cos\left(\frac{\pi \,
n}{2}\right)
\sin( n \, \pi \, \frac{p+2}{2 \, p} )= \nonumber\\
&=& -2 \, \delta_1\sum_{n=1}^\infty\frac{q^{2 n}}{n \, (1-q^{4
n})}\sin^2\left( \frac{2 \, \pi \, n}{p} \right) \ ,
\end{eqnarray}
where the second equality holds due to the identities
\begin{eqnarray} \label{Cidentities}
v_\infty-v_0 &=& \frac{\pi}{2 K(k)}(F(\nu_\infty,k)-F(\nu_0,k))=
\frac{\pi}{p} \ ,
\nonumber\\
v_\infty+v_0&=&\frac{\pi}{2
K(k)}(F(\nu_\infty,k)+F(\nu_0,k))=\frac{\pi}{2} \ .
\end{eqnarray}
and the last equality is a consequence of the fact that only even
$n$ contributes.

Finally $\delta_1$ in eq.~(\ref{Celliptic2}) will appear in
eq.~(\ref{pirippo}) only in the combination
\begin{eqnarray} \label{Celliptic3}
\delta_1\frac{k^\prime\cn(x,k)}{1+\sn(x,k)}=\frac{k^\prime\cn(x,k)}{1+\sn(x,k)}
\frac{\sn\left(\frac{\hat x}{2},k\right)}{\dn\left(\frac{\hat
x}{2},k\right) \cn\left(\frac{\hat x}{2},k\right)}=
\frac{{k^\prime}\left(\frac{1}{\cn(x,k)}-\frac{\sn(x,k)}{\cn(x,k)}\right)}{
\frac{\dn\left(\frac{\hat x}{2},k\right) \cn\left(\frac{\hat
x}{2},k\right)}{\sn\left(\frac{\hat x}{2},k\right)}} \ .
\end{eqnarray}
To expand in powers of $q$ we expand separately the numerator and
the denominator of the last expression in eq.~(\ref{Celliptic3}):
\begin{eqnarray} \frac{\dn\left(\frac{\hat
x}{2},k\right) \cn\left(\frac{\hat
x}{2},k\right)}{\sn\left(\frac{\hat x}{2},k\right)} &=&
\frac{\pi}{2K(k^2)}\left[\cot\frac{\pi(p+2)}{4 p}-
4\sum_{n=1}^\infty \frac{q^n}{1+q^n}\sin \frac{n\pi(p+2)}{2
p}\right] \ , \nonumber\\
%
%
k^\prime\left(\frac{1}{\cn(x,k)}-\frac{\sn(x,k)}{\cn(x,k)}\right)
&=& \frac{\pi}{2K(k^2)}\left[\frac{1}{\cos\frac{\pi}{p}}
-\tan\frac{\pi}{p}-
4\sum_{n=1}^\infty \frac{q^{n}}{1+q^{n}}\sin \frac{n\pi(p+2)}{2p}\right] \nonumber\\
&=& \cot\frac{\pi(p+2)}{4 p}- 4\sum_{n=1}^\infty
\frac{q^{n}}{1+q^{n}}\sin \frac{n\pi(p+2)}{2p} \ .
\end{eqnarray}
Collecting all together, we have\begin{eqnarray}
&&\frac{\displaystyle{{k^\prime}\left(\frac{1}{\cn(x,k)}-\frac{\sn(x,k)}{\cn(x,k)}\right)}}{\displaystyle{
\frac{\dn\left(\frac{\hat x}{2},k\right) \cn\left(\frac{\hat
x}{2},k\right)}{\sn\left(\frac{\hat x}{2},k\right)}}}=1 \ .
\end{eqnarray}
Finally we conclude that the second saddle point equation
(\ref{pirippo}) admits the following $q$-expansion
\begin{eqnarray}
\frac{t}{4}&=&\frac{p}{4}\log\left(\frac{\dn(x,k)}{\cn^2(x,k)}\right)+\frac{p}{2}\frac{k^\prime\cn(x,k)}{1+\sn(x,k)}\left[\Pi\left(
\nu_\infty,n,k\right)-  \Pi\left(
\nu_0,n,k\right)-\frac{2}{p}  \Pi\left(n,k\right) \right] \nonumber\\
&=&\frac{t_c}{4}+p \sum_{n=1}^\infty  \frac{1}{n}\frac{q^{2 n}}{1+
q^{2 n}}\sin^2\left(\frac{2 \, \pi \, n}{p}\right)+ 2 p
\sum_{n=1}^\infty  \frac{1}{2 n-1}\frac{q^{2(2 n-1)}}{1- q^{2(2n-1)}}\sin^2
\left( \frac{(2 n-1) \, \pi}{p} \right) \nonumber\\
&&-p\sum_{n=1}^\infty\frac{q^{2 n}}{n(1-q^{4 n})}\sin^2 \left(
\frac{2 \, \pi \, n}{p} \right) \nonumber\\
&=&\frac{t_c}{4}-2 p \sum_{n=1}^\infty  \frac{1}{2 \, n}\frac{q^{4
n}}{1-q^{4 n}}\sin^2\left(\frac{2 \, \pi \, n}{p}\right)+2 p
\sum_{n=1}^\infty  \frac{1}{2 n-1}\frac{q^{2(2 n-1)}}{1-
q^{2(2n-1)}}\sin^2\left(\frac{(2 n-1) \, \pi}{p}\right) \nonumber\\
&=& \frac{t_c}{4}-2 p \sum_{n=1}^\infty \frac{(-1)^n}{n}\frac{q^{2
n}}{1-q^{2 n}}\sin^2\left(\frac{\pi \, n}{p}\right) , \nonumber
\end{eqnarray}
that leads us to eq.~(\ref{nos1}).

\subsection{Endpoints}

We want to provide an expansion of the endpoints $c$ and $d$ in
terms of the modular parameter $q$. This can be obtained by
considering the system of equations given by eq.~(\ref{fava})
\begin{eqnarray}
\label{favabis}
c+d&=&\frac{2}{t}\mathrm{arctanh}\left(\mathrm{sn}\left(\frac{2 t K(k)}{a},k\right)\right) \nonumber\\
&=&\frac{2}{t}\mathrm{arctanh}\left( \frac{2\pi}{k \, K(k)}
\sum_{n=1}^\infty \frac{q^{n-1/2}}{1-q^{2 n-1}}
\sin\left(\frac{(2n-1) \, \pi}{p}\right)\right) \ ,
\end{eqnarray}
and by
\begin{eqnarray} \label{favater}
c-d&=& \frac{2}{t}\mathrm{arctanh}\left(\cn\left(\frac{p+2}{p}K(k),k\right)\right) \nonumber\\
&=& \frac{2}{t}\mathrm{arctanh}\left( \frac{2\pi}{k K(k)}
\sum_{n=1}^\infty\frac{ (-1)^n q^{n-1/2}}{1+q^{2
n-1}}\sin\left(\frac{(2n-1)\pi}{p}\right)\right) \ .
\end{eqnarray}
Solving for the endpoints we obtain
\begin{eqnarray}
d&=&\frac{1}{t} \, \mathrm{arctanh}\left( \frac{2\pi}{k K(k)}
\sum_{n=1}^\infty \frac{q^{n-1/2}}{1-q^{2 n-1}}
\sin\left(\frac{(2n-1)\pi}{p}\right)\right)
\nonumber\\
&&  - \frac{1}{t} \, \mathrm{arctanh}\left( \frac{2\pi}{k K(k)}
\sum_{n=1}^\infty\frac{ (-1)^n q^{n-1/2}}{1+q^{2
n-1}}\sin\left(\frac{(2n-1)\pi}{p}\right)\right) \ , \\
c&=&\frac{1}{t} \, \mathrm{arctanh}\left( \frac{2\pi}{k K(k)}
\sum_{n=1}^\infty \frac{q^{n-1/2}}{1-q^{2 n-1}}
\sin\left(\frac{(2n-1)\pi}{p}\right)\right)
\nonumber\\
&&  + \frac{1}{t} \, \mathrm{arctanh}\left( \frac{2\pi}{k K(k)}
\sum_{n=1}^\infty\frac{ (-1)^n q^{n-1/2}}{1+q^{2
n-1}}\sin\left(\frac{(2n-1)\pi}{p}\right)\right) \ .
\end{eqnarray}
These are expansions in powers of $q$ since
\begin{equation}
\frac{k \, K(k)}{2\pi}=\left(\sum_{n=1}^\infty q^{\left(\frac{2
n-1}{2}\right)^2}\right)^2 \ .
\end{equation}

\subsection{Distribution function}

The goal of this appendix is to provide an expansion in terms of
the modular parameter $q$ for the distribution function, that is
to derive eq.~(\ref{rhoexpansion}) from eq.~(\ref{rhoelliptic}).

Let us start with the difference between the two integrals
\begin{equation} \label{difference}
 \Pi\left(\nu_\infty,\frac{(\e^{d_+}-\e^{d_-})
(z-\e^{c_-})}{(\e^{d_+}-\e^{c_-})(z-\e^{d_-})},k\right)-
\Pi\left(\nu_0,\frac{(\e^{d_+}-\e^{d_-})(z-\e^{c_-})}{(\e^{d_+}-\e^{c_-})(z-\e^{d_-})},k\right),
\end{equation}
where $z\in[\e^{c_+},\e^{d_+}]$. For these values of $z$ they are
elliptic integrals of circular type since the parameter
\begin{equation}
n(z)=\frac{(\e^{d_+}-\e^{d_-})(z-\e^{c_-})}{(\e^{d_+}-\e^{c_-})(z-\e^{d_-})}
\end{equation}
always belongs to the region $k^2\le n\le 1$.

They admit the following $q$-expansion \cite{AS}:
\begin{equation}
\Pi(\nu, n,k)=\delta_2 \left(\lambda(\beta) -\mu(\beta) \,
v\right) \ ,
\end{equation}
where
\begin{eqnarray}
\lambda(\beta) &=& \arctan\left(\tanh\beta\tan
v\right)-2\sum_{n=1}^\infty \frac{(-1)^s}{s}\frac{q^{2 s}}{1-q^{2
s}} \sin (2 \, s \, v)\sinh(2 \, s \, \beta) \ , \nonumber\\
\mu(\beta) &=& \frac{\theta_3^\prime(\ii \,
\beta,q)}{4\theta_3(\ii \, \beta,q)} =
\frac{\displaystyle{\sum_{s=1}^\infty \, s \, q^{s^2}\sinh(2 \, s
\, \beta)}}{\displaystyle{\sum_{s=-\infty}^\infty \,
q^{s^2}\cosh(2 \, s \, \beta) }} \ .
\end{eqnarray}
We have introduced the following set of parameters
\begin{eqnarray}
\epsilon &=& \arcsin\left(\sqrt{\frac{1-n(z)}{1-k^2}}\right) \ ,
\nonumber\\
\beta &=& \frac{\pi}{2}\frac{F(\epsilon,k^\prime)}{K(k)} \ , \nonumber\\
v &=& \frac{\pi}{2} \frac{F(\nu,k)}{K(k)} \ , \nonumber\\
\delta_2 &=& \sqrt{\frac{n(z)}{(1-n(z))(n-k^2)}} \ .
\end{eqnarray}
From the explicit form
\begin{equation}
\delta_2 = \frac{\sqrt{\left( {\e^{c_+}} - {\e^{d_-}}
\right)\left( {\e^{d_+}}-{\e^{c_-}}
      \right)}}{{\left(
         {\e^{c_-}} - {\e^{d_-}} \right) }}\sqrt{\frac{
      \, \,\left( z-{\e^{c_-}} \right) \,
    \left( z-{\e^{d_-}} \right) }{\
     \left( \e^{d_+} -z\right) \,
    \left(z -\e^{c_+} \right) }} \ ,
\end{equation}
we see that the inverse of $\delta_2$ appears as a common
prefactor in the distribution function (\ref{rhoelliptic}).
 Moreover in our case
\begin{eqnarray}
\beta&=&\frac{\pi}{2\K(k)}F\left(\arcsin\left(\sqrt{\frac{\left(
{\e^{c_+}} - {\e^{d_-}} \right)
      \,\left( {\e^{d_+}} - z \right) }{
    \left({\e^{d_+}}- {\e^{c_+}}  \right) \,
    \left( z-{\e^{d_-}}  \right) }}\right),k^\prime\right) \nonumber\\
    &=&\frac{\pi}{2\K(k)}\sn^{-1}\left(\sqrt{\frac{\left( {\e^{c_+}} - {\e^{d_-}} \right)
      \,\left( {\e^{d_+}} - z \right) }{
    \left({\e^{d_+}}- {\e^{c_+}}  \right) \,
    \left( z-{\e^{d_-}}  \right) }},k^\prime\right) \equiv \beta_1 \ .
\end{eqnarray}
Collecting all these ingredients we can rewrite the
eq.~(\ref{difference}) as
\begin{eqnarray}
&&\!\!\!\!\!\!\!\!\!\!\!\!\!\!\Pi\left(\nu_\infty,\frac{(\e^{d_+}-\e^{d_-})
(z-\e^{c_-})}{(\e^{d_+}-\e^{c_-})(z-\e^{d_-})},k\right)-
\Pi\left(\nu_0,\frac{(\e^{d_+}-\e^{d_-})(z-\e^{c_-})}{(\e^{d_+}-\e^{c_-})(z-\e^{d_-})},k\right)  \nonumber\\
&=& \delta_2 \biggl(\arctan  \left(\tanh\beta_1\tan
v_\infty\right)-\arctan\left(\tanh\beta_1\tan v_0\right)
- 4 \, \mu \, (v_\infty-v_0)  \nonumber\\
&&  - \left.2\sum_{n=1}^\infty \frac{(-1)^s}{s}\frac{q^{2
s}}{1-q^{2 s}} \, \big( \sin (2 \, s \, v_\infty)-\sin (2 \, s \,
v_0) \big) \,
\sinh(2 \, s \, \beta_1) \right) \\
&=&  \delta_2\left( \arctan \left( \tan\frac{\pi}{p}\tanh(2 \,
\beta_1)\right)-4 \frac{\pi}{p} \, \mu(\beta_1)
-2\sum_{s=1}^\infty \frac{1}{s}\frac{q^{4 s}}{1-q^{4 s}} \sin
\left(\frac{2 \, \pi \, s}{p}\right)\sinh(2 \, s \, \beta_1)
\right) \ ,\nonumber
\end{eqnarray}
where we have used eqs.~(\ref{Cidentities}).

The last term that we have to expand is the complete elliptic
integral $\Pi(n,k)$
\begin{equation}
\Pi \left(
\frac{(\e^{c_+}-\e^{c_-})(z-\e^{d_-})}{(\e^{c_+}-\e^{d_-})(z-\e^{c_-})},k
\right) - K(k) = \hat\delta_2 \, \frac{\pi}{2} \, \left(
1-\Lambda(\eta,k) \right) \ ,
\end{equation}
where
\begin{equation}
\eta = \arcsin \left( \, \sqrt{\frac{1-\hat n(z)}{1-k^2}} \,
\right) = \arcsin\left( \, \sqrt{
\frac{(\e^{d_+}-\e^{c_-})(z-\e^{c_+})}
{(\e^{d_+}-\e^{c_+})(z-\e^{c_-})}} \, \right) \ ,
\end{equation}
with $\hat
n(z)=\frac{(\e^{c_+}-\e^{c_-})(z-\e^{d_-})}{(\e^{c_+}-\e^{d_-})(z-\e^{c_-})}$
. Since $\hat n(z)=k^2/n(z)$, we immediately see that
\begin{equation}
 \hat \delta_2=\sqrt{\frac{\hat
n(z)}{(1-\hat n(z))(\hat n(z)-k^2)}}=\delta_2 \ .
\end{equation}
The function $\Lambda(\eta,k)$ appearing in the expansion of $\Pi$
is the Heuman lambda-function defined by \cite{AS}
\begin{equation}
\Lambda(\eta,k)=\frac{2}{\pi}\left(E(\eta,k^\prime) \,
K(k)-(K(k)-E(k)) \, F(\eta,k^\prime)\right) \ .
\end{equation}
The distribution function $\rho$ can be finally written in the
form of eq.~(\ref{rhoexpansion})
\begin{eqnarray}
\rho(z) &=& \frac{2}{t \, \pi \,
z}\left(\frac{\pi}{2}-\frac{\pi}{2}
\Lambda(\eta,k)+\frac{p}{2}\arctan\left(\tan\frac{\pi}{p}\tanh(2
\, \beta_1)\right)- 2 \, \pi \, \mu(\beta_1)
\right. \nonumber\\
&&\left.-\sum_{s=1}^\infty \frac{p}{s}\frac{q^{4 s}}{1-q^{4 s}}
\sin \left(\frac{2\pi s}{p}\right)\sinh(2 s\beta_1)\right).
\end{eqnarray}
This representation obviously holds for $z\in[\e^{c_+},\e^{d_+}]$.
We are interested in the behavior at the transition point $t_c$,
where $q=k=c=0$. The critical values of the parameters that enter
in the distribution function are
\begin{eqnarray}
\beta^c_1 &=& F\left(\arcsin\left(\e^{-d_c t_c/2}\sqrt{\frac{
      \,\left( {\e^{d_{c+}}} - z \right) }{ \,
    \left( z-{\e^{d_{c-}}}  \right) }}\right),1\right) \nonumber\\
    &=& \mathrm{arctanh}\left(\e^{-d_c t_c/2}\sqrt{\frac{
      \,\left( {\e^{d_{c+}}} - z \right) }{ \,
    \left( z-{\e^{d_{c-}}}  \right) }}\right) \, \nonumber\\
\eta^c &=& \frac{\pi}{2} \ .
\end{eqnarray}
The second identity implies that $\Lambda(\eta^c,k)=1$. From these
results it is possible to derive the distribution function at the
critical point, eq.~(\ref{rhocritical}).


\end{document}